\documentclass[aps,prd,longbibliography,nofootinbib]{revtex4-2}

\usepackage{graphicx}
\usepackage{amsmath,amssymb,bm}
\usepackage{siunitx}
\DeclareSIUnit\year{yr}
\DeclareSIUnit\liter{L}
	
\usepackage{comment}
\usepackage{booktabs}
\usepackage{microtype}
\usepackage{natbib}
\usepackage{hyperref}
\hypersetup{colorlinks=true,allcolors=blue}

\usepackage{array,tabularx}
\usepackage{makecell}
\newcolumntype{Y}{>{\raggedright\arraybackslash}X}
\newcolumntype{L}{>{\raggedright\arraybackslash}X}

\newcommand{\Rmars}{R_{\mars}}
\newcommand{\mars}{\mathrm{Mars}}
\newcommand{\gMars}{g_{\mars}}
\newcommand{\SMars}{S_{\mars}}
\newcommand{\Abond}{A_{\mathrm{B}}}
\newcommand{\sigSB}{\sigma_{\mathrm{SB}}}

\newcommand{\Ps}{P_{\mathrm{s}}}
\newcommand{\Te}{T_{\mathrm{e}}}
\newcommand{\Ts}{T_{\mathrm{s}}}
\newcommand{\DeltaF}{\Delta F}
\newcommand{\Mair}{M_{\mathrm{atm}}}

\newcommand{\tauIR}{\tau_{\mathrm{IR}}}
\newcommand{\tauIReff}{\tau_{\mathrm{IR,eff}}}
\newcommand{\winfac}{w_{\mathrm{win}}}      
\newcommand{\deff}{d_{\mathrm{eff}}}       

\usepackage{pgfplots}
\pgfplotsset{compat=1.18}
\usepackage{threeparttable}
\usepackage{multirow}
\usepackage{mathtools}

\newcommand{\PiM}{\Pi_M}
\newcommand{\PiF}{\Pi_F}
\newcommand{\PiQ}{\Pi_{\dot M}}
\newcommand{\PiP}{\Pi_P}
\newcommand{\PiS}{\Pi_S}
\newcommand{\Pimin}{\Pi_{\min}}
\newcommand{\Lmaint}{\Lambda_{\mathrm{maint}}}
\newcommand{\tbuild}{t_{\mathrm{build}}}
\newcommand{\tloss}{\tau_{\mathrm{loss}}}

\newcommand{\Kmars}{\mathcal{K}_{\mars}}
\newcommand{\Mreq}{M_{\mathrm{req}}}
\newcommand{\Mavail}{M_{\mathrm{avail}}}
\newcommand{\Preq}{P_{\mathrm{req}}}
\newcommand{\Pavail}{P_{\mathrm{avail}}}
\newcommand{\Qreq}{\dot M_{\mathrm{req}}}
\newcommand{\Qavail}{\dot M_{\mathrm{avail}}}
\newcommand{\Qloss}{\dot M_{\mathrm{loss}}}
\newcommand{\Qrepl}{\dot M_{\mathrm{repl}}}
\newcommand{\Areg}{A_{\mathrm{reg}}}
\newcommand{\Preg}{P_{\mathrm{reg}}}

\begin{document}

\title{Terraforming Mars: Mass, Forcing, and Industrial Throughput Constraints}

\author{Slava G. Turyshev}
\affiliation{
Jet Propulsion Laboratory, California Institute of Technology,\\
4800 Oak Grove Drive, Pasadena, CA 91109-0899, USA
}%

\date{\today}

\begin{abstract}
Terraforming Mars can be evaluated with a small set of system-level feasibility constraints linking (i) target pressures and compositions to required atmospheric inventories, (ii) target surface temperatures to the required radiative control, (iii) inventories and climate agents to sustained industrial throughput and power over a build time, and (iv) persistence against collapse, escape, and geochemical sinks. We use transparent order-of-magnitude scalings to compare endogenous CO$_2$ release, synthetic super-greenhouse gases, CO$_2$--H$_2$ collision-induced absorption, engineered aerosols and nanoparticles, orbital mirrors, and regional paraterraforming. We find: (1) human-relevant pressures imply exaton-class inventories, because Mars requires \(3.89\times10^{15}\,\mathrm{kg}\) of atmosphere per mbar of global mean surface pressure; (2) accessible endogenous CO$_2$ is best treated as an order-of-tens-of-mbar resource, with a representative \(20\) mbar case yielding less than \(10\) K warming under present insolation; (3) mean surface temperatures of \(250\)--\(273\) K require either effective infrared opacity of order \(2\)--\(4\) or direct absorbed-solar forcing of order \(10^2~\mathrm{W\,m^{-2}}\), the latter implying reflector areas of order \(10^{13}\)--\(10^{14}\,\mathrm{m^2}\) together with large deployment, control, durability, and replacement burdens; and (4) breathable open-surface endpoints are dominated by oxygen and buffer-gas inventories and by a minimum oxygenation work exceeding \(10^{25}\) J, implying average build rates of \(10^{7}\)--\(10^{8}\,\mathrm{kg\,s^{-1}}\) and average power from multi-\(10^2\) TW to PW for century-to-millennial build times before inefficiencies and sink filling. We conclude that regional habitability gains via paraterraforming and covered-area strategies are plausible on near-term industrial scales, whereas no currently credible open-atmosphere pathway is identified to a Mars permitting pressure-unassisted human exposure or a breathable surface atmosphere without exaton-class volatile supply, multi-century planetary industry, and sustained climate actuation, retention, durability management, and replacement against sinks and loss.

\end{abstract}

\maketitle


\section{Introduction}
\label{sec:intro}

Terraforming Mars denotes deliberate, sustained modification of Mars' surface environment---atmospheric pressure and composition, radiative balance, and water stability---toward states that support progressively less technologically mediated surface habitability. The concept is widely discussed, but feasibility is governed by a compact set of planet-scale feasibility constraints that any proposed pathway must satisfy simultaneously.

We use four governing constraints as a common quantitative basis for comparing mechanisms:
(i) \emph{mass inventory constraint}, linking target pressures and partial pressures to required volatile inventories;
(ii) \emph{radiative-balance constraint}, linking target surface temperatures to required top-of-atmosphere (TOA) forcing $\Delta F_{\rm TOA}$ and/or effective longwave optical depth $\tau_{\rm IR}$;
(iii) \emph{throughput-and-power constraint}, linking inventories and radiative agents to required sustained throughput $\dot M$ and power $P$ over specified build times; and
(iv) \emph{stability/retention constraint}, requiring persistence against atmospheric collapse/condensation, escape to space, and geochemical sequestration. These simplified scalings are intentionally simple: they provide optimistic order-of-magnitude bounds that expose dominant bottlenecks and feasibility floors. Table~\ref{tab:acronyms} provides a list of acronyms used here.

This paper is an engineering-level synthesis. We avoid black-box climate and mission-design optimization tools and instead use transparent scalings to map proposals onto common, easily checked metrics $\{M,\ \tau_{\rm IR}/\Delta F_{\rm TOA},\ \dot M,\ P\}$. The results are therefore lower bounds and feasibility discriminators, not predictions of a specific Mars climate state; quantitative state prediction requires coupled 3-D GCM, aerosol microphysics, and photochemical/geochemical cycling.

Classic terraforming treatments \cite{McKay1982Terraforming,McKay1991MakingMarsHabitable,ZubrinMcKay1993Terraforming,Fogg1995Terraforming}
introduced key concepts and qualitative feasibility arguments, including early emphasis on warming Mars to enable oxygenic photosynthesis and longer-timescale biological transformation. A synthesis in \cite{McKayMarinova2001Habitable} explicitly linked the physics of warming Mars, the expected role of biology in longer-timescale oxygenation, and the environmental-ethics questions raised by making Mars habitable.
More recent work has expanded the design space to include limits on accessible CO$_2$ inventories \cite{JakoskyEdwards2018CO2Inventory,Tutolo2025Carbonates,Buhler2020Coevolution,BuhlerPiqueux2021Obliquity,Broquet2021SouthPolarCap}, regional solid-state greenhouse paraterraforming \cite{Wordsworth2019Aerogel}, self-sustaining living-habitat concepts and biologically generated enclosure materials \cite{WordsworthCockell2024LivingHabitats,Wordsworth2025Biomaterials}, engineered aerosol warming proposals \cite{Ansari2024Nanoparticles,Richardson2026IRParticles}, and renewed discussion of Mars-terraforming research as a coupled climate--biosphere--engineering problem \cite{DeBenedictis2025Case}. Work has also begun to formulate explicit \emph{research roadmaps} for assessing non-biological Mars-warming pathways, including local membranes, orbiting reflectors, and engineered aerosols \cite{Kite2026Roadmap}. 

The present paper is complementary: rather than prioritizing research sequencing or downselect criteria, it asks what mass, radiative, throughput, power, and retention conditions would have to be satisfied for specific atmospheric end states to be achieved and maintained. The contribution here is not a new warming mechanism; it is a unified systems/architecture trade framework that (a) keeps the governing constraints explicit and (b) reduces disparate proposals to common quantitative requirements in mass, forcing/opacity, throughput, and power. This framing highlights which constraint dominates for a given end state and makes key uncertainties (e.g., accessible nitrogen inventory, aerosol lifetimes, and sink capacities) explicit.

Three quantitative facts motivate our constraint-based framework. First, accessible endogenous CO$_2$ inventories are currently best described as being of \emph{order tens of mbar}, rather than as a secure multi-bar reservoir; a representative \(\sim 20\) mbar case yields $\lesssim 10$~K warming under present insolation \cite{JakoskyEdwards2018CO2Inventory,Tutolo2025Carbonates,Buhler2020Coevolution,BuhlerPiqueux2021Obliquity,Broquet2021SouthPolarCap}. Second, mechanisms with high radiative leverage at low added atmospheric mass split into two distinct industrial classes: short-lived aerosol pathways are maintenance-limited, whereas CO$_2$--H$_2$ CIA can be fill-dominated during $10^2$--$10^3$~yr global buildout even though it still requires replenishment on long hold times. Third, breathable endpoints are dominated by the required masses of O$_2$ and buffer gas (N$_2$/Ar) and by $\gtrsim 10^{25}$~J-class minimum energy requirements for oxygen production even at reversible limits.

The scope of this paper is deliberately narrower than the full terraforming literature.  That two-stage logic---abiotic environmental modification followed by longer-timescale biological transformation---was emphasized in \cite{McKayMarinova2001Habitable}; the present paper addresses the first stage quantitatively and treats the second mainly as a bounding comparator. In particular, biologically mediated oxygenation and ecosystem bootstrapping are important pathways in classic and recent Mars-terraforming discussions \cite{McKay1982Terraforming,Graham2004Biological,DeBenedictis2025Case}, but their quantitative analysis requires additional variables---net areal productivity, nutrient supply, burial efficiency, ecological stability, and sink competition---that are not developed here. The oxygenation numbers in this paper should therefore be read as lower bounds for \emph{abiotic} industrial oxygenation, not as a claim that biology is irrelevant.

\begin{table}[t]
\caption{Acronyms used in this paper.}
\label{tab:acronyms}
\centering
\small
\begin{tabular}{ll}
\toprule
Acronym & Meaning \\
\midrule
CIA & collision-induced absorption \\
GCM & general circulation model \\
IR & infrared \\
ISRU & in-situ resource utilization \\
LW & longwave \\
PFC & perfluorocarbon (super-greenhouse gas class) \\
PV & photovoltaics \\
SW & shortwave \\
TOA & top-of-atmosphere \\
\bottomrule
\end{tabular}
\end{table}

This paper is organized as follows: Section~\ref{sec:def} defines end states (E0--E4) and baseline Mars parameters. Section~\ref{sec:mass} develops  atmospheric mass bookkeeping and buffer gas requirements. Section~\ref{sec:thermal} establishes radiative requirements, including a minimal greenhouse mapping linking $\Te$ and $\Ts$. Section~\ref{sec:endogenous} bounds what can be achieved by mobilizing endogenous CO$_2$ and H$_2$O. Section~\ref{sec:warming} evaluates engineered and exogenous warming options (PFC-class gases, CO$_2$--H$_2$ CIA, engineered aerosols, mirrors/albedo).
Section~\ref{sec:oxygen} quantifies the mass and minimum energy requirements for oxygenation and breathable endpoints. 
Section~\ref{sec:loss} discusses loss, collapse, and sequestration constraints and introduces a minimal state-space control model. Section~\ref{sec:architecture} treats energy delivery and system architecture. Section~\ref{sec:discussion} discusses implications, industrial scaling, timescale--power trade, and cost floors; and Section~\ref{sec:concl} concludes.

\section{What ``terraforming Mars'' means in quantitative terms}
\label{sec:def}

\subsection{End states and success metrics}
\label{sec:Ei}

The term \emph{terraforming} is used loosely in the literature, but technical feasibility depends strongly on the endpoint being targeted. We therefore distinguish several \emph{end states}, labeled E0--E4, where the letter ``E'' denotes an \emph{end state} (equivalently, an endpoint label). These labels are bookkeeping shorthand, not additional physical variables. For technical analysis it is useful to define them with measurable criteria:
{}
\begin{itemize}
\item \textit{{\rm E0:} ``Robotic/human assisted operations'' (today)}. Current mean $\Ps \sim \SI{610}{Pa}$ and $\Ts \sim \SI{210}{K}$, requiring pressure suits and thermal control.

\item \textit{{\rm E1:} ``Recurrent open-surface liquid water''}. In the present climate, Mars can already cross the triple-point condition locally and transiently, so merely exceeding \SI{611.657}{Pa} is not by itself a useful progression criterion. In this paper, E1 denotes a more useful engineering target: recurrent or seasonally accessible \emph{surface} liquid water in some regions under open-sky conditions, which requires both pressure above the triple-point threshold and a local energy balance that can sustain liquid water for longer than brief instantaneous episodes \cite{Hecht2002Metastability,Schorghofer2020Crocus,LangeForget2026Gullies}. A minimum necessary condition is
\begin{equation}
\Ps > \SI{611.657}{Pa} \quad \text{at} \quad \SI{273.16}{K},
\end{equation}
following the triple-point relation for water \cite{IAPWS2011}, but this is far from sufficient for useful surface-water stability (Sec.~\ref{sec:water}).

\item \textit{{\rm E2:} ``Protected agriculture in enclosed/paraterraformed volumes''}. A representative benchmark is an \emph{internal} habitat pressure $P_{\rm in}\sim \SI{10}{-}\SI{30}{kPa}$, with the remaining thermal and compositional margins supplied by enclosure design and environmental control. For present Mars ambient pressure, such operating envelopes reduce the differential structural load by roughly a factor of \(\sim 3\) to \(\sim 10\) relative to a \SI{101}{kPa} habitat, depending on the chosen internal pressure. E2 is therefore primarily an enclosure and area-deployment problem rather than a global atmospheric-pressure problem.

\item \textit{{\rm E3:} ``Pressure-unassisted human exposure''}. The hard lower bound for avoiding ebullism is total pressure $\Ps \ge \SI{6.27}{kPa}$ (the Armstrong limit) \cite{NASAHabitableAtmosphere2023}. E3 is therefore a \emph{pressure threshold}, not a breathable-atmosphere criterion: a person may still require supplemental oxygen and low ambient $p_{\rm CO_2}$ to remain functional. Hypoxia, CO$_2$ toxicity, and decompression sickness impose stricter requirements on composition and partial pressures.

\item \textit{{\rm E4:} ``Breathable surface atmosphere''}. A useful distinction is between (i) lower-pressure O$_2$-rich operating envelopes that can be adequate for humans in controlled settings, and (ii) buffered, Earth-like atmospheres with $p_{\mathrm{O_2}} \sim \SI{21}{kPa}$ and tens of kPa of inert gas to control physiology, flammability, and broader ecological function. The latter remains the representative ``Earth-like'' benchmark used here.
A common benchmark is $p_{\mathrm{O_2}} \sim \SI{21}{kPa}$
with a buffer gas (N$_2$/Ar) to reach $\Ps \sim \SI{50}{-}\SI{100}{kPa}$ and to control fire risk and physiology. This endpoint is dominated by the required mass of O$_2$ and buffer gases (Sec.~\ref{sec:oxygen}).
\end{itemize}

Unless explicitly qualified otherwise, ``habitability'' in this paper means \emph{human-relevant surface habitability or protected biosphere utility}, not the broader astrobiological definition of any environment capable of supporting some form of life. Microbial or phototrophic habitability can occur under substantially different pressure, temperature, radiation, and compositional requirements \cite{WordsworthCockell2024LivingHabitats}.

For feasibility accounting, E4 should be decomposed rather than treated as a single colloquial endpoint. At minimum it combines: (i) a \emph{pressure} requirement, (ii) an \emph{oxygen} requirement, (iii) a \emph{buffer-gas} requirement, and (iv) a \emph{thermal} requirement. This decomposition matters because the dominant bottleneck is not the same for each sub-target. Warming can, in principle, be assisted by low-mass radiative agents, whereas breathable composition remains controlled by exaton-class O$_2$ and buffer-gas inventories. In other words, a proposal can satisfy the warm-state criterion while still failing the breathable-atmosphere criterion by many orders of magnitude in mass and energy.

Classic terraforming proposals and feasibility discussions span several decades and include early quantitative treatments of Mars warming and atmospheric engineering \cite{McKay1991MakingMarsHabitable,ZubrinMcKay1993Terraforming,Fogg1995Terraforming}. This paper emphasizes the distinction between (i) \emph{global atmospheric modification} and (ii) \emph{regional/paraterraforming} strategies that create habitable microclimates without altering
planetary-scale atmospheric mass.

We use representative mean values (Table~\ref{tab:marsparams}) for Mars physical and radiative parameters. Solar irradiance at Mars varies substantially between perihelion and aphelion; we use the
orbit-averaged top-of-atmosphere (TOA) value $\SMars \approx \SI{589}{W\,m^{-2}}$ and a mean planetary albedo $\Abond \approx 0.25$ \cite{Alexander2001MarsEnv}. These yield an effective radiating temperature ${\Te}_0 \approx \SI{210}{K}$, consistent with Mars' cold climate.

\begin{table}[t]
\caption{Nominal Mars parameters used in first-order calculations. }
\label{tab:marsparams}
\begin{tabular}{lcc}
\toprule
Quantity & Symbol & Value \\
\midrule
Mean radius & $\Rmars$ & $\SI{3389.5}{km}$ \\
Surface gravity & $\gMars$ & $\SI{3.71}{m\,s^{-2}}$ \\
Surface area & $4\pi\Rmars^2$ & $1.44\times 10^{14}\,\mathrm{m^2}$ \\
Mean TOA solar irradiance & $\SMars$ & $\SI{589}{W\,m^{-2}}$ \\
Bond albedo (mean) & $\Abond$ & $\sim 0.25$ \\
Effective radiating temperature\footnote{\(\sigma_{\rm SB}\) is the Stefan--Boltzmann constant.} & ${\Te}_0$ &
$\left[\SMars(1-\Abond)/(4\sigSB)\right]^{1/4}\approx \SI{210}{K}$ \\
\bottomrule
\end{tabular}
\end{table}

\subsection{Key results at a glance}
\label{sec:keyresults}

Table~\ref{tab:pressure_endpoints} lists reference pressure thresholds and representative atmospheric inventories relevant to the endpoint hierarchy developed in Sec.~\ref{sec:Ei}. Table~\ref{tab:keyresults} summarizes the requirement targets used throughout this paper as a compact ``dashboard'' linking end states (E0--E4) to (i) required inventories, (ii) radiative control targets, and (iii) implied industrial rates. This table makes explicit a central theme of the paper: as the endpoint advances from E2 to E4, the dominant constraint shifts from local radiative control and deployment area to exaton-class inventories and multi-century industrial throughput and power.

\begin{table}[t]
\caption{Reference pressure thresholds and representative atmospheric inventories on Mars. Atmospheric mass uses $\Mair \simeq (4\pi R_{\rm Mars}^2/g_{\rm Mars})\,P_s$ from Eq.~(\ref{eq:mass_pressure}) and column mass uses $m_{\rm col}=P_s/g_{\rm Mars}$. [Notes: (a) 1 bar is shown as a convenient Earth sea-level reference. (b) The E1 row tabulates only the triple-point pressure threshold associated with E1; the full E1 end state additionally requires recurrent or seasonally accessible open-surface liquid water under favorable local energy balance (Sec.~\ref{sec:water}). (c) The end state E2 is omitted here because it is typically implemented regionally inside habitats (paraterraforming); if applied globally, $P_s\sim 10$\,kPa would imply $M_{\rm atm}\approx 3.9 \times 10^{17}$\,kg, Eq.~(\ref{eq:mass_pressure}). (d) The CO$_2$ row is a representative accessible endogenous reference case consistent with current \emph{order-of-tens-of-mbar} estimates, not a secure hard ceiling. (e) The E4 row is a pressure reference only; the full breathable endpoint is decomposed in Table~\ref{tab:keyresults} into oxygen, buffer-gas, pressure, and thermal requirements.]}
\label{tab:pressure_endpoints}
\centering
\small
\setlength{\tabcolsep}{5pt}
\renewcommand{\arraystretch}{1.15}
\begin{tabular}{l c c c c c}
\toprule
Label & Meaning & $\Ps$ (Pa) & $P_s$ (mbar) & $M_{\rm atm}$ (kg) & $m_{\rm col}$ (kg\,m$^{-2}$) \\
\midrule

E0 & Present mean & $6.1\times 10^{2}$ & 6.1   & $2.37\times 10^{16}$ & $1.64\times 10^{2}$ \\
E1 (threshold) & Triple-point threshold & $6.12\times 10^{2}$ & 6.12 & $2.38\times 10^{16}$ & $1.65\times 10^{2}$ \\
--- & Representative accessible CO$_2$ case & $2.0\times 10^{3}$  & 20 & $7.78\times 10^{16}$ & $5.39\times 10^{2}$ \\
E3 (threshold) & Armstrong limit (no ebullism) & $6.27\times 10^{3}$ & 62.7 & $2.44\times 10^{17}$ & $1.69\times 10^{3}$ \\
--- & 0.2 bar (thick but not Earthlike) & $2.0\times 10^{4}$ & 200 & $7.78\times 10^{17}$ & $5.39\times 10^{3}$ \\
E4 (pressure ref.) & 1 bar (Earth sea-level reference) & $1.0\times 10^{5}$ & 1000 & $3.89\times 10^{18}$ & $2.70\times 10^{4}$ \\
\bottomrule
\end{tabular}
\end{table}

\begin{table*}[t]
\centering
\setlength{\tabcolsep}{1.5pt}
\renewcommand{\arraystretch}{0.90}
\caption{Requirements dashboard for representative endpoints. Inventories use
$M \simeq (4\pi R_{\rm Mars}^2/g_{\rm Mars})\,P$.
Radiative targets assume $T_{e0}\approx 210$~K unless noted. Where ranges are shown, they correspond to build times $t_{\rm build}=10^3~\mathrm{yr}$ (left) to $10^2~\mathrm{yr}$ (right). All power numbers are thermodynamic minima unless stated otherwise. $\dot M$ assumes global deployment; E2 is typically regional.}
\label{tab:keyresults}
\begin{tabular}{c c c c l l}
\toprule
End state &
Success criterion  &
Pressure / $p_i$ &
Inventory $M$ [kg] &
Radiative/control  &
Industrial implication \\

 & (global unless noted) & & &
target &
\\
\midrule
E0 &
Current Mars baseline &
$P_s\approx$ \SI{0.611}{kPa} &
$2.37\times 10^{16}$ &
$T_{e0}\approx 210$~K; &
None (baseline) \\

 & & (6.1 mbar)  & &
$T_s\approx 210$~K &  \\[4pt]

E1 &
Recurrent open-surface&
$P_s \gtrsim 0.612$~kPa &
$\approx 2.38\times 10^{16}$ &
Local pressure + energy    &
Regional heat/insulation    \\

 &liquid water (regional)& & &
balance; triple-point crossing  &
and water-vapor control;  \\

 && & &
alone is insufficient  &
not inventory-limited  \\[4pt]

E2 &
Protected agriculture  &
$P_s\sim 10$~kPa  &
--- &
Local greenhouse/insulation   &
Deployment area + local \\

 &
(typically regional) &
 (inside habitats) & &
/ enclosure control  &
 power; scalable in phases \\[4pt]

E3 &
No ebullism  &
$P_s=6.27$~kPa &
$2.44\times 10^{17}$ &
To approach  $T_s\gtrsim 250$~K: &
If pursued globally: \\

&
(Armstrong limit) &
& &
 $\tau_{\rm IR}\sim 2$--3 (or high-authority &
$\dot M\sim 7.7\times 10^{6}$--$7.7\times 10^{7}$  \\
& & & &
$\Delta F_{\rm TOA}$) &
kg\,s$^{-1}$  (inventory-limited) \\

\midrule
\multicolumn{6}{@{}l}{\textit{E4 (breathable): inventories + energy dominate; warming still required as in E3.}}\\

E4a &
O$_2$ partial pressure  &
$p_{O_2}=21$~kPa &
$8.2\times 10^{17}$ &
Non-collapse warm-state   &
$\dot M_{O_2}\sim 2.6\times 10^{7}$--$2.6\times 10^{8}$  \\

&
 target & & &
 climate + sink management  &
kg\,s$^{-1}$; $\bar P_{\min}\sim 0.38$--$3.8$~PW\\

&  & & & (Sec.~\ref{sec:O2_sinks}) &
 \\[4pt]

E4b &
Buffer gas example &
$p_{N_2}=50$~kPa &
$1.9\times 10^{18}$ &
Inventory  &
$\dot M_{N_2}\sim 6.0\times 10^{7}$--$6.0\times 10^{8}$ \\

 & & & &
 (climate-independent) &
 kg\,s$^{-1}$;\;
if  exogenous,   \\

 & & & & &
momentum/transport  \\

 & & & & &
dominates \\

\bottomrule
\end{tabular}

\emph{Notes:} (a) If $P_s\sim 10$~kPa were applied \emph{globally}, Eq.~(\ref{eq:mass_pressure}) implies $M_{\rm atm}\approx 3.9\times 10^{17}$~kg; E2 is typically implemented regionally (paraterraforming/habitats), so global inventory is not required.\\
(b) E3 is a pressure-only threshold; human habitability additionally requires low $p_{\rm CO_2}$ and, depending on operating envelope, either supplemental oxygen or a breathable O$_2$ partial pressure with an appropriate buffer gas (treated explicitly in E4).
\end{table*}

\subsection{Endpoint-normalized lower bounds and dimensionless feasibility numbers}
\label{sec:pi_groups}

A central difficulty in comparing terraforming pathways is that they are usually presented in incompatible variables:
pressure release, trace-gas radiative forcing, aerosol optical depth, mirror area, or industrial power. A more reusable formulation is to express every pathway relative to the same endpoint-normalized lower bounds (see Table~\ref{tab:pi_groups}).

For any global mean partial-pressure target $p_i$, define
\begin{equation}
\Mreq(p_i)=\Kmars\,p_i,
\qquad
\Kmars \equiv \frac{4\pi R_{\rm Mars}^2}{g_{\rm Mars}}
= 3.89\times 10^{13}\ {\rm kg\,Pa^{-1}}
= 3.89\times 10^{16}\ {\rm kg\,kPa^{-1}}.
\label{eq:Kmars_def}
\end{equation}
Hence the corresponding minimum build throughput over a build time $\tbuild$ is
\begin{equation}
\Qreq(p_i,\tbuild)=\frac{\Kmars\,p_i}{\tbuild}
\approx 1.23\times 10^{6}
\left(\frac{p_i}{1~{\rm kPa}}\right)
\left(\frac{10^3~{\rm yr}}{\tbuild}\right)
{\rm kg\,s^{-1}}.
\label{eq:Qreq_perkPa}
\end{equation}

For a constituent $i$ embedded in a well-mixed multicomponent atmosphere, the inventory associated with a target partial pressure $p_i$ is not, in general, species-independent. Instead,
\begin{equation}
M_i \simeq \Kmars\,\frac{\mu_i}{\bar\mu}\,p_i,
\label{eq:Mi_multicomp_early}
\end{equation}
where $\mu_i$ is the molecular mass of constituent $i$ and $\bar\mu$ is the mean molecular mass of the atmospheric mixture. Eqs.~(\ref{eq:Kmars_def})--(\ref{eq:Qreq_perkPa}) therefore apply directly to \emph{total pressure} or to a dominant constituent; for light minor constituents such as H$_2$ in a CO$_2$-dominated atmosphere, inventory and fill-rate requirements are reduced by the factor $\mu_i/\bar\mu$.

For oxygen made from water electrolysis, using the reversible minimum specific work
$\varepsilon_{O_2}\simeq 14.8~{\rm MJ\,kg^{-1}}$ of O$_2$,
\begin{equation}
E_{O_2,\min}(p_{O_2})=\varepsilon_{O_2}\Kmars\,p_{O_2}
\approx 5.76\times 10^{23}
\left(\frac{p_{O_2}}{1~{\rm kPa}}\right)\ {\rm J},
\label{eq:E_O2_perkPa}
\end{equation}
and
\begin{equation}
\bar P_{O_2,\min}(p_{O_2},\tbuild)
\approx 18.2
\left(\frac{p_{O_2}}{1~{\rm kPa}}\right)
\left(\frac{10^3~{\rm yr}}{\tbuild}\right)\ {\rm TW}.
\label{eq:P_O2_perkPa}
\end{equation}

If a buffer gas is imported exogenously with characteristic delivery speed $\Delta v$, a useful \emph{transport-energy benchmark} is
\begin{equation}
E_{\rm imp,\min}(p_i,\Delta v)
\simeq \frac{1}{2}\Kmars\,p_i\,\Delta v^2
\approx 4.86\times 10^{23}
\left(\frac{p_i}{1~{\rm kPa}}\right)
\left(\frac{\Delta v}{5~{\rm km\,s^{-1}}}\right)^2 {\rm J},
\label{eq:E_imp_perkPa}
\end{equation}
with corresponding average power
\begin{equation}
\bar P_{\rm imp,\min}(p_i,\Delta v,\tbuild)
\approx 15.4
\left(\frac{p_i}{1~{\rm kPa}}\right)
\left(\frac{\Delta v}{5~{\rm km\,s^{-1}}}\right)^2
\left(\frac{10^3~{\rm yr}}{\tbuild}\right)\ {\rm TW}.
\label{eq:P_imp_perkPa}
\end{equation}
The intent of Eqs.~(\ref{eq:E_imp_perkPa})--(\ref{eq:P_imp_perkPa}) is not to define a universal mission-energy floor, because gravity assists, aerocapture, and sail-assisted transfers can greatly reduce propulsive expenditure. Rather, they provide a compact normalization for the transport-scale energy associated with moving planetary-atmosphere-class mass inventories.

These endpoint-normalized lower bounds motivate five dimensionless feasibility numbers:
\begin{align}
\PiM &\equiv \frac{\Mavail}{\Mreq},
&
\PiQ &\equiv \frac{\Qavail}{\Qreq},
&
\PiP &\equiv \frac{\Pavail}{\Preq},
\label{eq:Pi_MQP}
\\
\PiF &\equiv
\begin{cases}
\Delta F_{\rm avail}/\Delta F_{\rm req}, & \text{TOA-forcing pathway},\\[3pt]
\tau_{\rm IR,eff}/\tau_{\rm IR,req}, & \text{greenhouse pathway},
\end{cases}
&
\PiS &\equiv \frac{\Qrepl}{\Qloss}.
\label{eq:Pi_FS}
\end{align}
Because planetary modification is a conjunctive problem rather than a weighted-average problem, the relevant
bottleneck metric is
\begin{equation}
\Pimin \equiv \min\{\PiM,\PiF,\PiQ,\PiP,\PiS\}.
\label{eq:Pimin}
\end{equation}
A pathway is feasibility-complete for a given endpoint only if $\Pimin \gtrsim 1$. This minimum-operator formulation is deliberate: failure in any one constraint vetoes the endpoint, regardless of strength in the others.

\begin{table}[t]
\caption{Dimensionless feasibility numbers introduced in Sec.~\ref{sec:pi_groups}. The paper's mechanism survey can be re-expressed in this language, making the dominant bottleneck explicit and falsifiable.}
\label{tab:pi_groups}
\setlength{\tabcolsep}{3pt}
\renewcommand{\arraystretch}{1.00}
\centering
\small
\begin{tabular}{c l c l}
\toprule
Quantity & Definition & Feasible regime & Meaning \\
\midrule
$\Pi_M$ & $\Mavail/\Mreq$ & $\gtrsim 1$ & Inventory sufficiency \\
$\Pi_F$ & $\Delta F_{\rm avail}/\Delta F_{\rm req}$ or $\tau_{\rm IR,eff}/\tau_{\rm IR,req}$ & $\gtrsim 1$ & Radiative control sufficiency \\
$\Pi_{\dot M}$ & $\Qavail/\Qreq$ & $\gtrsim 1$ & Industrial throughput sufficiency \\
$\Pi_P$ & $\Pavail/\Preq$ & $\gtrsim 1$ & Power sufficiency \\
$\Pi_S$ & $\Qrepl/\Qloss$ & $\gtrsim 1$ & Ability to hold the engineered state against loss \\
$\Pi_{\min}$ & $\min\{\Pi_M,\Pi_F,\Pi_{\dot M},\Pi_P,\Pi_S\}$ & $\gtrsim 1$ & Conjunctive feasibility criterion \\

$\Lambda_{\rm maint}$ & $\tbuild/\tloss$ & $\ll 1$ fill-dominated;  & Distinguishes one-shot from continuously replen-\\
 & & $\gtrsim 1$ maintenance- & ished pathways \\
&  &  dominated &\\
\bottomrule
\end{tabular}
\end{table}

To distinguish one-shot fill operations from truly maintenance-limited pathways, define
\begin{equation}
\Lmaint \equiv \frac{\tbuild}{\tloss}
= \frac{\Qloss}{M^\star/\tbuild},
\label{eq:Lmaint}
\end{equation}
where $M^\star$ is the maintained inventory of the active climate agent.
Pathways with $\Lmaint\ll 1$ are fill-dominated over the build phase; pathways with $\Lmaint\gtrsim 1$ are already maintenance-dominated during buildup.

\subsection{Worked examples under the \texorpdfstring{$\Pi$}{Pi} rubric}
\label{sec:pi_worked_examples}

The utility of Eqs.~(\ref{eq:Pi_MQP})--(\ref{eq:Lmaint}) is that they permit immediate rejection or classification of a pathway without a full climate simulation. Three examples show how the framework is intended to be used.

\paragraph{Example 1: endogenous CO$_2$ cannot reach a global E3 atmosphere.} Taking a representative accessible endogenous CO$_2$ reference case of $P_s\approx 20$ mbar (Sec.~\ref{sec:endogenous}) and comparing with the E3 threshold $P_s=62.7$ mbar gives
{}
\begin{equation}
\Pi_M^{({\rm CO_2}\rightarrow E3)}\lesssim \frac{20}{62.7}\approx 0.32.
\end{equation}
Radiatively the same pathway also fails: the accessible CO$_2$ inventory yields $\lesssim 10$ K warming, whereas the paper's own thermal targets imply an \(\mathcal{O}(40\text{--}60~\mathrm{K})\) shortfall to warm global E3/E4-like surface states. Thus \(\Pi_F\ll 1\) as well. The important conclusion is therefore stronger than ``endogenous CO$_2$ helps but is insufficient'': it fails \emph{both} the inventory and radiative tests before throughput is even considered.

\paragraph{Example 2: global CO$_2$--H$_2$ CIA is fill-dominated on century-to-millennial build times.}
For the illustrative case $P_{\rm tot}=\SI{0.5}{bar}$ and $f_{H_2}=0.05$ (Sec.~\ref{sec:fill_vs_maint}), the required H$_2$ inventory is $M_{H_2}^\star\approx 4.7\times 10^{15}$ kg once the multicomponent correction $\mu_{H_2}/\bar\mu$ is included. Building that inventory over $10^3$ yr requires an average fill throughput
\begin{equation}
\dot M_{H_2,{\rm fill}}\simeq \frac{M_{H_2}^\star}{t_{\rm build}}\approx 1.5\times 10^5~\mathrm{kg\,s^{-1}},
\end{equation}
whereas the diffusion-limited replenishment floor is only $\dot M_{H_2,{\rm loss}}\sim 6\times 10^3~\mathrm{kg\,s^{-1}}$. Equivalently, using Eq.~(\ref{eq:tauH2_loss_revised}),
\begin{equation}
\Lambda_{\rm maint}=\frac{t_{\rm build}}{\tau_{H_2,\rm loss}}\approx 4\times 10^{-2}
\qquad (t_{\rm build}=10^3~\mathrm{yr}).
\end{equation}
Hence CO$_2$--H$_2$ CIA is not primarily a replenishment-limited pathway during buildout; it is primarily a global-fill pathway with a long-duration maintenance tail.

\paragraph{Example 3: particle warming is maintenance-dominated from the outset.}
If a particle pathway has atmospheric residence time $\tau_p\sim 30$--100 days, then for any century-scale build schedule
\begin{equation}
\Lambda_{\rm maint}^{(p)}=\frac{t_{\rm build}}{\tau_p}\sim 10^3\text{--}10^4 \gg 1.
\end{equation}
Thus the dominant requirement is not cumulative injected mass but the sustained industrial rate needed to hold the atmospheric column against removal. This is why particle pathways and H$_2$-CIA pathways should not be grouped together merely because both are ``mass-efficient warming levers'': their industrial signatures are fundamentally different.

\section{Atmospheric mass bookkeeping: pressure targets imply exaton-scale gases}
\label{sec:mass}

\subsection{Pressure--mass relation}
\label{sec:mass_relation}

\begin{figure}[t!]
\includegraphics[width=0.50\linewidth]{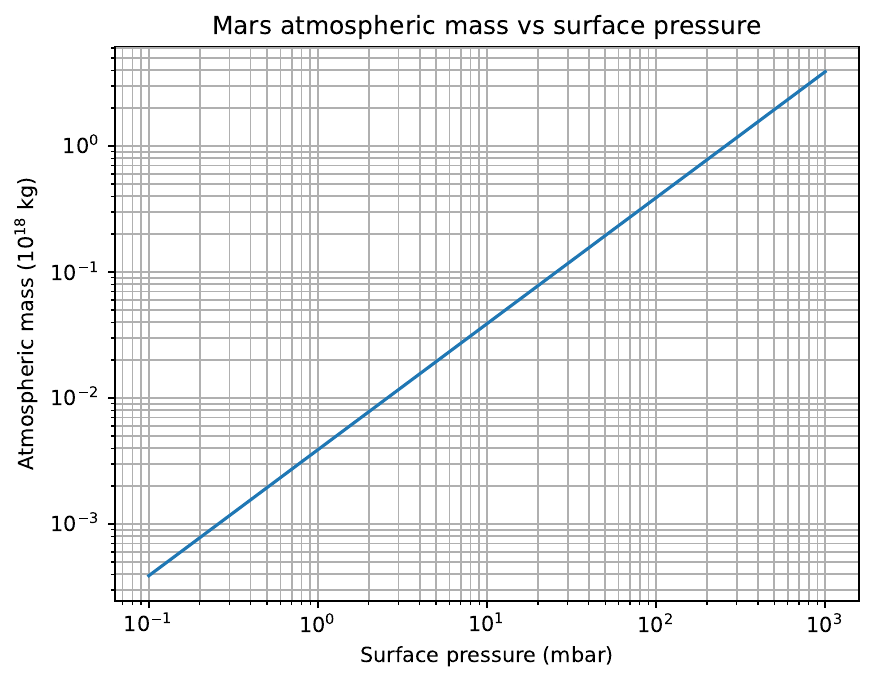}
\caption{Atmospheric mass required for a given mean surface pressure on Mars from
Eq.~(\ref{eq:mass_pressure}). A \SI{1}{bar} atmosphere corresponds to
$\Mair \approx 3.89\times 10^{18}\,\mathrm{kg}$.}
\label{fig:mass_pressure}
\end{figure}

For a thin atmosphere in hydrostatic equilibrium, the global atmospheric mass required to produce a
mean surface pressure $\Ps$ is approximately
\begin{equation}
\Mair \;\simeq\; \frac{4\pi \Rmars^2}{\gMars}\,\Ps .
\label{eq:mass_pressure}
\end{equation}
Using Table~\ref{tab:marsparams}, the conversion is
\begin{equation}
\Mair \approx 3.89\times 10^{15}\ \mathrm{kg\ per\ mbar}
\;\;\;\;\;(\text{Mars}).
\label{eq:mass_per_mbar}
\end{equation}
Figure~\ref{fig:mass_pressure} plots Eq.~(\ref{eq:mass_pressure}) over $0.1$--$1000$ mbar.

Table~\ref{tab:pressure_endpoints} highlights several pressure endpoints relevant to ``success'' criteria.
The central message is scale: moving from present-day Mars ($\sim 6$ mbar) to even modest human-relevant
pressures is an \emph{exaton} ($10^{18}$ kg) class problem.

Also, from Eq.~(\ref{eq:mass_pressure}), we derive atmospheric mass per unit pressure
\begin{equation}
\frac{d\Mair}{d\Ps} = \frac{4\pi \Rmars^2}{\gMars}
\approx 3.89\times 10^{13}\,\mathrm{kg\,Pa^{-1}}
\approx 3.89\times 10^{15}\,\mathrm{kg\,mbar^{-1}}.
\end{equation}

 Eqs.~(\ref{eq:mass_pressure})--(\ref{eq:mass_per_mbar})   are more than bookkeeping. They show that every additional \SI{1}{mbar} of global mean surface pressure on Mars costs about $3.89\times 10^{15}$ kg of atmosphere, so discussions framed only in pressure units can hide the true industrial scale of the problem. In particular, the step from present-day Mars to the Armstrong-limit threshold is not a modest atmospheric adjustment but a jump to $\sim 2.4\times 10^{17}$ kg, and the step to an Earth-sea-level atmosphere is nearly two orders of magnitude larger again. The hydrostatic conversion therefore already fixes the class of the problem: once the goal is global E3 or E4, the atmosphere is an exaton-scale engineering object.

\begin{table*}[t]
\caption{Endpoint-normalized lower bounds on Mars, expressed per \SI{1}{kPa} of target pressure. The inventory/throughput rows apply directly to \emph{total pressure} or to a dominant constituent. For a minor constituent $i$ in a multicomponent atmosphere, the corresponding inventory and fill-rate must be multiplied by $\mu_i/\bar\mu$ [Eq.~(\ref{eq:Mi_multicomp_early})]. Species-specific production/import energies enter through the relevant specific work or delivery architecture.}
\label{tab:per_kPa_scalings}
\setlength{\tabcolsep}{3pt}
\renewcommand{\arraystretch}{1.00}
\centering
\small
\begin{tabular}{l l l}
\toprule
Quantity & Expression & Numerical value \\
\midrule
Global inventory per \SI{1}{kPa} & $\Kmars (1~{\rm kPa})$ & $3.89\times 10^{16}\ {\rm kg}$ \\[2pt]

Global fill rate per \SI{1}{kPa} over $\tbuild$ & $\Kmars (1~{\rm kPa})/\tbuild$ &
$1.23\times 10^{6}\,(10^3{\rm yr}/\tbuild)\ {\rm kg\,s^{-1}}$ \\[2pt]

Reversible O$_2$ production energy per \SI{1}{kPa} & $\varepsilon_{O_2}\Kmars (1~{\rm kPa})$ &
$5.76\times 10^{23}\ {\rm J}$ \\[2pt]

Reversible O$_2$ production power floor per \SI{1}{kPa} & $\varepsilon_{O_2}\Kmars (1~{\rm kPa})/\tbuild$ &
$18.2\,(10^3{\rm yr}/\tbuild)\ {\rm TW}$ \\[2pt]

Kinetic-energy transport benchmark per  \SI{1}{kPa}  &
$\frac12 \Kmars (1~{\rm kPa})\Delta v^2$ &
$4.86\times 10^{23}\ {\rm J}$ \\
at $\Delta v=\SI{5}{km\,s^{-1}}$ & &  \\ [2pt]

Transport-power benchmark per \SI{1}{kPa}  &
$\frac12 \Kmars (1~{\rm kPa})\Delta v^2/\tbuild$ &
$15.4\,(10^3{\rm yr}/\tbuild)\ {\rm TW}$ \\
at $\Delta v=\SI{5}{km\,s^{-1}}$ & &  \\[2pt]

Electrolysis energy to produce H$_2$ for \SI{1}{kPa} H$_2$ partial   &
$\varepsilon_{H_2}\Kmars (1~{\rm kPa})$ &
$4.67\times 10^{24}\ {\rm J}$ \\
pressure in a multicomponent atmosphere & &  \\[2pt]

Electrolysis power floor for \SI{1}{kPa} H$_2$ partial pressure   &
$\varepsilon_{H_2}\Kmars (\mu_{H_2}/\bar\mu)(1~{\rm kPa})/\tbuild$ &
$148(\mu_{H_2}/\bar\mu)(10^3{\rm yr}/\tbuild)\ {\rm TW}$  \\
 in a multicomponent atmosphere  & & \\
\bottomrule
\end{tabular}

\vspace{2pt}
\footnotesize
Here $\varepsilon_{O_2}\simeq 14.8~{\rm MJ\,kg^{-1}}$ is the reversible minimum work to make O$_2$ from water, and $\varepsilon_{H_2}\simeq 1.2\times 10^8~{\rm J\,kg^{-1}}$ is the reversible minimum work to make H$_2$ from water. For a representative CO$_2$-dominated CIA atmosphere with $\bar\mu\approx 41.9$ g mol$^{-1}$, the H$_2$ factor is $\mu_{H_2}/\bar\mu\approx 0.0477$, so the H$_2$ energy/power requirements per kPa are about $\sim 21\times$ smaller than in pure H$_2$.
\end{table*}

\subsection{Composition matters: CO$_2$ is a poor buffer gas}
\label{sec:composition}

Table~\ref{tab:per_kPa_scalings} summarizes endpoint-normalized lower bounds on Mars.
Pressure targets do not specify composition, but composition determines physiology and climate.

\subsubsection{CO$_2$ toxicity constraint}

High CO$_2$ partial pressures are directly hazardous to humans and many plants. Therefore, even if CO$_2$ is the primary available endogenous greenhouse gas, a breathable endpoint cannot consist of ``CO$_2$ + O$_2$'' alone at tens of kPa.  A viable E3/E4 atmosphere requires a buffer gas (N$_2$ and/or Ar) at the tens of kPa level.

\subsubsection{Available buffer gas on Mars}

Mars' present atmosphere is $\sim 95\%$ CO$_2$ with minor N$_2$ and Ar. Scaling these minors to tens of kPa requires either (i) importing buffer gas from external sources or (ii) mining and releasing geochemically stored nitrogen (nitrates, ammonium minerals) if sufficient inventories exist (uncertain at planetary scale). This is a first-order roadblock for E4.

Not every human-use atmosphere requires tens of kPa of inert diluent. Historical spacecraft have used low-pressure O$_2$-rich atmospheres for crewed operations, and this design space remains relevant for controlled habitats and vehicles. However, such cases should be distinguished from an open planetary surface endpoint: removing the N$_2$/Ar burden can reduce the \emph{buffer-gas} mass requirement, but it does not eliminate the exaton-class O$_2$ requirement itself, and it introduces stronger flammability, decompression, and ecosystem-composition constraints. In this paper, buffered atmospheres remain the representative E4 open-surface benchmark, whereas low-buffer-gas cases are better interpreted as habitat/vehicle operating envelopes rather than planetary end states.

A useful quantitative comparison is that eliminating the inert diluent removes the \emph{buffer-gas} mass requirement, but it does not remove the oxygen mass requirement itself. Because $M_{O_2}\simeq \Kmars p_{O_2}$ for an open planetary atmosphere, a pure-O$_2$ operating envelope at $p_{O_2}\sim \SI{34.5}{kPa}$ would still imply $M_{O_2}\sim 1.34\times10^{18}$ kg if applied planet-wide, i.e. larger than the Earth-like $p_{O_2}=\SI{21}{kPa}$ benchmark. Low-buffer-gas cases therefore change the \emph{composition split} of E4, but they do not remove the exaton-class oxygen inventory if interpreted as an open-surface planetary endpoint \cite{NASAHabitableAtmosphere2023}.

\subsection{Column mass as a proxy for radiation and aerodynamic shielding}
\label{sec:column_mass}

Atmospheric column mass is $m_{\mathrm{col}}=P_s/g_{\mathrm{Mars}}$. This quantity is directly relevant to (i) attenuation of energetic particles and the mass loading relevant to entry/descent/aerodynamic performance, and only \emph{indirectly} relevant to UV shielding. In particular, energetic-particle shielding scales approximately with column mass, whereas surface UV depends strongly on atmospheric composition (e.g. ozone) and on dust/cloud opacity rather than on column mass alone \cite{Dartnell2007Radiation,Moores2007UVShielding}.
For representative endpoints:
\begin{equation}
m_{\mathrm{col}} \approx 2.7\times 10^{3}\ \mathrm{kg\,m^{-2}}\left(\frac{P_s}{10~\mathrm{kPa}}\right),
\end{equation}
so $P_s=6.27$~kPa (Armstrong limit) corresponds to $m_{\mathrm{col}}\approx 1.7\times 10^{3}$~kg\,m$^{-2}$,
while $P_s=101$~kPa corresponds to $m_{\mathrm{col}}\approx 2.7\times 10^{4}$~kg\,m$^{-2}$ on Mars
(because $g_{\mathrm{Mars}}<g_\oplus$).
Thus pressure-building strategies also directly modify shielding and entry/descent regimes, even before
composition is made breathable.

\subsection{Buffer gas mass scales and import energetics}
\label{sec:buffer}

A breathable E3/E4 atmosphere requires a \emph{buffer gas} (typically N$_2$ and/or Ar) at tens of kPa. The mass requirement follows directly from Eq.~(\ref{eq:mass_pressure}). For example, targeting $p_{\mathrm{N_2}}=\SI{50}{kPa}$ implies
\begin{equation}
\label{eq:MN2}
M_{\mathrm{N_2}} \simeq \frac{4\pi \Rmars^2}{\gMars}\,p_{\mathrm{N_2}}
\approx 1.9\times 10^{18}\,\mathrm{kg},
\end{equation}
comparable to the mass of a \SI{0.5}{bar} planetary atmosphere.

\subsubsection{External sourcing}

If buffer gas must be imported (as N$_2$, NH$_3$, or N-bearing ices), a useful first benchmark is the characteristic transport energy associated with delivering mass $M$ with characteristic hyperbolic excess $\Delta v$,
\begin{equation}
E_k \sim \frac{1}{2} M\,\Delta v^2.
\label{eq:import_energy}
\end{equation}
For $M\sim 10^{18}\,\mathrm{kg}$ and $\Delta v\sim \SI{5}{km\,s^{-1}}$, one finds
$E_k\sim 10^{25}\,\mathrm{J}$, comparable to the oxygenation energy scale in Eq.~(\ref{eq:Emin}). This should \emph{not} be interpreted as a universal lower bound on mission energy expenditure: gravity assists, aerocapture, sail-assisted trajectories, and staged capture architectures can all reduce the propulsive component substantially.

The more robust conclusion is logistical rather than propulsive. Even when the transfer architecture is energetically favorable, importing \(\sim 10^{18}\,\mathrm{kg}\) of volatiles remains a planetary-scale \emph{mass-handling} problem. The hard questions are capture of volatile-bearing bodies or cargo streams, controlled breakup or processing where needed, storage/transfer into Mars-bound logistics, and finally global atmospheric delivery and distribution. For feasibility classification, the dominant issue is therefore not a single mission \(\Delta v\), but whether a Mars-centered industrial system can repeatedly process and redistribute planetary-atmosphere-class mass inventories.

\subsubsection{NH$_3$ as an N-carrier}

If nitrogen is imported primarily as ammonia, photolysis and chemistry can convert it into N$_2$ while hydrogen escapes, leaving N$_2$ as a buffer gas. Stoichiometrically, producing one mole of N$_2$ (28 g) requires two moles of NH$_3$ (34 g) to supply two N atoms, so the imported mass is at least $\sim 1.21\times$ the desired N$_2$ mass, before losses.

These mass and momentum scales imply that E4 atmospheres are difficult to reach without either enormous
exogenous mass logistics or a yet-unquantified endogenous nitrogen inventory.

\subsection{How much fixed nitrogen would Mars need to supply a buffer gas?}
\label{sec:nitrogen_bound}

If Mars were to supply buffer gas endogenously, the limiting question is the global inventory of fixed nitrogen
(e.g., nitrates) accessible to mining and release.
Let an accessible regolith layer of depth $d$ and bulk density $\rho$ contain a mean nitrogen mass fraction $w_N$.
The total nitrogen mass is
\begin{equation}
M_N \simeq 4\pi R_{\mathrm{Mars}}^{2}\,\rho\,d\,w_N,
\end{equation}
and because N$_2$ is nitrogen by mass, the achievable buffer-gas pressure is
\begin{equation}
p_{N_2} \simeq \frac{g_{\mathrm{Mars}}\,M_N}{4\pi R_{\mathrm{Mars}}^{2}}
\simeq g_{\mathrm{Mars}}\,\rho\,d\,w_N.
\label{eq:pN2_regolith}
\end{equation}
Solving for the required nitrogen mass fraction,
\begin{equation}
w_N \simeq \frac{p_{N_2}}{g_{\mathrm{Mars}}\rho d}
\approx 6.7\times 10^{-2}
\left(\frac{p_{N_2}}{\SI{50}{kPa}}\right)
\left(\frac{\rho}{\SI{2000}{kg\,m^{-3}}}\right)^{-1}
\left(\frac{d}{\SI{100}{m}}\right)^{-1}.
\end{equation}
Thus, supplying $p_{N_2}=\SI{50}{kPa}$ from the top $\sim\SI{100}{m}$ would require an implausibly high
planet-wide mean nitrogen fraction (percent-level by mass).
Even $d=\SI{1}{km}$ requires $w_N\sim 6.7\times 10^{-3}$, still extremely large on a global average. This does not rule out regional nitrogen extraction, but it strongly suggests that E4 buffer-gas inventories are more naturally met by exogenous sourcing unless deep, nitrogen-rich deposits exist. In situ observations do indicate fixed nitrogen (nitrate) in Martian sediments, but the global inventory remains
poorly constrained \cite{Stern2015Nitrates}.

Eq.~\eqref{eq:pN2_regolith} demonstrates that supplying \(\mathcal{O}(10\text{--}50)\,\mathrm{kPa}\) of buffer gas as a
\emph{global mean} from shallow regolith requires implausibly large planet-wide \(w_N\). This does not exclude the possibility that \emph{localized} nitrogen-rich deposits could support \emph{regional} E2/E3 domes or paraterraforming zones, because required mass scales with covered area rather than planetary surface area. For a region of area \(A_{\rm reg}\), the buffer-gas mass required to achieve \(p_{N_2}\) over that region is
\begin{equation}
M_{N_2,{\rm reg}} \simeq \frac{A_{\rm reg}}{g_{\rm Mars}}\,p_{N_2}
\simeq 1.35\times 10^{16}\ {\rm kg}\,
\left(\frac{A_{\rm reg}}{10^{6}\ {\rm km^2}}\right)
\left(\frac{p_{N_2}}{50\ {\rm kPa}}\right),
\end{equation}
which is \(\sim 140\times\) smaller than the global requirement for the same \(p_{N_2}\).

As we have seen, Section~\ref{sec:mass} showed that E3 and above are intrinsically exaton-class mass problems:
even $P_s=\SI{6.27}{kPa}$ implies $M_{\rm atm}\approx 2.4\times 10^{17}$~kg (Table~\ref{tab:pressure_endpoints}),
and E4 buffer+oxygen inventories exceed $10^{18}$~kg.
However, pressure alone does not imply habitability; the next question is whether Mars can be warmed to
$T_s\gtrsim 250$--273~K under current insolation.
Section~\ref{sec:thermal} therefore derives radiative targets in the two natural control variables:
$\Delta F_{\rm TOA}$ for insolation changes and $\tau_{\rm IR}$ for greenhouse pathways.

\subsection{Local log-sensitivities and uncertainty propagation}
\label{sec:log_sensitivity}

Because the paper's objective is feasibility discrimination rather than high-fidelity state prediction, the most useful uncertainty metric is not a raw parameter range but the \emph{local logarithmic sensitivity} of each derived requirement to each uncertain input.
For a quantity $Y(\{x_a\})$, define
\begin{equation}
S_{x_a}^{(Y)} \equiv \frac{\partial \ln Y}{\partial \ln x_a}.
\label{eq:log_sensitivity}
\end{equation}
For independent fractional input uncertainties, first-order propagation gives
\begin{equation}
\left(\frac{\sigma_Y}{Y}\right)^2
\approx
\sum_a \left[S_{x_a}^{(Y)}\right]^2
\left(\frac{\sigma_{x_a}}{x_a}\right)^2.
\label{eq:uncertainty_prop}
\end{equation}

This representation is especially useful here because many requirement scalings are approximately separable power laws. For example, if engineered particles are treated through an effective forcing-per-column parameter $\gamma_p \equiv \partial \Delta F/\partial \Sigma_p$, then
\begin{equation}
\Sigma_p \propto \frac{\Delta F_{\rm req}}{\gamma_p},
\qquad
\dot M_p \propto \frac{\Delta F_{\rm req}}{\gamma_p\,\tau_p},
\qquad
A_m \propto \frac{\Delta F_{\rm TOA}}{\eta_m S_{\rm Mars}},
\qquad
p_{N_2,\rm achievable}\propto \rho\,d\,w_N.
\end{equation}
Thus the dominant uncertainties differ by mechanism: IR-active particle pathways are most sensitive to residence time $\tau_p$ and effective radiative efficacy $\gamma_p$; mirror pathways are linear in required forcing and inverse in optical/geometric efficiency; endogenous N$_2$ pathways are linear in accessible depth and nitrogen mass fraction; and O$_2$ sink estimates are linear in the effective oxidizable depth and FeO fraction. Table~\ref{tab:log_sensitivity} provides relevant details.

Reporting these sensitivities makes the uncertainty structure explicit and prevents false equivalence between uncertainties that merely move a coefficient by a factor of a few and uncertainties that change the feasibility class.

\begin{table*}[t]
\caption{Local log-sensitivity coefficients for key requirement quantities. Nonzero coefficients are shown explicitly. This table identifies which uncertain inputs can realistically change a pathway classification and which only perturb prefactors.}
\label{tab:log_sensitivity}
\setlength{\tabcolsep}{3pt}
\renewcommand{\arraystretch}{1.10}
\centering
\small
\begin{tabular}{l l l}
\toprule
Derived quantity $Y$ & Leading scaling & Nonzero log-sensitivities $S_x^{(Y)}=\partial \ln Y/\partial \ln x$ \\
\midrule

Maintained particle column $\Sigma_p$ &
$\Delta F_{\rm req}/\gamma_p$ &
$S_{\Delta F}=+1$, $S_{\gamma_p}=-1$ \\

Particle injection rate $\dot M_p$ &
$\Delta F_{\rm req}/(\gamma_p\tau_p)$ &
$S_{\Delta F}=+1$, $S_{\gamma_p}=-1$, $S_{\tau_p}=-1$ \\

Mirror area $A_m$ &
$\Delta F_{\rm TOA}/(\eta_m S_{\rm Mars})$ &
$S_{\Delta F}=+1$, $S_{\eta_m}=-1$, $S_{S_{\rm Mars}}=-1$ \\

Achievable endogenous $p_{N_2}$ &
$\rho\,d\,w_N$ &
$S_\rho=+1$, $S_d=+1$, $S_{w_N}=+1$ \\

O$_2$ sink capacity $M_{O_2,\rm sink}$ &
$w_{\rm FeO}\rho d_{\rm eff}$ &
$S_{w_{\rm FeO}}=+1$, $S_\rho=+1$, $S_{d_{\rm eff}}=+1$ \\

H$_2$ replenishment power $P_{H_2,\min}$ &
$\Phi_0 f_{H_2}$ &
$S_{\Phi_0}=+1$, $S_{f_{H_2}}=+1$ \\

H$_2$ loss time $\tau_{H_2,\rm loss}$ &
$P_{\rm tot}\Phi_0^{-1}$ &
$S_{P_{\rm tot}}=+1$, $S_{\Phi_0}=-1$ \\
\bottomrule
\end{tabular}
\end{table*}

\section{Thermal requirements: radiative forcing and the \texorpdfstring{$\sim 60$ K}{60 K} problem}
\label{sec:thermal}

\subsection{Global mean forcing needed for large temperature changes}

A minimal, transparent way to relate temperature targets to energy balance is through the effective
radiating temperature $\Te$ at TOA:
\begin{equation}
\sigSB \Te^4 = \frac{\SMars}{4}(1-\Abond) + \DeltaF_{\mathrm{TOA}},
\label{eq:Te_balance}
\end{equation}
where $\DeltaF_{\mathrm{TOA}}$ is an imposed perturbation to the net absorbed flux
(e.g., from orbital mirrors or albedo change).
In this simple model, raising $\Te$ from ${\Te}_0$ to $\Te$ requires
\begin{equation}
\DeltaF_{\mathrm{TOA}} \;=\; \sigSB\left(\Te^4-{\Te}_0^4\right).
\label{eq:forcing_Te}
\end{equation}
For Mars, increasing $\Te$ by \SI{30}{K} (from 210 K to 240 K) requires
$\DeltaF_{\mathrm{TOA}}\approx \SI{78}{W\,m^{-2}}$, while \SI{60}{K} requires
$\DeltaF_{\mathrm{TOA}}\approx \SI{191}{W\,m^{-2}}$.
These are extremely large global perturbations.

\subsubsection{Surface temperature $T_s$ vs. $T_e$: why the same ``60 K'' implies different requirements}

Eqs.~(\ref{eq:Te_balance})--(\ref{eq:forcing_Te}) quantify the TOA forcing required to raise the effective radiating temperature $T_e$ directly (e.g., via mirrors or albedo changes). This is the correct metric for \emph{insolation-modifying} approaches. However, greenhouse approaches primarily seek to raise the surface temperature $T_s$ at approximately fixed absorbed solar power by increasing longwave optical depth, thereby increasing $T_s$ relative to $T_e$. Accordingly, the $\Delta F_{\mathrm{TOA}}$ values computed above should be interpreted as bounds for \emph{direct TOA forcing} methods, whereas greenhouse methods are more naturally expressed in terms of the required infrared optical depth $\tau_{\mathrm{IR}}$ (Sec.~\ref{sec:greenhouse_closure}).

\subsubsection{Why ``waste heat'' is not a global terraforming lever}

A useful sanity check is to compare industrial waste heat to planetary radiative fluxes. A global mean forcing of just $\DeltaF=\SI{10}{W\,m^{-2}}$ corresponds to total power
$\Delta P = 4\pi \Rmars^2\DeltaF \approx 1.4\times 10^{15}\,\mathrm{W}$, i.e.\ \(\sim\SI{1400}{\tera\watt}\).
Therefore, even a civilization dissipating \(\sim \SI{10}{\tera\watt}\) of waste heat would supply only
$\sim 0.07~\mathrm{W\,m^{-2}}$ globally. Direct heat injection can matter locally, but \emph{global} warming requires radiative (albedo/insolation) or greenhouse modifications.

The surface temperature exceeds $\Te$ when the atmosphere provides greenhouse effect. Terraforming proposals therefore focus on increasing greenhouse opacity (longwave trapping) rather than directly increasing absorbed solar flux. Greenhouse pathways are better compared through the additional longwave opacity required to obtain a target $T_s$ at nearly fixed absorbed solar power, whereas mirror/albedo pathways are naturally compared through explicit $\Delta F_{\rm TOA}$.

\subsection{Relating TOA forcing to surface temperature: a minimal greenhouse mapping}
\label{sec:greenhouse_closure}

Terraforming objectives are typically stated in terms of surface temperature $\Ts$, whereas Eqs.~(\ref{eq:Te_balance})--(\ref{eq:tau_kappa}) constrain the effective radiating temperature $\Te$ at TOA. Greenhouse interventions primarily alter the mapping $\Ts(\Te)$ by increasing longwave optical depth, rather than by imposing a direct TOA forcing $\DeltaF_{\mathrm{TOA}}$. The window/saturation language is most relevant for thickened or engineered atmospheres; it is not intended to imply that present-day \(\sim 6\) mbar Mars already behaves like a dense, band-saturated CO$_2$ atmosphere outside the strongest 15-\(\mu\)m region.

A standard first-order approximation is the Eddington grey-atmosphere relation for radiative equilibrium:
\begin{equation}
\Ts^4 \;\simeq\; \frac{3}{4}\,\Te^4\left(\tau_{\mathrm{IR}} + \frac{2}{3}\right),
\label{eq:eddington_grey}
\end{equation}
where $\tau_{\mathrm{IR}}$ is the broadband infrared optical depth from the surface to space.
Solving Eq.~(\ref{eq:eddington_grey}) for the optical depth required to reach a target $\Ts$ gives
\begin{equation}
\tau_{\mathrm{IR}} \;\simeq\; \frac{4}{3}\left(\frac{\Ts}{\Te}\right)^4 - \frac{2}{3}.
\label{eq:tau_required}
\end{equation}

Note that the grey-atmosphere relation in Eq.~\eqref{eq:eddington_grey} is used here only to define an \emph{architecture-level, order-of-magnitude target} for the longwave opacity required to raise $T_s$ at fixed absorbed solar power; it is \emph{not} a substitute for line-by-line radiative transfer or 3-D GCM predictions of a specific Mars climate state.

For Mars' present ${\Te}_0 \approx \SI{210}{K}$ (Table~\ref{tab:marsparams}), achieving $\Ts = \SI{273}{K}$ implies $\tau_{\mathrm{IR}} \approx 3.1$, while $\Ts=\SI{250}{K}$ implies $\tau_{\mathrm{IR}}\approx 2.0$.
These values provide a transparent target for greenhouse strategies: CO$_2$ alone at $\Ps\sim \mathcal{O}(10~\mathrm{mbar})$ does not supply sufficient $\tau_{\mathrm{IR}}$ under current insolation,
consistent with detailed inventory and climate constraints \cite{JakoskyEdwards2018CO2Inventory}.

A further useful mapping relates $\tau_{\mathrm{IR}}$ to pressure through a mass absorption coefficient
$\kappa_{\mathrm{IR}}$:
\begin{equation}
\tau_{\mathrm{IR}} \;\sim\; \kappa_{\mathrm{IR}}\,\frac{\Ps}{\gMars},
\label{eq:tau_kappa}
\end{equation}
highlighting the trade between (i) increasing $\Ps$ (at large gas mass cost) and (ii) increasing $\kappa_{\mathrm{IR}}$
via trace super-greenhouse gases or aerosol absorption in spectral window regions. Eq.~(\ref{eq:tau_kappa}) is not intended as a substitute for line-by-line radiative transfer, but as a first-order engineering scaling relation that exposes which proposals require exaton-class pressure increases versus those that require high-$\kappa_{\mathrm{IR}}$ engineered constituents.

For real gases such as CO$_2$, Eq.~(\ref{eq:tau_kappa}) should be read as an \emph{effective bookkeeping relation}, not as a literal statement that forcing grows linearly with pressure at all $P_s$. In spectrally saturated bands, additional warming is controlled by line wings, window closure, and pressure broadening, so the incremental greenhouse effect can be strongly sublinear in $P_s$ even when an effective $\kappa_{\rm IR,eff}$ is convenient for architecture-level trades.

Uncertainties in translating a target \(T_s\) into a required \(\tauIReff\) are generically
\(\mathcal{O}(\pm 30\text{--}50\%)\), driven by non-grey spectral structure (windows vs.\ saturated bands),
dust/cloud feedbacks, lapse-rate changes, and orbital/seasonal variability. This uncertainty does \emph{not} affect the exaton-scale mass conclusions for E3/E4, which follow directly
from hydrostatic pressure--mass relation (Sec.~\ref{sec:mass}) and composition requirements (Sec.~\ref{sec:oxygen}).

\subsubsection{Spectral interpretation of the grey \texorpdfstring{$\tauIR$}{tauIR}}

In a non-grey Mars atmosphere, incremental surface warming is often controlled by
(i) \emph{spectral window regions} that are weakly absorbing in the baseline state,
(ii) \emph{pressure broadening} that extends absorption into line wings, and
(iii) cloud/dust feedbacks that redistribute LW opacity and SW heating.
Accordingly, the grey \(\tauIR\) in Eq.~\eqref{eq:eddington_grey} should be interpreted as an
\emph{effective, window-weighted longwave optical depth} required to obtain a target \(T_s/T_e\),
not as a literal spectrally uniform opacity.

For architecture trades it is therefore useful to introduce a ``window-filling efficacy'' factor
\(\winfac\in(0,1]\) such that a mechanism with mass absorption coefficient \(\kappa_{\rm IR}\) produces
\begin{equation}
\tauIReff \;\sim\; \winfac\,\kappa_{\rm IR}\,\frac{P_s}{g_{\rm Mars}},
\label{eq:tau_eff_window}
\end{equation}
where \(\winfac\sim 1\) for constituents that add opacity primarily in LW window regions and \(\winfac\ll 1\) for constituents that add opacity mainly in already-saturated bands. Eq.~\eqref{eq:tau_eff_window} makes explicit that the relevant engineering question is not ``how much \(\tauIR\) is added,'' but ``how much \emph{window-weighted} opacity can be added per kg\,m\(^{-2}\) column.''

\subsubsection{Sensitivity to orbital, seasonal, latitudinal, and diurnal radiative state}

The required optical depth depends on $T_e$ through Eq.~(\ref{eq:tau_required}). Using $T_e=[S(1-A_B)/(4\sigma_{\mathrm{SB}})]^{1/4}$, Mars' orbit-averaged value is $T_{e0}\approx 210$~K (Table~\ref{tab:marsparams}),
but $T_e$ varies with heliocentric distance and albedo.
For illustration, if $T_e$ spans $\sim 201$--$221$~K across Mars' orbit for fixed $A_B=0.25$, then the required
$\tau_{\mathrm{IR}}$ for $T_s=273$~K spans
\begin{equation}
\tau_{\mathrm{IR}}(273~\mathrm{K}) \approx 2.5\text{--}3.9,
\end{equation}
compared to $\tau_{\mathrm{IR}}\approx 3.1$ at $T_e=210$~K. This sensitivity underscores that the greenhouse mapping is an engineering-scale target range rather than a single predictive number; a full treatment would additionally resolve latitude, local time, and cloud/dust structure with a 3-D GCM.

\subsection{Why the water triple point is not a sufficient criterion}
\label{sec:water}

Exceeding the triple point of water does not ensure stable surface liquid water. Even at $\Ps \gtrsim \SI{6.1}{mbar}$, liquid water can remain metastable only briefly because:

\begin{enumerate}
\item \textit{Evaporation into an unsaturated atmosphere.} Mars' atmosphere is extremely dry; evaporation rates
can be high even when boiling is suppressed.
\item \textit{Radiative cooling and freezing.} Without sustained warming, transient melt refreezes quickly.
\item \textit{Local energy balance dominates.} Insolation varies strongly with latitude, season, time of day,
and dust loading; liquid water stability is therefore spatially and temporally localized.
\end{enumerate}

The conclusion that crossing the triple point is insufficient for useful \emph{surface} liquid water is consistent with a large literature on transient melting, evaporation, brines, and slope microclimates on Mars \cite{Hecht2002Metastability,Schorghofer2020Crocus,LangeForget2026Gullies}. Even if polar CO$_2$ mobilization raises pressure modestly above the triple point, the temperature shortfall remains large and evaporation/sublimation cooling is severe \cite{JakoskyEdwards2018CO2Inventory}.
Even if polar CO$_2$ is mobilized to exceed the triple point, liquid water would not be stable at the surface
because the temperature shortfall remains $\mathcal{O}(60\,\mathrm{K})$ and evaporation is rapid.

\subsection{Model validity and uncertainty: what the simplified scalings do and do not claim}
\label{sec:uncertainty}

The constraints used here are intended as feasibility discriminators, not detailed climate predictions.
The grey-atmosphere mapping from $T_e$ to $T_s$ [Eq.~(\ref{eq:eddington_grey})] compresses spectral structure (window regions, line saturation, pressure broadening) and dynamical feedbacks (dust, clouds, lapse-rate changes) into a single effective $\tau_{\rm IR}$. Consequently, inferred $\tau_{\rm IR}$ targets should be interpreted as order-of-unity ranges (e.g., $\tau_{\rm IR}\sim 2$--4 for $T_s\sim 250$--273 K at present $T_e$), not point values. Mechanism feasibility is therefore assessed by whether a proposal can plausibly supply $\mathcal{O}(1)$ optical-depth changes (or $\mathcal{O}(10^2)\,\mathrm{W\,m^{-2}}$ TOA forcing for mirror/albedo approaches) at the required timescale and maintenance burden.

Where the scaling indicates feasibility, higher-fidelity coupled 3-D GCM + aerosol microphysics + photochemistry is required to determine stability, spatial structure, and control authority. Where the scaling indicates severe shortfall (e.g., endogenous CO$_2$ inventories vs.\ E3/E4 pressures), additional modeling cannot remove the mass deficit.

\subsection{Synthesis: the feasibility rubric implied by the governing constraints}
\label{sec:rubric}

Sections~\ref{sec:mass}--\ref{sec:thermal} reduce terraforming to a small set of quantitative targets.
For any endpoint E$i$ (Sec.~\ref{sec:Ei}), a proposal must simultaneously satisfy:

\begin{enumerate}
\item \textit{Mass inventory constraint:} achieve the required pressure and composition inventories.
For E3 ($P_s=\SI{6.27}{kPa}$), Eq.~(\ref{eq:mass_pressure}) implies $M_{\rm atm}\approx 2.4\times 10^{17}$~kg (Table~\ref{tab:pressure_endpoints}).
For E4, $p_{O_2}=\SI{21}{kPa}$ implies $M_{O_2}\approx 8.2\times 10^{17}$~kg [Eq.~(\ref{eq:MO2})],
and a modest $p_{N_2}=\SI{50}{kPa}$ buffer implies $M_{N_2}\approx 1.9\times 10^{18}$~kg [Eq.~(\ref{eq:MN2})].

\item \textit{Radiative requirement:} provide either (a) direct TOA forcing $\Delta F_{\rm TOA}$ (mirrors/albedo) or
(b) sufficient longwave optical depth $\tau_{\rm IR}$ to raise $T_s$ at fixed absorbed solar power. At current insolation, Eq.~(\ref{eq:tau_required}) implies $\tau_{\rm IR}\sim 2$--4 for $T_s\sim 250$--273~K.

\item \textit{Industrial throughput/power constraint:} supply required inventories/absorbers at rate $\dot M=M/t_{\rm build}$ [Eq.~(\ref{eq:mdot_required})] and power $P=E/t_{\rm build}$ [Eq.~(\ref{eq:Pavg_required})]. Even the reversible minimum for oxygenation is $E_{\min}\approx 1.2\times 10^{25}$~J [Eq.~(\ref{eq:Emin})],
corresponding to $\sim 380$~TW averaged over $10^3$~yr.

\item \textit{Stability/maintenance constraint:} ensure the engineered state is stable against loss and sinks. For H$_2$-based warming, diffusion-limited escape implies a long-duration replenishment requirement [Eqs.~(\ref{eq:phiDL_H})--(\ref{eq:P_H2}); Table~\ref{tab:H2_maint}], even if the initial global buildout is fill-dominated for $t_{\rm build}\ll \tau_{H_2,\rm loss}$.  For CO$_2$-centric pathways, collapse/condensation introduces hysteresis risk (Sec.~\ref{sec:CO2_collapse}).
\end{enumerate}

This rubric turns the mechanism survey (Secs.~\ref{sec:endogenous}--\ref{sec:warming}) into a controlled comparison: each mechanism is evaluated by \emph{which constraint dominates} (inventory-limited, radiatively-limited, throughput/power-limited, or maintenance-limited) and whether that dominance shifts with the target endpoint. Here ``control authority'' means the achievable change in a planetary output variable (e.g., $T_s$, $P_s$, or $p_{\rm CO_2}$) per unit deployed actuation, together with the ability to maintain that output against seasonal forcing, stochastic disturbances, and internal feedbacks.

We have seen that Section~\ref{sec:thermal} provided the radiative yardstick: global ``melt-class'' climates (i.e., a climate warm enough for recurrent or regionally stable open-surface liquid water) require either very large direct forcing ($\Delta F_{\rm TOA}\sim 10^2~\mathrm{W\,m^{-2}}$ for large $T_e$ shifts) or a greenhouse optical depth $\tau_{\rm IR}\sim 2$--4 at current absorbed solar power. The remainder of the paper evaluates candidate mechanisms by whether they can plausibly deliver these targets given the mass constraints of  Sec.~\ref{sec:mass} and the industrial constraints developed later in Sec.~\ref{sec:architecture}. We begin with what Mars can supply endogenously (Sec.~\ref{sec:endogenous}).

\section{Endogenous volatile mobilization: what Mars can provide}
\label{sec:endogenous}

\subsection{CO$_2$ reservoirs and the \texorpdfstring{$\sim 20$ mbar}{20 mbar} reference case within an order-of-tens-of-mbar budget}

The most cited quantitative constraint on Mars terraforming is the inventory of accessible CO$_2$. Jakosky and Edwards \cite{JakoskyEdwards2018CO2Inventory} argued that readily accessible reservoirs (polar deposits, adsorbed regolith CO$_2$, and near-surface carbonates) plausibly sum to only \(\sim\)0.02 bar under conservative assumptions. More recent work has refined the inventory picture, including buried south-polar CO$_2$ exchange and stratigraphic constraints \cite{Buhler2020Coevolution,BuhlerPiqueux2021Obliquity,Broquet2021SouthPolarCap} and newly identified carbonate-bearing strata observed by Curiosity \cite{Tutolo2025Carbonates}. Taken together, the present evidence supports an \emph{order-of-tens-of-mbar} accessible endogenous CO$_2$ budget rather than a secure multi-bar reservoir. In what follows, \(\sim 20\) mbar is used as a representative reference case for scaling.

Two consequences follow:
\begin{enumerate}

\item \textit{Pressure shortfall.} Even if one adopts the most optimistic currently discussed \emph{accessible} endogenous inventories, the implied pressures remain far below global E3/E4 targets. A representative \(\sim 20\) mbar case corresponds to \(\Mair \sim 7.8\times 10^{16}\) kg (Table~\ref{tab:pressure_endpoints}), whereas the Armstrong threshold is 62.7 mbar and breathable endpoints require far more.

\item \textit{Warming shortfall.} For a \SI{20}{mbar} CO$_2$ atmosphere, climate models predict warming \emph{less than} \SI{10}{K} at current solar output \cite{JakoskyEdwards2018CO2Inventory}. Reaching temperatures close to melting would require $\sim \SI{1}{bar}$ CO$_2$, well beyond currently discussed accessible inventories \cite{JakoskyEdwards2018CO2Inventory}.

\end{enumerate}

\paragraph{Implied grey $\kappa_{\mathrm{IR}}$ scale for CO$_2$ under present insolation.}
Eq.~(\ref{eq:tau_kappa}) provides a useful inverse inference.
If detailed studies indicate that approaching melting under present solar output would require
$p_{\mathrm{CO_2}}\sim \mathcal{O}(1~\mathrm{bar})$ (as discussed in \cite{JakoskyEdwards2018CO2Inventory}),
then achieving a representative $\tau_{\mathrm{IR}}\sim 3$ at $P_s\sim 10^5$~Pa implies an effective grey
\begin{equation}
\kappa_{\mathrm{IR,eff}} \sim \frac{\tau_{\mathrm{IR}} g_{\mathrm{Mars}}}{P_s}
\sim 1\times 10^{-4}\ \mathrm{m^2\,kg^{-1}}.
\end{equation}
With this $\kappa_{\mathrm{IR,eff}}$, a $P_s=20$~mbar CO$_2$ atmosphere corresponds to $\tau_{\mathrm{IR}}\sim 0.05$, consistent with the conclusion that mobilizable CO$_2$ inventories cannot provide near-melting global climates.

\subsection{Energetics of CO$_2$ release: sublimation versus mining}

Releasing polar CO$_2$ ice is energetically straightforward in principle (sublimation), but not necessarily easy in practice. If the buried south polar CO$_2$ deposit corresponds to $\sim 0.006$ bar (6 mbar) \cite{JakoskyEdwards2018CO2Inventory},
its mass is $M\!\approx\!2.3\times 10^{16}$ kg.
Using a representative CO$_2$ latent heat of sublimation
$L_{\mathrm{sub}}\!\sim\!6\times 10^5\,\mathrm{J\,kg^{-1}}$ \cite{NISTCO2WebBook}, the ideal energy to sublimate this reservoir is $\sim 10^{22}$ J, or several TW sustained over a century. Mining carbonates is harder in an \emph{industrial-throughput} sense: although the per-kilogram heating energy for carbonate feedstock need not exceed the latent heat of CO$_2$ sublimation, carbonate release requires excavation, beneficiation, bulk solids handling, reactor processing, and heat delivery to geologic material rather than direct phase change of a volatile deposit. The difficulty is therefore dominated by material handling and processing logistics, not by the simple per-kilogram enthalpy comparison alone.

The representative accessible endogenous CO$_2$ reference case (\(\sim 20\)~mbar) implies both a pressure shortfall relative to E3/E4 and a warming shortfall relative to melt-class climates. Therefore, any global strategy must rely on engineered opacity/absorbers (PFCs, aerosols, CIA) and/or direct insolation modification (mirrors/albedo), which are treated in Sec.~\ref{sec:warming}.

\section{Exogenous and engineered warming mechanisms}
\label{sec:warming}

Given the representative accessible endogenous CO$_2$ reference case discussed in Sec.~\ref{sec:endogenous}, any pathway toward $T_s\gtrsim 250$~K or toward E3/E4 endpoints must rely on engineered radiative levers and/or imported volatiles. Table~\ref{tab:mechanism_matrix} summarizes the mechanism classes considered here and identifies whether each is (i) one-shot/fill-dominated, (ii) maintenance-dominated, or (iii) hybrid, with buildout and long-duration hold controlled by different constraints.

\begin{table*}[t]
\caption{Mechanism comparison matrix (order-of-magnitude). ``One-shot'' indicates a fill/build operation dominated by initial deployment, ``maintenance'' indicates a sustained industrial rate set by loss/removal timescales, and ``hybrid'' denotes pathways for which buildout is fill-dominated but long-duration hold requires replenishment. Values indicate which constraint dominates: inventory-limited (I), radiatively-limited (R), throughput/power-limited (T), or maintenance-limited (M).}
\label{tab:mechanism_matrix}
\setlength{\tabcolsep}{1pt}
\renewcommand{\arraystretch}{1.00}
\begin{tabular}{l l l l l}
\toprule
Mechanism & Control variable / efficacy  & Required scale  & Lifetime / replenishment  & Dominant bottleneck / \\
 &  & (illustrative)  &   &  limiting constraint\\

\midrule
Endogenous CO$_2$  & $\Ps$ increase via released   & Representative accessible  & One-shot (mostly), but &
Accessible-inventory limit; \\

mobilization &  CO$_2$, modest greenhouse  & 
case $\Ps\lesssim \mathcal{O}(10$--$20~\mathrm{mbar})$ &  collapse risk & limited warming \cite{JakoskyEdwards2018CO2Inventory,Tutolo2025Carbonates,Buhler2020Coevolution,BuhlerPiqueux2021Obliquity,Broquet2021SouthPolarCap}; I + R \\ [3pt]

PFC-class super- & Strong window absorption;  & Feedstock + synthesis  & Long-lived but chemistry- &
Industrial chemistry/ feed-  \\

GHG &$\kappa_{\mathrm{IR}}$ in window bands &  at extreme scale & dependent;  maintenance &
stock; lifetime \cite{Gerstell2001SuperGHG,Marinova2005PFCMars}; T  \\

 &  &  & &
(+M if short-lived)  \\[3pt]

CO$_2$--H$_2$ CIA & Requires $f_{H_2}$ at percent  & Multi-mbar to multi-kPa  & Hybrid: fill-dominated for   &
Buildout: inventory/fill +  \\

&  level (model-dependent); & global H$_2$ inventory,  & $t_{\rm build}\ll \tau_{H_2,\rm loss}$; mainte- &
power; hold phase: replen- \\

& $p_{\mathrm{H_2}}=f_{H_2}P_{\rm tot}$ & 
depending on $P_{\rm tot}$ and &  nance on long hold times &
ment against escape \cite{Ramirez2014WarmEarlyMars};\\

&  & $f_{H_2}$,  plus replenishment  &   & I/T during buildout, M+T \\
&  & against escape  &   & in hold phase  \\[3pt]

Engineered  & IR-active particles with  & 
\(\Sigma_p\) at mg\,m$^{-2}$ to sub-  & Residence time months   &
Lifetime, lofting, and radia-   \\

aerosols/  &  high thermal-IR / visible  & g\,m$^{-2}$ class depending &
to years (design- and  & 
tive efficacy \cite{Ansari2024Nanoparticles,Richardson2026IRParticles}; \\
 
nanoparticles  &  extinction ratio;    & on lifetime/design  & dynamics-dependent);  & M + T \\

 &maintained  column \(\Sigma_p\)  &  & maintenance &  \\[2pt]

Orbital mirrors /& Direct $\DeltaF_{\mathrm{TOA}}$; aggregate  & $A_m$ continent-scale for  
& Long-duration control,   &
Structure mass + deployment   \\

reflector  & area + control  &  moderate global forcing; & degradation, and partial  &
count + pointing + orbit con- \\

constellations & & $10^{13}$ m$^2$ class for  & 
replacement over hold & trol + replacement (global);  R \\

 & &  melt-class forcing &  times& 
 + T (+M on long hold times)\\[3pt]

Regional aerogel & Local radiative transfer & cm-scale layers over  & One-shot (regional) +  &
Manufacturing/placement \\

 paraterraforming &  & km$^2$--continent regions & maintenance of coverage & at area;  
 local operations \cite{Wordsworth2019Aerogel}; \\
 
&  &  &  &  T (area/logistics) \\
\bottomrule
\end{tabular}
\end{table*}

\subsection{Synthetic super-greenhouse gases (PFC-class)}
\label{sec:pfc}

Artificial greenhouse gases (perfluorocarbons and related species) have been proposed to warm Mars efficiently
because of strong absorption in infrared window regions and long chemical lifetimes.
Quantitative radiative-convective modeling has been performed for this class of gases
\cite{Gerstell2001SuperGHG,Marinova2005PFCMars}.
Two engineering constraints dominate:

\begin{enumerate}
\item \textit{Feedstock and synthesis scale.}
Even optimistic assessments imply fluorine availability and industrial throughput far beyond near-term capability.
For context, recent analysis of alternative proposals notes that PFC-based warming may require volatilizing
$\sim 10^5$ megatons of fluorine \cite{Ansari2024Nanoparticles}, i.e.\ $\sim 10^{14}$ kg of feedstock element.
\item \textit{Photochemistry and loss.}
Some candidate gases are long-lived, but lifetimes depend on UV flux, atmospheric composition, and catalytic
cycles; therefore the required production rate is set by \emph{replacement} in addition to initial fill.
\end{enumerate}

Because PFC warming does not directly provide buffer gas, it must be coupled to a separate pressure-building
pathway to achieve E3/E4.

\subsection{CO$_2$--H$_2$ collision-induced absorption}
\label{sec:cia}

CO$_2$--H$_2$ collision-induced absorption (CIA) can provide substantial greenhouse warming \emph{only when embedded in a sufficiently dense CO$_2$ background atmosphere}. In the published early-Mars literature, tens of kelvin of warming generally require percent-level H$_2$ in \(\gtrsim\)0.5--1 bar CO$_2$ atmospheres rather than in present-day thin Mars conditions \cite{Ramirez2014WarmEarlyMars,Turbet2019FarIR,Godin2020CIA}. For terraforming, CIA is attractive because H$_2$ can in principle be produced \emph{in situ} from water, with O$_2$ as a coproduct. However, hydrogen escapes efficiently from Mars, so any H$_2$-assisted warm state eventually requires replenishment.
The key architecture question is whether the dominant industrial burden is the initial global H$_2$ fill or the steady replenishment needed to hold the warm state against escape. Accordingly, a complete system analysis couples
\begin{equation}
\text{required } p_{\mathrm{H_2}} \quad \Longleftrightarrow \quad \text{required inventory } M_{\mathrm{H_2}}
\quad \Longleftrightarrow \quad \text{required production rate } \dot M_{\mathrm{H_2}}
\quad \Longleftrightarrow \quad \text{power } P.
\end{equation}
As shown below, CO$_2$--H$_2$ CIA is best treated as a coupled fill-plus-maintenance pathway: fill-dominated during century-to-millennial buildout, but replenishment-limited on long hold times.

\subsubsection{Bracketed mapping: CIA warming implies a required $f_{H_2}$ and an eventual replenishment load}

CO$_2$--H$_2$ collision-induced absorption (CIA) scales with the frequency of CO$_2$--H$_2$ collisions and therefore
requires both (i) a sufficiently dense background atmosphere and (ii) an H$_2$ mixing ratio $f_{H_2}$ at the percent-to-tens-of-percent level in published early-Mars radiative--convective studies \cite{Ramirez2014WarmEarlyMars}.
Here and throughout this subsection, \(f_{H_2}\) denotes the \emph{mole/number mixing ratio} of H$_2$ in a well-mixed multicomponent atmosphere.
For architecture-level accounting we treat $f_{H_2}$ as the primary control variable and translate it into a
diffusion-limited replenishment requirement:
\begin{equation}
\dot M_{H_2}\approx 1.2\times 10^{5}\, f_{H_2}\ {\rm kg\,s^{-1}},
\end{equation}
which gives $\dot M_{H_2}\sim 6\times 10^{3}$~kg\,s$^{-1}$ for $f_{H_2}=0.05$ and
$\dot M_{H_2}\sim 1.2\times 10^{4}$~kg\,s$^{-1}$ for $f_{H_2}=0.10$. If replenishment is provided by water electrolysis, the reversible minimum work per kg of H$_2$ is
$\sim 1.2\times 10^{8}$~J\,kg$^{-1}$, implying a minimum maintenance power
\begin{equation}
P_{H_2,\min}\sim (0.7\text{--}1.4)~{\rm TW}\quad \text{for}\quad f_{H_2}\sim 0.05\text{--}0.10,
\end{equation}
before plant inefficiencies and compression/storage.
Thus, any CIA pathway that requires $f_{H_2}$ in the few-to-ten percent range couples directly to TW-class
power and multi-$10^3$~kg\,s$^{-1}$ hydrogen production capability, in addition to the background atmospheric
mass inventory constraint required to reach the CIA-effective pressure regime.

\subsubsection{Closing the loop: CIA warming implies a required \(p_{H_2}\) and a replenishment power}

For a well-mixed multicomponent atmosphere, the mass of constituent \(i\) associated with a target partial pressure \(p_i\) is not, in general, species-independent. Instead,
\begin{equation}
M_i \simeq \frac{4\pi R_{\rm Mars}^2}{g_{\rm Mars}}\frac{\mu_i}{\bar\mu}\,p_i
= \Kmars\,\frac{\mu_i}{\bar\mu}\,p_i,
\label{eq:Mi_multicomp}
\end{equation}
where \(\mu_i\) is the molecular mass of species \(i\) and \(\bar\mu\) is the mean molecular mass of the atmospheric mixture. Thus, for an engineered atmosphere with total pressure \(P_{\rm tot}\) and hydrogen mole fraction \(f_{H_2}\), one has \(p_{H_2}=f_{H_2}P_{\rm tot}\) and
\begin{equation}
M_{H_2}\simeq \Kmars\,\frac{\mu_{H_2}}{\bar\mu}\,f_{H_2}P_{\rm tot},
\qquad
\bar\mu \simeq (1-f_{H_2})\mu_{\rm CO_2}+f_{H_2}\mu_{H_2}
\quad
\text{(for a CO$_2$--H$_2$ atmosphere)}.
\label{eq:MH2_multicomp}
\end{equation}
For the representative case \(P_{\rm tot}=\SI{0.5}{bar}\) and \(f_{H_2}=0.05\), \(\bar\mu\simeq 41.9\) g mol\(^{-1}\), \(p_{H_2}=\SI{25}{mbar}\), and
\begin{equation}
M_{H_2}\approx 4.7\times 10^{15}\,\mathrm{kg},
\end{equation}
about \(21\times\) smaller than the single-component estimate.

The reversible minimum electrolysis work to create this H$_2$ inventory is then
\begin{equation}
E_{H_2,\rm fill,min}\simeq \varepsilon_{H_2}M_{H_2}
\approx (1.2\times 10^{8}\,\mathrm{J\,kg^{-1}})(4.7\times 10^{15}\,\mathrm{kg})
\approx 5.6\times 10^{23}\,\mathrm{J}.
\end{equation}
This is still extremely large, but it is no longer comparable to the global O$_2$-production floor in Eq.~(\ref{eq:Emin}). Over a \(10^3\) yr build time it corresponds to an average reversible fill power of \(\sim 18\) TW.

Hydrogen escape then sets the ongoing replenishment requirement. Using the diffusion-limited upper bound [Eq.~(\ref{eq:MdotH2})], \(f_{H_2}=0.05\) implies
\(\dot M_{H_2}\sim 6\times 10^{3}\,\mathrm{kg\,s^{-1}}\), corresponding to a reversible electrolysis power floor
\(P_{H_2,\min}\sim 0.7~\mathrm{TW}\) (Table~\ref{tab:H2_maint}), before real inefficiencies and compression. Thus, even if CIA provides high radiative leverage, it implies (i) a substantial H$_2$ inventory tied directly to the total pressure, and (ii) a persistent TW-class industrial metabolism to maintain \(f_{H_2}\) against loss.

\subsubsection{Fill-dominated versus maintenance-dominated climate agents}
\label{sec:fill_vs_maint}

A useful architecture-level distinction is whether the industrial burden is dominated by the initial fill of the active climate agent or by its continuous replenishment against loss.
For an agent with target inventory $M^\star$ and steady loss rate $\Qloss$ at that inventory, define
\begin{equation}
\tloss \equiv \frac{M^\star}{\Qloss},
\qquad
\Lmaint \equiv \frac{\tbuild}{\tloss}
= \frac{\Qloss}{M^\star/\tbuild}.
\label{eq:fill_maint_def}
\end{equation}
If $\Lmaint\ll 1$, the pathway is fill-dominated during buildup; if $\Lmaint\gtrsim 1$, it is maintenance-dominated already during buildout.

For H$_2$-based CIA warming, take a target atmosphere with total pressure \(P_{\rm tot}\) and H$_2$ mole fraction \(f_{H_2}\), so that \(p_{H_2}=f_{H_2}P_{\rm tot}\) and
\begin{equation}
M^\star_{H_2}=\Kmars\,\frac{\mu_{H_2}}{\bar\mu}\,f_{H_2}P_{\rm tot}.
\label{eq:MH2_star_revised}
\end{equation}
Using the diffusion-limited escape form \(\Qloss{}_{H_2}\simeq 4\pi R_{\rm Mars}^2 m_{H_2}\Phi_0 f_{H_2}\), the H$_2$ loss time is
\begin{equation}
{\tau}_{H_2,\rm loss}
= \frac{M^\star_{H_2}}{\Qloss{}_{H_2}}
= \frac{P_{\rm tot}}{g_{\rm Mars}\,\bar m\,\Phi_0},
\label{eq:tauH2_loss_revised}
\end{equation}
where \(\bar m=\bar\mu m_u\) is the mean molecular mass in kilograms. For the representative \(P_{\rm tot}=\SI{0.5}{bar}\), \(f_{H_2}=0.05\) case, this gives \(\tau_{H_2,\rm loss}\approx 2.5\times 10^4\) yr, \(\Qloss{}_{H_2}\approx 6\times10^3\) kg s\(^{-1}\), and therefore
\begin{equation}
\Lambda_{\rm maint}\approx
\begin{cases}
4\times10^{-3}, & t_{\rm build}=10^2~{\rm yr},\\
4\times10^{-2}, & t_{\rm build}=10^3~{\rm yr}.
\end{cases}
\end{equation}
Therefore, for century-to-millennial global build schedules, a CIA pathway is \emph{fill-dominated} rather than maintenance-dominated: the dominant burden is creating the required global H$_2$ inventory, which can range from multi-mbar to multi-kPa depending on $P_{\rm tot}$ and $f_{
H_2}$. Maintenance remains important for long hold times, but it is not the primary scaling bottleneck during buildout.

By contrast, for an aerosol pathway with residence time $\tau_p\sim 30$--100 days,
\begin{equation}
\Lmaint^{(p)}=\frac{\tbuild}{\tau_p}\sim 10^2\text{--}10^4
\qquad (\tbuild=10^2\text{--}10^3~{\rm yr}),
\end{equation}
so aerosol warming is maintenance-dominated from the outset. This distinction sharpens the mechanism taxonomy: aerosols are truly replenishment-limited, whereas global H$_2$ CIA is typically inventory/fill-limited on civilization timescales and only secondarily maintenance-limited.

\begin{figure}[t]
\centering
\begin{tikzpicture}
\begin{loglogaxis}[
width=0.50\linewidth,
xlabel={Build time $\tbuild$ [yr]},
ylabel={H$_2$ throughput [kg\,s$^{-1}$]},
xmin=1e2, xmax=1e6,
ymin=1e2, ymax=1e7,
grid=both,
legend style={font=\footnotesize, draw=none, fill=none},
legend pos=north east
]
\addplot+[thin, samples=25, domain=1e2:1e6] {1.47e8/x};
\addlegendentry{Initial fill to $p_{H_2}=\SI{25}{mbar}$}
\addplot+[thick, dashed, samples=2, domain=1e2:1e6] {6.0e3};
\addlegendentry{Maintenance at $f_{H_2}=0.05$}
\addplot+[only marks, mark=*, mark size=2.2pt] coordinates {(2.45e4,6.0e3)};
\addlegendentry{$t_\times=\tau_{H_2,{\rm loss}}$}
\end{loglogaxis}
\end{tikzpicture}
\caption{Fill-versus-maintenance crossover for a representative global H$_2$ CIA pathway. The example assumes $P_{\rm tot}=\SI{0.5}{bar}$ and $f_{H_2}=0.05$, so that $p_{H_2}=\SI{25}{mbar}$ and $M^\star_{H_2}\approx 4.7\times10^{15}$ kg once the multicomponent correction $\mu_{H_2}/\bar\mu$ is included. The horizontal dashed line is the diffusion-limited replenishment rate $\Qloss{}_{H_2}\approx 6\times10^3$ kg\,s$^{-1}$; the solid line is the fill rate required to build the inventory over time $\tbuild$. The crossover occurs at $t_\times\simeq \tau_{H_2,\rm loss}\approx 2.5\times 10^4$ yr. Century-to-millennial build schedules lie well to the left of this crossover, so global H$_2$ CIA is still fill-dominated during buildout even though long-term replenishment remains a TW-class obligation.}
\label{fig:H2_fill_vs_maint}
\end{figure}

\subsubsection{Electrolysis-based H$_2$ fill is intrinsically coupled to O$_2$ management}
\label{sec:H2_O2_coupling}

If the required H$_2$ inventory is produced from water electrolysis, O$_2$ is generated stoichiometrically as a coproduct:
\begin{equation}
2\,\mathrm{H_2O} \rightarrow 2\,\mathrm{H_2} + \mathrm{O_2}.
\end{equation}
By mass, $M_{O_2,{\rm co}} = 8 M_{H_2}$, but the corresponding partial-pressure relation is controlled by \emph{mole number}, not by the 8:1 mass ratio. If both gases are retained in the atmosphere, one has $n_{O_2}=n_{H_2}/2$, implying to first approximation
\begin{equation}
p_{O_2,{\rm co}} \sim \frac{1}{2} p_{H_2},
\end{equation}
prior to accounting for total-pressure changes and chemical sinks. For the representative $p_{H_2}=\SI{25}{mbar}$ case, the retained coproduct is therefore of order \SI{12.5}{mbar} of O$_2$, not an Earth-like O$_2$ partial pressure. The systems point remains important: large-scale H$_2$ generation from water is inseparable from large-scale O$_2$ management and associated safety, sink-filling, storage, export, and compositional-control issues.

\subsection{Engineered IR-active aerosols and nanoparticles}
\label{sec:nanoparticles}

A recently quantified proposal is to introduce engineered \emph{IR-active} particles that interact strongly with upwelling thermal radiation while incurring a comparatively smaller visible penalty \cite{Ansari2024Nanoparticles}. The attraction of this pathway is high radiative leverage at low added atmospheric mass; the cost is that the particles must remain aloft and dispersed globally. Recent 3-D modeling further suggests that radiative--dynamical feedbacks can assist lofting and global spread for suitably designed particles \cite{Richardson2026IRParticles}.

As an illustrative conversion, \cite{Ansari2024Nanoparticles} discuss sustained particle injection of order tens of liters per second. A volumetric injection rate \(\dot V \sim \SI{30}{\liter\per\second}\) corresponds to \(\dot V\approx \SI{0.03}{\cubic\meter\per\second}\). For a representative particle bulk density \(\rho \sim \SI{3e3}{kg\,m^{-3}}\),
\begin{equation}
\dot M_{\mathrm{p}} \sim \rho \dot V \sim \SI{90}{kg\,s^{-1}}
\sim 2.8\times 10^{9}\,\mathrm{kg\,yr^{-1}}.
\end{equation}
This is \emph{small} compared to atmospheric mass requirements (Sec.~\ref{sec:mass}), but large as an industrial mass flow, comparable to terrestrial mining/processing streams. The power required depends on mining, comminution, transport, and lofting energy, which are architecture-specific. 
The maintained global-mean atmospheric column is then
\begin{equation}
\Sigma_p \simeq \frac{\dot M_p\,\tau_p}{4\pi R_{\rm Mars}^2}.
\label{eq:sigma_p_revised}
\end{equation}
For \(\tau_p\sim 30\)–100 days, \(\Sigma_p\sim 1.6\)–\(5.4~\mathrm{mg\,m^{-2}}\); for \(\tau_p\sim 10\) yr, the same injection sustains \(\Sigma_p\sim 0.2~\mathrm{g\,m^{-2}}\), comparable to the maintained columns invoked for tens of kelvin of warming \cite{Ansari2024Nanoparticles}. This provides a direct bridge between industrial injection rate and the ``mg\,m$^{-2}$ class'' entries in Table~\ref{tab:mechanism_matrix}.

Particle-induced warming is a \emph{maintenance} mechanism. If particles have an atmospheric residence time $\tau_p$ set by sedimentation and scavenging, maintaining a
global column mass $\Sigma_p$ requires a source flux $\dot\Sigma_p \approx \Sigma_p/\tau_p$. At Mars, dust residence times can range from days (local storms) to months (global events), and engineered particles must be designed to balance (i) optical efficiency, (ii) coagulation resistance, and (iii) acceptable health and environmental impacts. Thus, the required injection rates scale inversely with achievable residence time.

Detailed climate outcomes depend on particle optical properties and atmospheric dynamics, but the appropriate figure of merit is not shortwave absorption alone. Instead, the key design objective is a particle population with strong thermal-IR interaction, limited solar penalty, and a long effective atmospheric residence time. For architecture-level bookkeeping, the practical conclusion is simple: particle pathways are maintenance-limited, with industrial requirements set primarily by the pair \((\tau_p,\ \Sigma_p)\), and therefore by lifetime extension mechanisms such as slow sedimentation, photophoretic support, and radiative self-lofting \cite{Richardson2026IRParticles}. The literature values should therefore be read as highly lifetime-sensitive rather than as one-size-fits-all injection prescriptions.

For carbon-based particle concepts, the feedstock itself need not come from mineral mining: in principle it can be derived from atmospheric CO$_2$ processing, which shifts the industrial burden from excavation toward gas processing, particle fabrication, and lofting. This does not remove the maintenance character of the pathway, but it can materially change the supply-chain architecture \cite{Richardson2026IRParticles}.

\subsection{Orbital mirrors and albedo modification}
\label{sec:mirrors}

Directly increasing absorbed solar flux is conceptually simple but scale-limited. Solar-reflector and sail-based climate-engineering concepts for Mars have a substantial prior literature, including displaced non-Keplerian reflector orbits and mass-produced sail architectures launched from Earth orbit \cite{McInnes2010MarsReflectors,Handmer2024MarsSails}. The scaling developed here is therefore not intended to claim novelty for the mirror concept itself, but to place such approaches on the same forcing/area architecture axis as the other mechanisms.

\begin{figure}[t]
\includegraphics[width=0.50\linewidth]{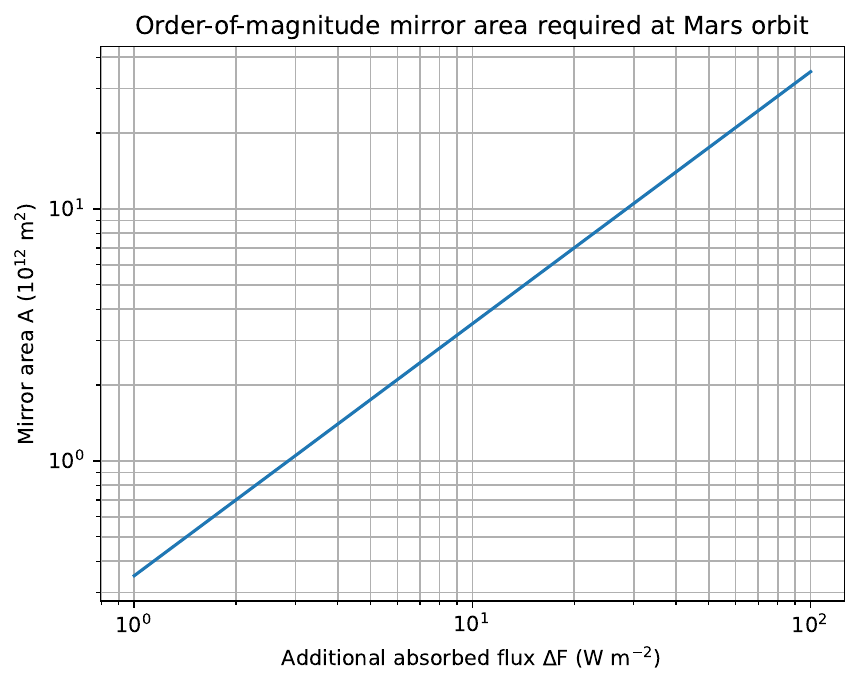}
\caption{Order-of-magnitude mirror area required to supply a global mean additional absorbed flux 
$\DeltaF_{\mathrm{TOA}}$ at Mars orbit, using Eq.~(\ref{eq:mirror_area}) with $\eta_m=0.7$ and
$\SMars=\SI{589}{W\,m^{-2}}$ \cite{Alexander2001MarsEnv}.}
\label{fig:mirror_area}
\end{figure}

Suppose one seeks an average TOA forcing $\DeltaF_{\mathrm{TOA}}$ by adding reflected sunlight.
The additional absorbed power is
\begin{equation}
\Delta P = 4\pi \Rmars^2\,\DeltaF_{\mathrm{TOA}}.
\end{equation}
A mirror of area $A_m$ at Mars orbit intercepts solar power $\SMars A_m$. If an overall efficiency $\eta_m$ accounts for reflectivity, pointing, and geometric losses, then
\begin{equation}
A_m \simeq \frac{4\pi \Rmars^2\,\DeltaF_{\mathrm{TOA}}}{\eta_m \SMars}.
\label{eq:mirror_area}
\end{equation}
Combining the definition of global mean forcing with intercepted solar flux yields Eq.~(\ref{eq:mirror_area}).
For a quick estimate, the mirror area scaling is given  by
\begin{equation}
A_m \approx 7.0\times 10^{12}\left(\frac{\DeltaF_{\mathrm{TOA}}}{20\,\mathrm{W\,m^{-2}}}\right)
\left(\frac{0.7}{\eta_m}\right)
\left(\frac{589\,\mathrm{W\,m^{-2}}}{\SMars}\right)\ \mathrm{m^2}.
\end{equation}
Figure~\ref{fig:mirror_area} shows this scaling for $\eta_m=0.7$. Even a modest global forcing $\DeltaF_{\mathrm{TOA}}=\SI{20}{W\,m^{-2}}$ implies
$A_m \sim 7\times 10^{12}\,\mathrm{m^2}$ (\(\sim 7\times 10^{6}\,\mathrm{km^2}\)), a continent-scale structure.

Albedo reduction is a related lever.
Because absorbed solar is $\SMars(1-\Abond)/4$, a change $\Delta \Abond$ yields
\begin{equation}
\DeltaF_{\mathrm{TOA}} \approx -\frac{\SMars}{4}\,\Delta \Abond .
\end{equation}
Reducing $\Abond$ by 0.05 provides only $\sim \SI{7}{W\,m^{-2}}$, illustrating that large temperature
changes require either very large albedo changes, very large mirrors, or strong greenhouse feedbacks.

\subsubsection{Mirror mass scaling}
Mirror area is not the only constraint: total mirror mass scales as
\begin{equation}
M_m \sim \sigma_m A_m,
\label{eq:mirror_mass}
\end{equation}
where $\sigma_m$ is areal density.
For $A_m\sim 7\times 10^{12}$~m$^2$ (Fig.~\ref{fig:mirror_area}) and $\sigma_m\sim 1$--$10$~g\,m$^{-2}$, the mirror mass is
$M_m\sim 10^{10}$--$10^{11}$~kg, comparable to a large terrestrial megaproject and likely requiring in-space
manufacturing for credibility.

\subsubsection{Mirror scale for melt-class forcing}

The $\sim 60$ K ``melt-class'' deficit corresponds to $\Delta F_{\rm TOA}\approx 191~\mathrm{W\,m^{-2}}$
for a direct $T_e$ increase from 210 K to 270 K [Eq.~(\ref{eq:forcing_Te})]. Using Eq.~(\ref{eq:mirror_area}), this implies a mirror area larger by a factor $\sim 191/20\approx 9.6$ than the $\Delta F_{\rm TOA}=20~\mathrm{W\,m^{-2}}$ example, i.e.\ $A_m\sim 7\times 10^{13}~\mathrm{m^2}\sim 7\times 10^{7}~\mathrm{km^2}$,
before additional geometric and control losses. This highlights why mirrors are best viewed as a high-authority but extreme-scale reflector architecture. The burden need not be a single monolithic structure---distributed reflector or sail constellations are possible---but the aggregate area, mass, deployment, and control requirements remain extreme at global-forcing levels.

As we have seen, Sec.~\ref{sec:warming} has shown that substantial warming can, in principle, be achieved with relatively small added atmospheric mass, but typically at the cost of either large initial fills plus long-term replenishment (CO$_2$--H$_2$ CIA), sustained maintenance (aerosol injection), or extreme structures (mirrors). Importantly, warming alone does not deliver a breathable atmosphere: E4 is dominated by O$_2$ and buffer-gas inventories and their associated energy and throughput requirements.
Section~\ref{sec:oxygen} quantifies these composition-driven requirements.

\subsubsection{Operational realism: reflector control, station-keeping, degradation, and replacement}

Area scaling alone is not a sufficient feasibility test for Mars reflectors. A viable reflector architecture must also be delivered to Mars space, deployed at large area, pointed continuously, kept in the required orbit family, monitored, and partially replaced as membranes and interfaces degrade over hold times of \(10^2\)--\(10^3\)~yr. Solar sails are relevant to this discussion because they can provide propellant-free transfer and non-Keplerian control authority for \emph{small} sailcraft \cite{Turyshev2023Smallsats,McInnes2010MarsReflectors,Turyshev2026SGLTrades}; they do \emph{not} by themselves close the aggregate area and operations problem for climate-scale Mars reflectors.

At Mars, the photon pressure on a perfectly reflecting sheet normal to the Sun is
\begin{equation}
p_{\rm rad,M} \simeq \frac{2S_{\rm Mars}}{c}
\approx 3.93\times10^{-6}\ {\rm N\,m^{-2}},
\label{eq:prad_mars_reflector}
\end{equation}
using \(S_{\rm Mars}\approx \SI{589}{W\,m^{-2}}\) (Table~\ref{tab:marsparams}). The corresponding characteristic acceleration for a sail or reflector of areal mass density \(\sigma\) is
\begin{equation}
a_{\rm SRP,M}\simeq \frac{p_{\rm rad,M}}{\sigma}
\approx 3.93\times10^{-4}
\left(\frac{10\ {\rm g\,m^{-2}}}{\sigma}\right)
{\rm m\,s^{-2}}.
\label{eq:a_srp_sigma}
\end{equation}
For comparison, NASA's ACS3 technology demonstration has sail area \(\sim \SI{80}{m^2}\), total spacecraft mass \(\sim \SI{16}{kg}\), and effective characteristic acceleration \(a_c\simeq \SI{0.045}{mm\,s^{-2}}\) at 1 AU \cite{Wilkie2021ACS3,Wilkie2023ACS3Update}; this corresponds to a total system areal density \(\sigma\sim 0.20\ {\rm kg\,m^{-2}}\), for which Eq.~(\ref{eq:a_srp_sigma}) gives \(a_{\rm SRP,M}\sim 2\times10^{-5}\ {\rm m\,s^{-2}}\) at Mars. NASA's current solar-sail roadmap discusses controlled sails with areas \(>2.5\times10^3\ {\rm m^2}\) (Generation~2) and \(>10^4\ {\rm m^2}\) (Generation~3a), still many orders of magnitude below the \(10^{12}\)--\(10^{13}\ {\rm m^2}\) areas implied by Eqs.~(\ref{eq:mirror_area})--(\ref{eq:forcing_Te}) \cite{Johnson2025SailRoadmap}.

It is important to separate displaced non-Keplerian orbits from literal ``hovering.'' To statically balance Mars surface gravity using photon pressure alone would require
\begin{equation}
\sigma_{\rm hover}\lesssim \frac{p_{\rm rad,M}}{g_{\rm Mars}}
\approx 1.1\times10^{-6}\ {\rm kg\,m^{-2}}
\approx 1.1\ {\rm mg\,m^{-2}},
\label{eq:sigma_hover_mars}
\end{equation}
which is far below current or roadmap-class sail areal densities. Thus low-altitude hovering or body-fixed stationkeeping is not a credible interpretation of Mars reflector concepts. What the solar-sail literature does support is the \emph{in-principle} existence of displaced reflector orbits and SRP-assisted control for appropriately light, modular systems \cite{McInnes2010MarsReflectors}.

The aggregate operations burden is also extreme. For the representative forcing levels already used in this paper,
\(A_m\sim 7\times10^{12}\ {\rm m^2}\) for \(\Delta F_{\rm TOA}=\SI{20}{W\,m^{-2}}\) and
\(A_m\sim 7\times10^{13}\ {\rm m^2}\) for melt-class forcing.
If one imagines \SI{80}{m^2} ACS3-class units, the required module count is
\begin{equation}
N_{\rm ACS3}\sim \frac{A_m}{80\ {\rm m^2}}
\sim 9\times10^{10}\text{--}9\times10^{11},
\label{eq:N_acs3_modules}
\end{equation}
while even \(\SI{1e4}{m^2}\)-class future sail/reflector units imply
\begin{equation}
N_{10^4}\sim \frac{A_m}{10^4\ {\rm m^2}}
\sim 7\times10^{8}\text{--}7\times10^{9}.
\label{eq:N_large_modules}
\end{equation}
At \(\sigma=\SI{10}{g\,m^{-2}}\), the associated reflector mass is
\(M_m\sim 7\times10^{10}\)--\(7\times10^{11}\ {\rm kg}\), consistent with Eq.~(\ref{eq:mirror_mass}). If an illustrative module lifetime of only \(\tau_{\rm mod}=\SI{30}{yr}\) is assumed, the steady replacement flow is
\begin{equation}
\dot M_{m,\rm repl}\sim \frac{M_m}{\tau_{\rm mod}}
\sim 7\times10^{1}\text{--}7\times10^{2}\ {\rm kg\,s^{-1}}
\qquad (\sigma=\SI{10}{g\,m^{-2}}),
\label{eq:mirror_replacement_rate}
\end{equation}
before accounting for failed deployment, attitude-control hardware, metrology, or disposal/reconfiguration.

A convenient operations metric is the required \emph{area replacement rate}:
\begin{equation}
\dot A_{m,\rm repl}\sim \frac{A_m}{\tau_{\rm mod}}
\sim 7\times10^3\text{--}7\times10^4\ {\rm m^2\,s^{-1}}
\qquad (\tau_{\rm mod}=\SI{30}{yr}),
\label{eq:mirror_area_replacement_rate}
\end{equation}
for the representative forcing range already considered in this paper. In module units, the same replacement burden corresponds to
\begin{equation}
\dot N_{10^4}\sim \frac{N_{10^4}}{\tau_{\rm mod}}
\sim 0.7\text{--}7\ {\rm s^{-1}},
\qquad
\dot N_{\rm ACS3}\sim \frac{N_{\rm ACS3}}{\tau_{\rm mod}}
\sim 10^2\text{--}10^3\ {\rm s^{-1}},
\label{eq:mirror_module_replacement_rate}
\end{equation}
for \SI{1e4}{m^2}-class and ACS3-class units, respectively. This makes explicit that the reflector problem is not merely a launch or transfer problem: it is a long-duration manufacturing, deployment, metrology, control, and retirement/replacement problem at rates that remain extreme even if the individual modules are assumed to be lightweight.

This local-versus-global distinction is also reflected in recent roadmap analyses: orbiting reflectors may be relevant for warming specific sites or volatile reservoirs, but their local-base examples already require 100s of km$^2$ of reflector area, far below yet qualitatively consistent with the extreme aggregate-area burden implied here for global forcing \cite{Kite2026Roadmap}.

Finally, long-duration reflector operations are limited by environmental durability as well as by orbital control. NASA solar-sail materials work emphasizes that mission success depends on membrane/interface durability over packaging, deployment, and operations; radiation-driven adhesive degradation and associated reflectivity loss are active engineering concerns rather than solved issues \cite{Kang2020SolarSailDurability}. The correct interpretation is therefore not that Mars reflectors are ruled out by physics, but that they belong to a \emph{far-horizon space-manufacturing and operations} regime: solar sails may help with transfer and trim of modular reflectors, yet the dominant bottlenecks remain aggregate area, mass fabrication, deployment count, pointing/control, durability, and replacement over centuries to millennia.

\section{Oxygenation and breathable endpoints}
\label{sec:oxygen}

\subsection{O$_2$ mass required for breathable partial pressures}

A breathable atmosphere requires substantial O$_2$ partial pressure.
Using Eq.~(\ref{eq:mass_pressure}), the mass of O$_2$ corresponding to a partial pressure
$p_{\mathrm{O_2}}$ is
\begin{equation}
M_{\mathrm{O_2}} \simeq \frac{4\pi \Rmars^2}{\gMars}\,p_{\mathrm{O_2}}.
\label{eq:MO20}
\end{equation}
For $p_{\mathrm{O_2}}=\SI{21}{kPa}$ (Earth-like),
\begin{equation}
M_{\mathrm{O_2}} \approx 8.2\times 10^{17}\,\mathrm{kg}.
\label{eq:MO2}
\end{equation}
This is already comparable to the total mass of a \SI{0.2}{bar} atmosphere.

\paragraph{Covered-area (``worldhouse'') crossover.}
A sealed enclosure trades the open-atmosphere hydrostatic mass requirement into a finite-volume gas requirement. For a covered area $A_{\rm cov}$ with mean interior height $H$, interior pressure $P_{\rm in}$, mean gas molecular mass $\bar\mu$, and temperature $T$, the enclosed gas mass is
\begin{equation}
M_{\rm encl}\simeq A_{\rm cov} H\,\frac{\bar\mu P_{\rm in}}{RT}
= A_{\rm cov}\,\frac{H}{H_s}\,\frac{P_{\rm in}}{g_{\rm Mars}},
\qquad
H_s\equiv \frac{RT}{\bar\mu g_{\rm Mars}}.
\label{eq:Mencl}
\end{equation}
Thus the ratio of enclosed-gas mass to open-atmosphere hydrostatic mass over the same area is simply $H/H_s$. For a representative breathable enclosure with $H=\SI{3}{m}$, $P_{\rm in}=\SI{50}{kPa}$, $T=\SI{250}{K}$, and $\bar\mu\approx 29$ g mol$^{-1}$, Mars gives $H_s\approx \SI{19}{km}$ and hence $H/H_s\approx 1.6\times 10^{-4}$. If applied over the full planetary surface area, this corresponds to
\begin{equation}
M_{\rm encl,global}\approx 3.0\times10^{14}\ \mathrm{kg},
\end{equation}
about $6.4\times10^3$ smaller than an open global \SI{50}{kPa} atmosphere. This does not make a global worldhouse easy—structural mass, support spacing, puncture tolerance, maintenance, and thermal control become the dominant burdens—but it reinforces the paper's broader conclusion that covered-area strategies can beat open-atmosphere strategies by trading atmospheric mass into engineered area and support infrastructure.

\subsection{Thermodynamic minimum energy to produce O$_2$ from water}

Producing O$_2$ by electrolyzing water,
\begin{equation} 2\,\mathrm{H_2O} \rightarrow 2\,\mathrm{H_2} + \mathrm{O_2},
\end{equation}
requires a minimum Gibbs free energy input $\Delta G^\circ$ per mole of O$_2$.
Using standard thermochemical data \cite{Chase1998JANAF}, the reversible minimum at \SI{298}{K} is
$\Delta G^\circ \approx \SI{474}{kJ\,mol^{-1}}$ per mole of O$_2$
(\(\approx \SI{14.8}{MJ\,kg^{-1}}\) of O$_2$).
Therefore, the idealized minimum energy to produce Eq.~(\ref{eq:MO2}) is
\begin{equation}
E_{\min} \approx \left(\frac{M_{\mathrm{O_2}}}{0.032\,\mathrm{kg\,mol^{-1}}}\right)
\Delta G^\circ \approx 1.2\times 10^{25}\,\mathrm{J}.
\label{eq:Emin}
\end{equation}
Real systems require larger energy due to overpotentials, compression/liquefaction, plant losses, and ancillary mining/processing. Even the theoretical minimum corresponds to an average power of $\sim \SI{380}{\tera\watt}$ sustained for $1000~\mathrm{yr}$, illustrating that E4 endpoints are
\emph{energy-dominated}.

Stoichiometrically, producing one kilogram of O$_2$ from water requires \(36/32=1.125\) kg of H$_2$O processed. Thus producing \(M_{O_2}\approx 8.2\times 10^{17}\) kg implies processing \(M_{H_2O}\approx 9.2\times 10^{17}\) kg of water, equivalent to a global water layer of thickness
\(\sim M_{H_2O}/(\rho_w\,4\pi R_{\rm Mars}^2)\approx 6~\mathrm{m}\) for \(\rho_w\approx 1000~\mathrm{kg\,m^{-3}}\).

As a benchmark against purely industrial oxygenation, one may compare with Earth-like biological productivity. Using an Earth-like global ${\rm NPP}_C\approx 105$ Pg C yr$^{-1}$ \cite{Field1998BiosphereNPP}$\approx 1.05\times10^{14}$ kg C yr$^{-1}$, the corresponding gross O$_2$-equivalent photosynthetic flux is
\begin{equation}
\dot M_{O_2,{\rm gross}}\sim \frac{32}{12}\,{\rm NPP}_C \approx 2.8\times10^{14}\ \mathrm{kg\,yr^{-1}}.
\end{equation}
However, only the fraction that is not rapidly reconsumed by respiration, oxidation, and recycling contributes to long-term atmospheric oxygenation. Writing this effective burial/export fraction as $\beta$, the oxygenation time becomes
\begin{equation}
t_{O_2,{\rm bio}}\sim \frac{M_{O_2}}{\beta\,\dot M_{O_2,{\rm gross}}}
\approx 2.9\times10^3\,\beta^{-1}\ \mathrm{yr}
\left(\frac{M_{O_2}}{8.2\times10^{17}\,\mathrm{kg}}\right)
\left(\frac{105\,\mathrm{Pg\ C\,yr^{-1}}}{{\rm NPP}_C}\right).
\end{equation}
Thus biological oxygenation can relax the electrical-power requirement relative to abiotic electrolysis, but it does not remove the fundamental sink/burial bottleneck; for $\beta\ll 1$, timescales readily extend to $10^5$--$10^6$ yr.

\subsection{MOXIE as a scale anchor}

MOXIE demonstrated oxygen production from atmospheric CO$_2$ on Mars \cite{Hoffman2022MOXIE,NASAMOXIE2023}. Such demonstrations are crucial technology anchors, but MOXIE is not invoked here as a direct terraforming route; rather, it serves as a scale benchmark for the gulf between a successful Mars oxygen-production demonstration and the exaton-class oxygen inventories relevant to atmospheric transformation. A breathable-atmosphere inventory $M_{\mathrm{O_2}}\sim 10^{18}$~kg is $\sim 10^{19}$ times larger than an $\mathcal{O}(10^2)$~g-class technology demonstration output, even before accounting for compression, storage, distribution, and O$_2$ sinks into the regolith and crust. Therefore oxygenation is not a ``scale-up by a factor of ten'' problem; it is an orders-of-magnitude industrial-civilization problem.

\subsection{O$_2$ sinks: oxidation capacity of a basaltic regolith/crust can be comparable to atmospheric targets}
\label{sec:O2_sinks}

A breathable endpoint requires not only producing $M_{\mathrm{O_2}}$ (Sec.~VII), but also overcoming sinks as O$_2$ oxidizes reduced minerals. An upper-envelope \emph{capacity estimate} is obtained by assuming an accessible oxidizable layer of thickness \(d\) and density \(\rho\) with representative FeO mass fraction \(w_{\rm FeO}\)  (e.g., FeO $\sim 18$ wt\% in Mars-relevant silicates \cite{Taylor2013BulkMars,McSween2009Crust}).

If FeO is oxidized to Fe$_2$O$_3$ via $4\mathrm{FeO}+\mathrm{O_2}\rightarrow 2\mathrm{Fe_2O_3}$,
the required moles of O$_2$ per mole of FeO is $1/4$.
The corresponding O$_2$ mass sink capacity of a global layer is approximately
\begin{equation}
M_{\mathrm{O_2,sink}}(t) \sim \frac{1}{4}\left(\frac{w_{\mathrm{FeO}}\,4\pi R_{\mathrm{Mars}}^{2}\rho\,\deff(t)}{\mu_{\mathrm{FeO}}}\right)\mu_{\mathrm{O_2}},
\label{eq:MO2_sink}
\end{equation}
where $\mu_{\mathrm{FeO}}$ and $\mu_{\mathrm{O_2}}$ are molar masses.
Numerically,
\begin{equation}
M_{\mathrm{O_2,sink}}(t)
\sim 9\times 10^{17}~\mathrm{kg}
\left(\frac{w_{\mathrm{FeO}}}{0.18}\right)
\left(\frac{\rho}{\SI{3000}{kg\,m^{-3}}}\right)
\left(\frac{\deff(t)}{\SI{100}{m}}\right),
\end{equation}
comparable to the atmospheric inventory required for $p_{\mathrm{O_2}}\sim \SI{21}{kPa}$ (Sec.~\ref{sec:oxygen}). This illustrates that oxygenation is generically a coupled production-plus-sink-filling problem, and that the energy and timescales in Sec.~\ref{sec:oxygen} should be interpreted as optimistic lower bounds.

The normalization \(\deff=\SI{100}{m}\) is illustrative rather than a measured regolith-thickness claim. The physically relevant quantity is the \emph{effective oxidizable depth} actually accessed over the build/hold interval by permeability, weathering kinetics, impacts, and especially excavation and construction. Accordingly, Eq.~(\ref{eq:MO2_sink}) should be interpreted as a sink-capacity envelope, not as a century-scale oxidation forecast.

The relevant sink capacity is set by an \emph{effective oxidizable depth} \(\deff(t)\), not necessarily the full
geometric depth of regolith/crust, because oxygen uptake depends on kinetics, permeability, and how much fresh
reduced material is exposed by impacts, dust gardening, and (critically) human excavation and construction.
In a systems sense, oxygenation couples to an intentional strategy for managing \(\deff(t)\):
\emph{passivation} (oxidize reactive material in engineered reactors), \emph{sealing/vitrification} or
surface stabilization to reduce atmospheric access, and \emph{controlled excavation} to avoid continually
exposing fresh reduced minerals.
Naming \(\deff(t)\) makes explicit that sink-filling is not only a ``one scary number'' but a coupled design and
operations variable in a long-duration terraforming architecture.

As we have seen above, Section~\ref{sec:oxygen} demonstrates that breathable endpoints are fundamentally energy- and sink-limited: $M_{O_2}\sim 10^{18}$~kg and $E_{\min}\sim 10^{25}$~J even before inefficiencies, and regolith oxidation can absorb an O$_2$ inventory comparable to atmospheric targets. Whether such inventories persist requires explicit accounting of escape, condensation collapse, and geochemical sequestration, which motivates the stability analysis in Sec.~\ref{sec:loss}.

\section{Atmospheric loss, retention, and control}
\label{sec:loss}

A terraformed Mars must retain its engineered atmosphere against escape to space and against surface sequestration (carbonates, adsorption, polar condensation).

\subsection{Escape to space: why it matters more for H$_2$ than for CO$_2$}

MAVEN observations quantify present-day atmospheric loss and its dependence on solar conditions
\cite{Jakosky2018AtmosLoss}. For heavy species (CO$_2$, N$_2$), bar-level atmospheres have enormous mass and are likely to be lost on geological timescales even without a global magnetic field, whereas for hydrogen the relevant benchmark escape timescale can be much shorter. In this paper we use the diffusion-limited form as a conservative upper-bound maintenance benchmark for H$_2$-rich warm states; actual escape may be smaller if other limits intervene. Thus, H$_2$-based warming (Sec.~\ref{sec:cia}) implies a persistent replenishment requirement on long hold times, even if the initial global buildout is fill-dominated for $t_{\rm build}\ll \tau_{H_2,\rm loss}$.

\subsection{Artificial magnetic shielding}

Artificial magnetic shielding may reduce ion escape and alter the radiation environment \cite{Green2017FutureMarsEnv}, but it does not by itself remove the dominant neutral-H diffusion-limited escape benchmark used here for CO$_2$--H$_2$ maintenance. We therefore treat magnetic shielding as a secondary retention/radiation-environment concept rather than as a primary solution to the H$_2$ maintenance burden.

\subsection{Hydrogen escape sets the long-duration replenishment floor for CO$_2$--H$_2$ warming}
\label{sec:H2_escape}

A key distinction between CO$_2$-mass-based warming and CO$_2$--H$_2$ CIA warming is that hydrogen can be lost
rapidly, implying a persistent replenishment requirement. A standard upper bound on hydrogen escape is the diffusion-limited flux, often expressed in terms of the hydrogen
atom flux (``H equivalents'') (e.g., \cite{Hunten1973,CatlingKasting2017}):
\begin{equation}
\Phi_{\rm DL,H} \approx \Phi_0\, f_H,
\label{eq:phiDL_H}
\end{equation}
where $f_H$ is the total hydrogen mixing ratio expressed as H atoms (summing over all H-bearing species).
If hydrogen is supplied primarily as H$_2$, then $f_H \approx 2 f_{H_2}$ and the corresponding H$_2$ molecular flux is
\(\Phi_{\rm DL,H_2}\approx \tfrac{1}{2}\Phi_{\rm DL,H}\approx \Phi_0 f_{H_2}\). The corresponding global H$_2$ mass loss rate is then
\begin{equation}
\dot{M}_{H_2} \approx 4\pi R_{\rm Mars}^2\, m_{H_2}\, \Phi_{\rm DL,H_2}
\approx 1.2\times 10^{5}\left(\frac{\Phi_0}{2.5\times 10^{13}\ \mathrm{cm^{-2}\,s^{-1}}}\right) f_{H_2}\ \mathrm{kg\,s^{-1}},
\label{eq:MdotH2}
\end{equation}
where $m_{H_2}$ is the molecular mass of H$_2$ and we used Mars' surface area for scaling.

Even modest $f_{H_2}$ can therefore imply $\dot{M}_{H_2}\sim 10^3$--$10^4~\mathrm{kg\,s^{-1}}$ class replenishment. If H$_2$ is produced by water electrolysis, the reversible specific energy is $\sim 1.2\times 10^{8}$~J\,kg$^{-1}$ of H$_2$, implying a minimum maintenance power
\begin{equation}
\label{eq:P_H2}
P_{H_2,\min} \sim 0.1\text{--}1~\mathrm{TW}\quad \text{for}\quad \dot{M}_{H_2}\sim 10^3\text{--}10^4~\mathrm{kg\,s^{-1}},
\end{equation}
before compression, storage, and plant inefficiencies.
Therefore CO$_2$--H$_2$ warming should be analyzed as a coupled fill-plus-maintenance system: Sec.~\ref{sec:fill_vs_maint} shows that the initial global H$_2$ fill dominates for $t_{\rm build}\ll \tau_{H_2,\rm loss}$, while the equations in this subsection quantify the replenishment floor relevant to long-duration hold of the warm state.

\begin{table}[t]
\caption{Illustrative diffusion-limited H$_2$ replenishment requirements using Eq.~(\ref{eq:MdotH2}) and a reversible
electrolysis specific energy of \(1.2\times 10^8~\mathrm{J\,kg^{-1}}\) of H$_2$.}
\label{tab:H2_maint}
\begin{tabular}{cccc}
\toprule
$f_{H_2}$ & $\dot{M}_{H_2}$ (kg\,s$^{-1}$) & $P_{H_2,\min}$ (TW) \\
\midrule
0.01 & $1.2\times 10^{3}$ & 0.14 \\
0.05 & $6.0\times 10^{3}$ & 0.72 \\
0.10 & $1.2\times 10^{4}$ & 1.44 \\
\bottomrule
\end{tabular}
\end{table}

\subsection{CO$_2$ condensation and atmospheric collapse: a stability constraint on CO$_2$-centric pathways}
\label{sec:CO2_collapse}

CO$_2$-centric terraforming pathways must contend with the possibility of atmospheric collapse, i.e., condensation of CO$_2$ into permanent polar deposits that reduce $P_s$ and weaken greenhouse warming. Three-dimensional climate simulations of early Mars with 0.1--7 bar CO$_2$ atmospheres show that CO$_2$ ice clouds, obliquity, dust loading, and surface properties can lead to climates where CO$_2$ condensation and cold trapping limit warming and can induce collapse-like behavior under some conditions \cite{Forget2013EarlyMarsCO2}.

In a first-order control sense, CO$_2$ stability requires maintaining polar temperatures above the CO$_2$
condensation threshold for the prevailing $p_{\mathrm{CO_2}}$. Thus, even if CO$_2$ is available in principle, the engineered climate may require active control authority (e.g., directed insolation, albedo management, or supplemental absorbers) to avoid pressure-loss hysteresis
and to keep the system on a warm branch.

\subsection{Regional--global crossover law}
\label{sec:regional_crossover}

The principal scaling advantage of regional habitability is that required gas inventory scales with covered area rather than planetary area.
For a pressurized region of area $\Areg$ and target pressure $\Preg$,
\begin{equation}
M_{\rm reg}=\frac{\Areg\,\Preg}{g_{\rm Mars}}.
\label{eq:Mreg}
\end{equation}
If the net available gas-production or gas-delivery throughput is $\dot M_{\rm net}$, the fill time is
\begin{equation}
t_{\rm fill,reg}=\frac{\Areg\,\Preg}{g_{\rm Mars}\dot M_{\rm net}},
\label{eq:tfill_reg}
\end{equation}
or, equivalently, the maximum area that can be brought to pressure in a build time $\tbuild$ is
\begin{equation}
A_{\max}=\frac{g_{\rm Mars}\dot M_{\rm net}\tbuild}{\Preg}.
\label{eq:Amax_reg}
\end{equation}
Numerically,
\begin{equation}
A_{\max}\approx 1.17\times10^{6}\ {\rm km^2}
\left(\frac{\dot M_{\rm net}}{10^{6}\ {\rm kg\,s^{-1}}}\right)
\left(\frac{\tbuild}{100\ {\rm yr}}\right)
\left(\frac{10\ {\rm kPa}}{\Preg}\right).
\label{eq:Amax_reg_num}
\end{equation}

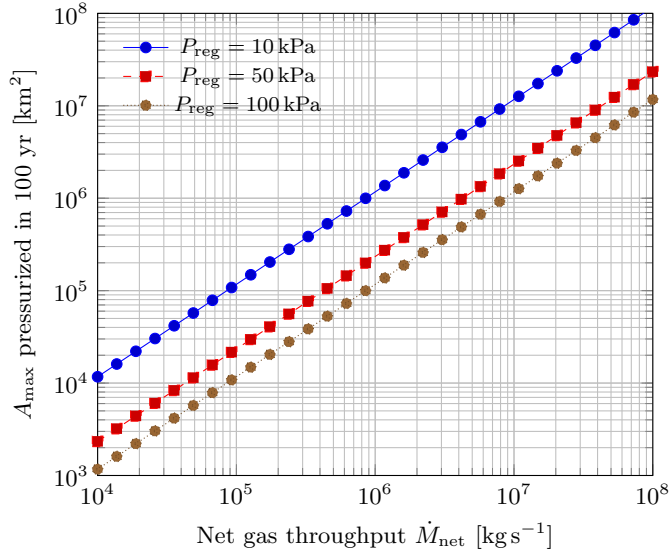
\begin{figure}[t]
\centering
\begin{tikzpicture}
\begin{loglogaxis}[
width=0.50\linewidth,
xlabel={Net gas throughput $\dot M_{\rm net}$ [kg\,s$^{-1}$]},
ylabel={$A_{\max}$ pressurized in 100 yr [km$^2$]},
xmin=1e4, xmax=1e8,
ymin=1e3, ymax=1e8,
grid=both,
legend style={font=\footnotesize, draw=none, fill=none},
legend pos=north west
]
\addplot+[thin, samples=30, domain=1e4:1e8] {1.17*x};
\addlegendentry{$P_{\rm reg}=\SI{10}{kPa}$}
\addplot+[thin, dashed, samples=30, domain=1e4:1e8] {0.234*x};
\addlegendentry{$P_{\rm reg}=\SI{50}{kPa}$}
\addplot+[thin, densely dotted, samples=30, domain=1e4:1e8] {0.117*x};
\addlegendentry{$P_{\rm reg}=\SI{100}{kPa}$}
\end{loglogaxis}
\end{tikzpicture}
\caption{Maximum regional area that can be pressurized within \SI{100}{yr} as a function of net gas throughput, from Eq.~(\ref{eq:Amax_reg}). The linear dependence in log--log space makes the regional/global crossover transparent: throughputs that are irrelevant at planetary scale can still support geographically large regional programs.}
\label{fig:Amax_regional}
\end{figure}

\begin{table}[t]
\caption{Maximum pressurized area $A_{\max}$ achievable in \SI{100}{yr} for representative net gas throughputs and target regional pressures, from Eq.~(\ref{eq:Amax_reg_num}).}
\label{tab:Amax_regional}
\centering
\small
\begin{tabular}{cccc}
\toprule
$\dot M_{\rm net}$ (kg\,s$^{-1}$) & $P_{\rm reg}=\SI{10}{kPa}$ & $P_{\rm reg}=\SI{50}{kPa}$ & $P_{\rm reg}=\SI{100}{kPa}$ \\
\midrule
$10^{5}$ & $1.17\times10^{5}$ km$^2$ & $2.34\times10^{4}$ km$^2$ & $1.17\times10^{4}$ km$^2$ \\
$10^{6}$ & $1.17\times10^{6}$ km$^2$ & $2.34\times10^{5}$ km$^2$ & $1.17\times10^{5}$ km$^2$ \\
$10^{7}$ & $1.17\times10^{7}$ km$^2$ & $2.34\times10^{6}$ km$^2$ & $1.17\times10^{6}$ km$^2$ \\
\bottomrule
\end{tabular}
\end{table}

Eq.~(\ref{eq:Amax_reg_num}) provides a useful crossover criterion. A throughput that is utterly insufficient for global E3/E4 can nevertheless support a regional E2/E3-scale program over areas comparable to large terrestrial countries or subcontinental provinces. This is the quantitative reason that regional paraterraforming and locally pressurized biospheres remain the most credible near- to mid-term pathway even when global atmospheric transformation does not. Figure~\ref{fig:Amax_regional} and Table~\ref{tab:Amax_regional} show the maximum regional area that can be pressurized within \SI{100}{yr} as a function of net gas throughput.  

\subsection{A minimal state-space model for Mars as a controlled climate--industrial system}
\label{sec:state_space}

The preceding sections motivate treating terraforming as a coupled dynamical system rather than as a sequence of isolated static budgets. To avoid redundancy, we treat the partial pressures as states and the total surface pressure as an algebraic output. A minimal slow state vector may therefore be written as
\begin{equation}
\mathbf{x} \equiv
\begin{bmatrix}
p_{\rm CO_2} &
p_{\rm H_2} &
p_{\rm O_2} &
p_{\rm N_2} &
M_p &
\alpha_{\rm pol} &
T_s
\end{bmatrix}^{\mathsf T},
\label{eq:state_vector}
\end{equation}
where $M_p$ is the atmospheric mass of engineered particles and $\alpha_{\rm pol}$ is the fraction of the accessible CO$_2$ reservoir that is cold-trapped in permanent polar deposits. A corresponding control vector is
\begin{equation}
\mathbf{u} \equiv
\begin{bmatrix}
\dot M_{\rm CO_2} &
\dot M_{\rm H_2} &
\dot M_{\rm O_2} &
\dot M_{\rm N_2} &
\dot M_{p,\rm inj} &
A_m
\end{bmatrix}^{\mathsf T}.
\label{eq:control_vector}
\end{equation}
The total surface pressure is then treated diagnostically as
\begin{equation}
P_s = p_{\rm CO_2}+p_{\rm H_2}+p_{\rm O_2}+p_{\rm N_2}+p_{\rm other},
\label{eq:Ps_algebraic}
\end{equation}
where $p_{\rm other}$ denotes any neglected minor constituents and may be set to zero in the minimal closure.

\begin{table*}[t]
\caption{Qualitative sign structure of the steady-state gain matrix $G_{kj}=\partial y_k/\partial u_j$ around a warm operating point, with $P_s$ treated as an algebraic output via Eq.~(\ref{eq:Ps_algebraic}) rather than as an independent state. Parentheses indicate a weaker or indirect coupling.}
\label{tab:gain_sign}
\centering
\small
\begin{tabular}{lcccccc}
\toprule
Control input $u_j$ & $T_s$ & $P_s$ & $p_{\rm CO_2}$ & $p_{\rm O_2}$ & $p_{\rm H_2}$ & $\alpha_{\rm pol}$ \\
\midrule
$\dot M_{\rm CO_2}$ & $+$ & $+$ & $+$ & $0$ & $0$ & $\pm$ \\
$\dot M_{\rm H_2}$  & $+$ & $(+)$ & $0$ & $0$ & $+$ & $-$ \\
$\dot M_{\rm O_2}$  & $(+)$ & $+$ & $0$ & $+$ & $0$ & $0$ \\
$\dot M_{\rm N_2}$  & $(+)$ & $+$ & $0$ & $0$ & $0$ & $(-)$ \\
$\dot M_{p,\rm inj}$ & $+$ & $0$ & $0$ & $0$ & $0$ & $-$ \\
$A_m$               & $+$ & $0$ & $0$ & $0$ & $0$ & $-$ \\
\bottomrule
\end{tabular}
\end{table*}

A first-order architecture model is then
\begin{align}
\dot p_{\rm CO_2} &=
\frac{g_{\rm Mars}}{4\pi R_{\rm Mars}^2}
\left(
\dot M_{\rm CO_2}
-\dot M_{\rm CO_2,loss}
-\dot M_{\rm CO_2,sink,chem}
-M_{\rm CO_2,acc}\,\dot\alpha_{\rm pol}
\right),
\label{eq:pCO2_dyn}
\\
\dot p_{\rm H_2} &=
\frac{g_{\rm Mars}}{4\pi R_{\rm Mars}^2}\dot M_{\rm H_2}
-\frac{p_{\rm H_2}}{\tau_{H_2,\rm loss}},
\label{eq:pH2_dyn}
\\
\dot p_{\rm O_2} &=
\frac{g_{\rm Mars}}{4\pi R_{\rm Mars}^2}\dot M_{\rm O_2}
-\frac{p_{\rm O_2}}{\tau_{O_2,\rm sink}},
\label{eq:pO2_dyn}
\\
\dot p_{\rm N_2} &=
\frac{g_{\rm Mars}}{4\pi R_{\rm Mars}^2}
\left(
\dot M_{\rm N_2}
-\dot M_{\rm N_2,loss}
-\dot M_{\rm N_2,sink}
\right),
\label{eq:pN2_dyn}
\\
\dot M_p &= \dot M_{p,\rm inj} - \frac{M_p}{\tau_p},
\label{eq:Mp_dyn}
\\
\dot\alpha_{\rm pol} &= \mathcal{C}(T_{\rm pol},p_{\rm CO_2})-\mathcal{S}(A_m,\dot M_{p,\rm inj},\dot M_{\rm H_2},\ldots),
\label{eq:alphapol_dyn}
\\
C_{\rm eff}\dot T_s &= F_{\rm in}(A_B,A_m,t)
+F_{\rm GH}(p_{\rm CO_2},p_{\rm H_2},M_p,\ldots)
-{\rm OLR}(T_s,\mathbf{x}),
\label{eq:Ts_dyn}
\end{align}
where $M_{\rm CO_2,acc}$ is the accessible CO$_2$ reservoir against which $\alpha_{\rm pol}$ is defined and $C_{\rm eff}$ is an effective surface--atmosphere heat capacity.

Linearizing about an operating point $(\mathbf{x}_0,\mathbf{u}_0)$ gives
\begin{equation}
\delta \dot{\mathbf{x}} = \mathbf{A}\,\delta\mathbf{x} + \mathbf{B}\,\delta\mathbf{u},
\qquad
\delta\mathbf{y}=\mathbf{C}\,\delta\mathbf{x},
\label{eq:linearized_system}
\end{equation}
where $\mathbf{y}$ collects the controlled outputs of interest, e.g.
\begin{equation}
\mathbf{y}=
\begin{bmatrix}
T_s &
P_s &
p_{\rm O_2} &
p_{\rm CO_2} &
\alpha_{\rm pol}
\end{bmatrix}^{\mathsf T},
\qquad
P_s \ \text{from Eq.~(\ref{eq:Ps_algebraic})}.
\end{equation}
The steady-state control-authority matrix is then
\begin{equation}
\mathbf{G}(0) = -\,\mathbf{C}\mathbf{A}^{-1}\mathbf{B},
\label{eq:G0}
\end{equation}
which quantifies how strongly each industrial control lever changes each planetary output near the operating point.

The value of Eqs.~(\ref{eq:pCO2_dyn})--(\ref{eq:G0}) is not that they replace high-fidelity climate modeling, but that they define the minimum dynamical structure that any credible terraforming architecture must close: source terms, loss terms, sink terms, reservoir bookkeeping, and control authority. This makes the phrase ``planetary-scale control system'' mathematically precise.

\section{Energy delivery and system architecture}
\label{sec:architecture}

\subsection{Energetic ``success'' thresholds}

Table~\ref{tab:energy} summarizes representative energy scales for several canonical operations. These numbers should be interpreted as \emph{lower bounds}; real architectures add large multiplicative factors for inefficiency, infrastructure, and operations.

\begin{table}[t]
\caption{Representative energetic scales (order-of-magnitude).}
\label{tab:energy}
\begin{tabular}{lll}
\toprule
Operation & Scale assumption & Energy / size \\
\midrule
Sublimate \(\sim 6\) mbar CO$_2$ polar deposit & $M\!\sim\!2.3\times 10^{16}$ kg, $L_{\rm sub}\!\sim\!6\times 10^5$ J/kg & $\sim 1\times 10^{22}$ J \\
Produce Earth-like $p_{\mathrm{O_2}}=\SI{21}{kPa}$ from water & $M_{\mathrm{O_2}}\!\sim\!8\times 10^{17}$ kg, reversible $\Delta G^\circ$ & $\sim 1\times 10^{25}$ J \\
Global forcing by orbital mirrors & $\DeltaF_{\mathrm{TOA}}=\SI{20}{W\,m^{-2}}$, $\eta_m=0.7$ & $A_m \sim 7\times 10^{12}$ m$^2$ \\
Decadal engineered nanoparticle injection & $\dot V=\SI{30}{\liter\per\second}$, $\rho=\SI{3e3}{kg\,m^{-3}}$ & $\sim 3\times 10^{10}$ kg/decade \\
\bottomrule
\end{tabular}
\end{table}

\subsection{Throughput requirement: required mass flow and average power versus build time}
\label{sec:throughput}

For any proposed endpoint with target inventory $M_{\mathrm{target}}$ achieved over a build time $t_{\mathrm{build}}$,
the implied mean industrial throughput is
\begin{equation}
\dot{M}_{\mathrm{req}} \;\equiv\; \frac{M_{\mathrm{target}}}{t_{\mathrm{build}}}.
\label{eq:mdot_required}
\end{equation}
Likewise, for an intervention requiring integrated energy $E$, the implied mean power is
\begin{equation}
P_{\mathrm{avg}} \;\equiv\; \frac{E}{t_{\mathrm{build}}}.
\label{eq:Pavg_required}
\end{equation}
These relations are decisive because they convert ``large but abstract'' quantities into concrete kg\,s$^{-1}$ and TW
requirements that can be compared with industrial analogs and with plausible Mars power architectures.

\begin{table*}[t]
\caption{Illustrative throughput and power requirements for representative endpoints, showing the scaling with assumed
build time. Values use $M_{O_2}\approx 8.2\times 10^{17}$~kg for $p_{O_2}=21$~kPa [Eqs.~(\ref{eq:MO20})--(\ref{eq:MO2})],
$M_{N_2}\approx 1.9\times 10^{18}$~kg for $p_{N_2}=50$~kPa [Eq.~(\ref{eq:MN2})],
and $E_{\min}\approx 1.2\times 10^{25}$~J for reversible electrolysis [Eq.~(\ref{eq:Emin})].}
\label{tab:throughput}
\begin{tabular}{lcccc}
\toprule
Target & $M_{\mathrm{target}}$ (kg) & $t_{\mathrm{build}}$ & $\dot{M}_{\mathrm{req}}$ (kg\,s$^{-1}$) & $P_{\mathrm{avg}}$ (TW) \\
\midrule
O$_2$ for $p_{\mathrm{O_2}}=\SI{21}{kPa}$  & $8.2\times 10^{17}$ & $10^3$~yr & $2.6\times 10^{7}$ & $380$ \\
O$_2$ for $p_{\mathrm{O_2}}=\SI{21}{kPa}$  & $8.2\times 10^{17}$ & $10^2$~yr & $2.6\times 10^{8}$ & $3800$ \\
N$_2$ buffer for $p_{\mathrm{N_2}}=\SI{50}{kPa}$ & $1.9\times 10^{18}$ & $10^3$~yr & $6.0\times 10^{7}$ & --- \\
N$_2$ buffer for $p_{\mathrm{N_2}}=\SI{50}{kPa}$ & $1.9\times 10^{18}$ & $10^2$~yr & $6.0\times 10^{8}$ & --- \\
\bottomrule
\end{tabular}
\end{table*}

Table~\ref{tab:throughput} emphasizes that breathable endpoints (E4) are not merely ``large'': they imply sustained, multi-decade to multi-millennial industrial mass flow rates orders of magnitude above the maintenance rates implied by aerosol-based warming proposals \cite{Ansari2024Nanoparticles}. This does not prove impossibility, but it places E4 in the category of long-duration planetary industry rather than ``scale-up of demonstrated ISRU.''

Equations of the form $\dot M \sim M/t_{\rm build}$ convert a thermodynamic/atmospheric end-state requirement into a required \emph{continuous industrial mass flow}. This is a useful discriminator for feasibility because it folds together mining, chemical processing, plant uptime, logistics, maintenance, and scaling laws for an industrial base capable of sustaining the required rates. To make these rates interpretable, Sec.~\ref{sec:industrial_scaling} benchmarks them against present-day terrestrial energy use and bulk material production.

\subsection{Power generation scaling: solar versus nuclear}
\label{sec:power}

Any sustained \emph{engineering-dominated atmospheric modification} activity is power-limited. Biological pathways can alter the oxygen-production bottleneck, but they do not remove the requirement for large-scale energy, materials handling, and environmental control in the regional/covered systems that are the only currently plausible path identified in this paper. Using the global mean absorbed solar flux on Mars, \(\langle F_{\odot}\rangle = \SMars(1-\Abond)/4 \approx \SI{110}{W\,m^{-2}}\) (Table~\ref{tab:marsparams}),
the \emph{global-mean} electrical power density from photovoltaics is at best
\(\eta_{\mathrm{pv}}\langle F_{\odot}\rangle\), where $\eta_{\mathrm{pv}}$ includes conversion efficiency and
system losses. Thus, the panel area required for an average electrical power $P$ is
\begin{equation}
A_{\mathrm{pv}} \approx \frac{P}{\eta_{\mathrm{pv}}\SMars(1-\Abond)/4}.
\label{eq:pv_area}
\end{equation}
For $\eta_{\mathrm{pv}}=0.2$, this implies $\sim \SI{4.5e4}{km^2}$ of panels per TW of average power. 
Table~\ref{tab:pv_area} reports representative values.

\begin{table}[t]
\caption{Photovoltaic (PV) area required to supply average electrical power on Mars using Eq.~(\ref{eq:pv_area})
with $\SMars=\SI{589}{W\,m^{-2}}$ and $\Abond=0.25$ \cite{Alexander2001MarsEnv}. Values are global-mean lower bounds; real installations require additional area for latitude, dust, night, and storage.}
\label{tab:pv_area}
\begin{tabular}{ccc}
\toprule
Average power & $A_{\mathrm{pv}}$ at $\eta_{\mathrm{pv}}=0.2$ & $A_{\mathrm{pv}}$ at $\eta_{\mathrm{pv}}=0.3$ \\
\midrule
\SI{1}{\tera\watt}   & $4.5\times 10^{4}\,\mathrm{km^2}$ & $3.0\times 10^{4}\,\mathrm{km^2}$ \\
\SI{10}{\tera\watt}  & $4.5\times 10^{5}\,\mathrm{km^2}$ & $3.0\times 10^{5}\,\mathrm{km^2}$ \\
\SI{100}{\tera\watt} & $4.5\times 10^{6}\,\mathrm{km^2}$ & $3.0\times 10^{6}\,\mathrm{km^2}$ \\
\SI{1000}{\tera\watt} & $4.5\times 10^{7}\,\mathrm{km^2}$ & $3.0\times 10^{7}\,\mathrm{km^2}$ \\
\bottomrule
\end{tabular}
\end{table}

All mirror and PV areas reported here are optimistic lower bounds; realistic designs increase required area by factors of a few to account for pointing losses, dust, latitude, diurnal cycling, seasonal insolation, and storage. These areas are not physically impossible, but they imply continent-scale deployment and maintenance, plus storage for night and multi-week dust storms. High-capacity fission or fusion provides an alternative with much higher power density but large development and safety overhead. Most plausible long-term architectures are therefore \emph{hybrid}: nuclear baseload plus solar where practical, with large-scale energy storage and distribution.

\subsection{Architecture as a coupled control problem}

Terraforming is not a single technology but a coupled planetary climate--industrial system whose source terms, loss terms, and feedbacks must be monitored and actively managed over long timescales. A credible architecture must include:

\begin{enumerate}
\item \textit{Energy generation and transmission.}
Required power is mechanism- and endpoint-dependent: GW-class source powers may be relevant for some warming-only particle pathways, whereas open-atmosphere E4 oxygenation requires multi-$10^2$~TW to PW-class average power. Energy delivery is limited by transmission, storage, and reliability (surface grids, beamed power, or distributed reactors).
\item \textit{Mass production and logistics.}
Industrial throughput must be stated by pathway rather than globally. Warming-only particle pathways can lie in the megaton-per-year class, whereas O$_2$/buffer-gas atmospheric buildout for E4 requires gigaton-to-teraton-per-year class mass flow sustained for centuries.
\item \textit{Climate monitoring and feedback control.}
The required degree of active control is mechanism-dependent. Some interventions may benefit from self-lofting or radiative--dynamical feedbacks, but any serious planetary-scale program still requires continuous sensing and at least some actuation authority (factory throttling, mirror pointing, gas-production control, or particle release management) to maintain targets and avoid collapse/hysteresis.
\item \textit{Planetary protection, biosafety, and biogeochemical coupling.}
Biological components can grow rapidly, but ecological turnover, nutrient cycling, burial/export efficiency, and sink competition introduce additional system variables and long biogeochemical response times that lie outside the present minimal engineering closure.
\end{enumerate}

\subsection{Staged roadmap: from regional habitability to global modification}

Given the scale barriers for E4 endpoints, a technically plausible progression is:

\begin{itemize}
\item \textit{Phase I (decades): regional paraterraforming.}
Deploy solid-state greenhouse layers (silica aerogel) over ice-rich terrain to enable persistent melt and
photosynthesis locally without global atmospheric change \cite{Wordsworth2019Aerogel}.
\item \textit{Phase II (century): climate nudging and resource mobilization.} Use localized warming to trigger limited volatile release, deploy pilot aerosol systems, and build power and manufacturing base.
\item \textit{Phase III (multi-century): sustained atmospheric engineering.} If pursued, combine an operationally sustained warming mechanism (aerosols/PFC/CIA) with buffer gas import or extraction and large-scale oxygen production, while simultaneously investing in retention and geochemical management.
\end{itemize}

This staged view recognizes that the fastest path to meaningful surface habitability is likely \emph{regional}
rather than \emph{planetary}.

\section{Discussion and implications}
\label{sec:discussion}

The constraint-based framework in Table~\ref{tab:keyresults} shows that the dominant feasibility constraint shifts sharply with the target end state. For regional endpoints (E1--E2), deployment area and local power dominate; for global pressure (E3), atmospheric inventory dominates; and for breathable endpoints (E4), composition inventories, minimum work, and sink filling dominate. For near-term habitability return per unit industrial scale, the results favor regional approaches: paraterraforming and contained biospheres, local thermal/insulation control, and build-out of power and autonomous manufacturing. If global modification is pursued, credible architectures must be evaluated as coupled control systems that explicitly close (i) inventory sourcing, (ii) radiative authority, (iii) sustained throughput and power, and (iv) retention/sink management over multi-century timescales.

\subsection{Which constraint dominates depends on the endpoint}

A central implication of the constraint-based framework is that the limiting factor shifts with the target end state:
\begin{itemize}
\item For E1--E2 (regional water stability, protected agriculture), the dominant constraints are local radiative control and deployment area; global exaton-scale gas inventories are not required.
\item For E3 (no ebullism), the dominant constraint becomes atmospheric mass: $M_{\rm atm}\approx 2.4\times 10^{17}$~kg (Table~\ref{tab:pressure_endpoints}), plus enough warming to avoid CO$_2$ collapse.
\item For E4 (breathable), the dominant constraints are composition inventories and energy:
$M_{O_2}\approx 8.2\times 10^{17}$~kg and $M_{N_2}\approx 1.9\times 10^{18}$~kg, with reversible oxygenation energy
$E_{\min}\approx 1.2\times 10^{25}$~J (Sec.~\ref{sec:oxygen}) and additional sink-filling (Sec.~\ref{sec:O2_sinks}).
\end{itemize}

Where the scaling suggests feasibility (aerosols/CIA), quantitative climate outcomes still require coupled 3‑D GCM + aerosol microphysics for a better than an order-of-magnitude accuracy.

\subsection{Design rules implied by the numbers}

The results above imply several robust design rules:
\begin{enumerate}
\item \textit{Warming-only proposals cannot shortcut buffer-gas and oxygen inventories.} Even if $T_s$ targets are met via aerosols or CIA, E4 remains dominated by $10^{18}$~kg-class O$_2$ and N$_2$/Ar.

\item \textit{Mass-efficient radiative levers split into maintenance-dominated and fill-dominated classes.}
For IR-active particles, the representative source rate is $\dot M_p\sim 90$~kg\,s$^{-1}$ (i.e.\ $\sim 3\times 10^9$~kg\,yr$^{-1}$ for the illustrative \SI{30}{L\,s}$^{-1}$ case), and the maintained column mass scales as $\Sigma_p\sim \dot M_p\tau_p/(4\pi R_{\rm Mars}^2)$ (Sec.~\ref{sec:nanoparticles}); these pathways are genuinely maintenance-limited. By contrast, Fig.~\ref{fig:H2_fill_vs_maint} and Sec.~\ref{sec:fill_vs_maint} show that global CO$_2$--H$_2$ CIA is typically fill-dominated during century-to-millennial buildout, with TW-class replenishment becoming the relevant floor for long hold times rather than for initial construction.

\item \textit{One-shot forcing is structure-mass-and-operations-limited.}
Eq.~(\ref{eq:mirror_area}) implies continent-scale mirrors for \(\Delta F_{\rm TOA}\) of order \(10\)–\(20~\mathrm{W\,m^{-2}}\), with total mass \(M_m\sim \sigma_m A_m\) in the \(10^{10}\)–\(10^{11}\)~kg range for \(\sigma_m\sim 1\)–10~g\,m\(^{-2}\). But the harder problem is not launch energy alone: it is aggregate deployment count, pointing/control, non-Keplerian orbit maintenance, environmental degradation, and replacement over long hold times. Solar sails can assist transfer and trim of modular reflectors, but current and roadmap sail systems remain many orders of magnitude below the aggregate area implied by global Mars forcing targets.

\item \textit{E4 is a planetary industry problem: power + throughput + control.} Even at thermodynamic minima, $E_{\min}$ implies $\sim 380$~TW averaged over $10^3$~yr, and Table~\ref{tab:throughput} implies $\dot M\sim 10^7$--$10^8$~kg\,s$^{-1}$ for century-to-millennial build times—far beyond ``scaled ISRU.''
\end{enumerate}

\subsection{Technology implications and a credible maturation path}

If the objective is to maximize near-term habitability impact per unit industrial scale, the results favor: (i) regional paraterraforming (solid-state greenhouse) and enclosed agriculture; (ii) MW--GW-class power and localized thermal control; and (iii) atmospheric monitoring/control infrastructure. If global endpoints are pursued, the pacing technologies become continent-scale energy generation, multi-gigaton/year materials handling, and long-duration closed-loop climate control (sensing + actuation), not a single “magic gas.”

A complementary class of concepts is quasi-global paraterraforming or ``worldhouse'' enclosure of the surface \cite{Taylor2001Worldhouse}. Such approaches largely avoid the exaton atmospheric-mass requirement by trading it into membrane/covering area, fabrication, puncture tolerance, and long-term maintenance. They therefore reinforce, rather than weaken, the paper’s main conclusion that covered-area strategies are more plausible than open global atmospheric transformation on near-term industrial scales.

A practical way to summarize the architecture result is therefore: \emph{use space industry to build habitable area before trying to build habitable planet}. The staging logic is area first, atmosphere later: local coverings, local water access, local biosystems, local pressure control, and only then any attempt at larger-scale atmospheric or orbital climate intervention. This ordering is not merely conservative; it is what the quantitative mass and power bookkeeping favors.

With current inventories, current solar-sail/reflector technology, and plausible 21st-century industrial assumptions, there is no currently credible path to a \emph{global open-atmosphere} E3/E4 Mars. The simultaneous requirements---exaton-class volatile supply, sustained \(\mathcal{O}(10^2)\)~TW to PW-class power, century-to-millennial climate control, and sink/retention management---are not jointly closed by any proposal surveyed in this paper.

A plausible path does exist, but it is narrower: staged \emph{regional} habitability via covered or partly covered systems. This includes local aerogel-style solid-state greenhouse deployment over ice-rich terrains \cite{Wordsworth2019Aerogel}, enclosed agriculture at \(P_{\rm in}\sim 10\)–30 kPa, and, at larger scales, worldhouse-type coverings that trade atmospheric mass into membrane area and repair logistics \cite{Taylor2001Worldhouse,WordsworthCockell2024LivingHabitats,Wordsworth2025Biomaterials}. In this regime, microbes and plants are useful because they operate inside controlled environments as oxygen recyclers, food producers, biomaterial sources, and bioprocessing agents, rather than as exposed planetary-scale terraforming agents.

The scale of this plausible path is already quantified by Eq.~(\ref{eq:Amax_reg_num}). For example, a net gas throughput of \(\dot M_{\rm net}=10^5\)–\(10^6\,\mathrm{kg\,s^{-1}}\) sustained for \SI{100}{yr} supports
\begin{equation}
A_{\max}\approx 1.17\times10^5\text{--}1.17\times10^6\ {\rm km^2}
\qquad (P_{\rm reg}=\SI{10}{kPa}),
\end{equation}
and five times less area at \SI{50}{kPa}. These are already country- to subcontinent-scale regional programs, yet they remain far short of the requirements for a breathable open planetary atmosphere. This is why the most defensible Mars-terraforming roadmap is a progressive expansion of \emph{habitable area}, not an early attempt at habitable planet.

The practical implication is that ``Mars terraforming'' should be disaggregated into two very different classes. Regional/paraterraformed Mars is a difficult but technically plausible long-horizon development. A global open-air breathable Mars is not currently a plausible engineering program on known inventories and present-to-forecast sail/manufacturing technology.

This endpoint-based feasibility split is broadly consistent with recent roadmap-style assessments of Mars warming \cite{Kite2026Roadmap}, which emphasize local membranes, reflectors for key sites, and engineered aerosols as early research priorities. Our contribution is to show that even if such warming pathways prove workable, they do not by themselves close the exaton-class inventory and oxygen/buffer-gas requirements associated with open-atmosphere global E3/E4 endpoints.

\subsection{Industrial scaling, timescale--power trade, and cost floors}
\label{sec:industrial_scaling}

A central result of this paper is that once the endpoint moves beyond E2--E3, feasibility is governed less by the existence of a mechanism and more by whether an industrial base can sustain the required \emph{mass flow} and \emph{power} for centuries to millennia. For the nominal E4 targets summarized above, the required atmospheric inventories are $M_{\mathrm{O}_2}\simeq 8.2\times 10^{17}\,\mathrm{kg}$ and $M_{\mathrm{N}_2}\simeq 1.9\times 10^{18}\,\mathrm{kg}$, with a minimum reversible electrochemical work $E_{\min}\simeq 1.2\times 10^{25}\,\mathrm{J}$ for oxygenation alone (exclusive of compression, losses,
and co-products).

\subsubsection{Biological oxygenation: upper-envelope and area-limited benchmarks}

The paper's oxygenation numbers are lower bounds for \emph{abiotic industrial} oxygenation. Biology provides a different type of benchmark: it can reduce the direct electrical-power burden, but only if a large productive biosphere exists and only to the extent that reduced products are buried or exported rather than rapidly reoxidized.

An optimistic upper-envelope comparison is to Earth's total modern net primary productivity, \({\rm NPP}_{C,\oplus}\approx 104.9\) Pg C yr\(^{-1}\) \cite{Field1998BiosphereNPP}. If a Martian biosphere ever achieved a comparable total fixed-carbon productivity, the gross O\(_2\)-equivalent flux would be
\begin{equation}
\dot M_{O_2,{\rm gross}} \simeq \frac{32}{12}\,{\rm NPP}_C
\approx 2.8\times10^{14}
\left(\frac{{\rm NPP}_C}{104.9\ {\rm Pg\,C\,yr^{-1}}}\right)
\mathrm{kg\,yr^{-1}}.
\label{eq:bio_o2_gross_upper}
\end{equation}
This is useful as an absolute upper-envelope scale anchor, but it is too optimistic as a default Mars comparison because it assumes an Earth-class global biosphere.

A more transferable benchmark is the \emph{mean areal} productivity. Defining
\begin{equation}
\phi_C \equiv \frac{{\rm NPP}_C}{A},
\label{eq:phiC_def}
\end{equation}
Earth's modern global-mean value is
\begin{equation}
\phi_{C,\oplus}^{\rm mean}\approx 0.206\ {\rm kg\,C\,m^{-2}\,yr^{-1}},
\label{eq:phiC_earth_mean}
\end{equation}
using \({\rm NPP}_{C,\oplus}\approx 104.9\) Pg C yr\(^{-1}\) over Earth's total surface area. If productive conditions are achieved over only a fraction \(f_{\rm prod}\) of Mars' surface area, then the gross O\(_2\)-equivalent production is
\begin{equation}
\dot M_{O_2,{\rm gross}}
\simeq \frac{32}{12}\,\phi_C\,f_{\rm prod}\,4\pi R_{\rm Mars}^2
\approx 7.9\times10^{13}
\left(\frac{\phi_C}{0.206\ {\rm kg\,C\,m^{-2}\,yr^{-1}}}\right)
\left(\frac{f_{\rm prod}}{1}\right)
\mathrm{kg\,yr^{-1}}.
\label{eq:bio_o2_gross_areal}
\end{equation}
Writing the long-term burial/export efficiency as \(\beta\ll 1\), the oxygenation time is
\begin{equation}
t_{O_2,{\rm bio}}
\approx \frac{M_{O_2}}{\beta\,\dot M_{O_2,{\rm gross}}}
\approx 1.0\times10^{4}\,\beta^{-1}
\left(\frac{M_{O_2}}{8.2\times10^{17}\,\mathrm{kg}}\right)
\left(\frac{0.206\ {\rm kg\,C\,m^{-2}\,yr^{-1}}}{\phi_C}\right)
\left(\frac{1}{f_{\rm prod}}\right)
\mathrm{yr}.
\label{eq:bio_o2_time_areal}
\end{equation}
Hence even with Earth-like \emph{mean areal} productivity across all of Mars, the timescale is \(\sim 10^4\,\beta^{-1}\) yr; for \(\beta=10^{-3}\) this is already \(\sim 10\) Myr, and if only \(f_{\rm prod}=0.1\) of the surface is productive it becomes \(\sim 10^8\) yr.

These numbers argue against treating microbes or plants as a near-term shortcut to open-planet E4. Biology is more plausibly leveraged in \emph{protected} systems: closed or semi-closed habitats, shallow-water bioreactors, oxygen recycling, food production, biomaterials, and possibly perchlorate-processing ISRU \cite{WordsworthCockell2024LivingHabitats,Wordsworth2025Biomaterials,Rzymski2024Perchlorates}. Present-day open-surface Mars remains especially hostile to exposed microorganisms because UV-activated perchlorates can be rapidly biocidal under Mars-analog conditions \cite{Wadsworth2017Perchlorates}. Accordingly, biological bootstrapping is best viewed as a regional, contained, or very long-timescale pathway, not as a near-term substitute for the mass, sink, and climate constraints on global atmospheric engineering.

\subsubsection{Dominant constraint as a function of end state}

For E1--E2 regional habitability, the dominant constraints are deployment area and local power; global volatile inventories are not required. For E3 global pressure, the mass inventory constraint becomes dominant ($M_{\rm atm}\sim 2.4\times 10^{17}$ kg at the Armstrong limit). For E4 breathable endpoints, composition inventories and energy dominate: $M_{O_2}\sim 8\times 10^{17}$ kg and buffer gases at the $10^{18}$ kg scale, with oxygenation work $E_{\min}\sim 10^{25}$ J even at reversible limits.

\subsubsection{Timescale--power--throughput trade}

For any target inventory $M$ achieved over build time $t_{\rm build}$,
\begin{equation}
\dot M \equiv \frac{M}{t_{\rm build}},\qquad \bar P \equiv \frac{E}{t_{\rm build}}.
\end{equation}
Here and below, an overbar denotes a build-time average (e.g., \(\bar P \equiv E/t_{\rm build}\)). These relations translate end states into continuous industrial requirements. For example, $p_{O_2}=21$ kPa implies $M_{O_2}\approx 8.2\times 10^{17}$ kg, so $\dot M_{O_2}\approx 2.6\times 10^7$--$2.6\times 10^8$ kg\,s$^{-1}$ for $t_{\rm build}=10^3$--$10^2$ yr. The reversible oxygenation floor implies $\bar P_{\min}\approx 0.38$--$3.8$ PW over the same range. Real systems require
$\bar P \gtrsim \bar P_{\min}/\eta_{\rm sys}$ with $\eta_{\rm sys}\ll 1$ once compression, separation, thermal losses, and downtime are included.

\subsubsection{Benchmarking E4 mass flows against terrestrial industry}

For a build time $t_{\rm build}$, the implied continuous production/import rates are
\begin{equation}
\dot M_{\mathrm{O}_2}\simeq \frac{M_{\mathrm{O}_2}}{t_{\rm build}},\qquad
\dot M_{\mathrm{N}_2}\simeq \frac{M_{\mathrm{N}_2}}{t_{\rm build}}.
\end{equation}
For $t_{\rm build}=10^3~\mathrm{yr}$ this corresponds to
$\dot M_{\mathrm{O}_2}\approx 8.2\times 10^{14}\,\mathrm{kg\,yr^{-1}}$ ($\approx 820~\mathrm{Gt\,yr^{-1}}$) and
$\dot M_{\mathrm{N}_2}\approx 1.9\times 10^{15}\,\mathrm{kg\,yr^{-1}}$ ($\approx 1900~\mathrm{Gt\,yr^{-1}}$),
i.e.\ $\mathcal{O}(10^4)$--$\mathcal{O}(10^5)$ tonnes per second.

As a reality check, global material extraction on Earth has grown from $\sim 30$ to $\sim 106$ billion tonnes
per year since 1970 \cite{UNEP_GRO2024}, and world crude steel production is $\sim 1.9$ billion tonnes per year \cite{worldsteel_figures_2024}. Thus, even under a $10^3$-yr schedule, the \emph{oxygen alone} production rate is of order
$\sim 8\times$ present-day total terrestrial material extraction, and $\sim 400\times$ present-day global steel output. The nitrogen buffer requirement is larger still.

Importantly, $\dot M_{\mathrm{O}_2}$ and $\dot M_{\mathrm{N}_2}$ are not directly mined ``as is''; they must
be extracted from feedstocks. If oxygen is produced from mined water ice, the required water throughput is
\begin{equation}
\dot M_{\mathrm{H_2O}}\simeq \frac{36}{32}\,\dot M_{\mathrm{O_2}}\approx 1.125\,\dot M_{\mathrm{O_2}}.
\end{equation}
while if oxygen (or nitrogen-bearing species) is extracted from regolith with effective yield $f$ by mass, the required regolith handling is $\dot M_{\rm reg}\sim \dot M/f$. For plausible $f\sim 0.01$--$0.1$, regolith throughput rises by one to two orders of magnitude, pushing the problem decisively into ``planetary industry'' rather than ``chemical plant'' scaling.

\begin{table}[t]
\caption{Order-of-magnitude comparison between Mars E4 build requirements (for $t_{\rm build}=10^3$ yr)
and representative present-day terrestrial industry magnitudes.}
\label{tab:industry_scale}
\begin{tabular}{lcc}
\toprule
Quantity & Mars E4 (1000 yr build) & Earth today\\
\midrule
$\dot M_{\mathrm{O}_2}$ & $8.2\times 10^{14}\,\mathrm{kg\,yr^{-1}}\approx 820~\mathrm{Gt\,yr^{-1}}$ & --- \\
$\dot M_{\mathrm{N}_2}$ & $1.9\times 10^{15}\,\mathrm{kg\,yr^{-1}}\approx 1900~\mathrm{Gt\,yr^{-1}}$ & --- \\
Global material extraction & --- & $\sim 106~\mathrm{Gt\,yr^{-1}}$ \cite{UNEP_GRO2024} \\
World crude steel production & --- & $\sim 1.89~\mathrm{Gt\,yr^{-1}}$ \cite{worldsteel_figures_2024} \\
Global primary energy (avg) & --- & $620~\mathrm{EJ\,yr^{-1}}\approx 20~\mathrm{TW}$ \cite{EI_StatReview_2024} \\
\bottomrule
\end{tabular}
\end{table}

\subsubsection{Timescale--power trade: what does ``accelerating'' E4 actually require?}

The minimum average power required to supply the reversible oxygenation work is
\begin{equation}
\bar P_{\min} \simeq \frac{E_{\min}}{t_{\rm build}}
\simeq 3.8\times 10^{14}\,\mathrm{W}\,
\left(\frac{10^3~\mathrm{yr}}{t_{\rm build}}\right)
\left(\frac{E_{\min}}{1.2\times 10^{25}~\mathrm{J}}\right).
\label{eq:power_time_trade}
\end{equation}
For $t_{\rm build}=10^3~\mathrm{yr}$, $\bar P_{\min}\approx 380~\mathrm{TW}$, which is $\sim 19\times$ today's global primary energy consumption rate ($\approx 620~\mathrm{EJ\,yr^{-1}}$) \cite{EI_StatReview_2024}.
Conversely, if one demands $t_{\rm build}\sim 100~\mathrm{yr}$, Eq.~(\ref{eq:power_time_trade})
implies $\bar P_{\min}\sim 3.8~\mathrm{PW}$ \emph{before} accounting for real inefficiencies. Because real systems incur conversion, compression, separation, thermal losses, and downtime, a more realistic requirement is $\bar P \sim \bar P_{\min}/\eta_{\rm sys}$ with $\eta_{\rm sys}\ll 1$,
pushing required power further upward.

Using Eq.~(\ref{eq:pv_area}), the PV area required to supply the reversible oxygenation power floor is
\begin{equation}
A_{\rm pv} \sim \left(4.5\times 10^{4}\,\mathrm{km^2/TW}\right)\left(\frac{\bar P_{\min}}{1~\mathrm{TW}}\right)
\left(\frac{0.2}{\eta_{\rm pv}}\right)\,\frac{1}{C_f},
\end{equation}
where $C_f\le 1$ is an effective capacity factor capturing diurnal cycling, latitude, dust storms, and storage limits.
For $\bar P_{\min}\approx 380~\mathrm{TW}$ (1000-yr build), even the optimistic $C_f=1$ bound implies
$A_{\rm pv}\sim 1.7\times 10^{7}~\mathrm{km^2}$ at $\eta_{\rm pv}=0.2$. For $t_{\rm build}=100$ yr, $\bar P_{\min}\sim 3.8~\mathrm{PW}$ implies $A_{\rm pv}$ comparable to or exceeding Mars' total surface area unless power is provided by higher-density sources (nuclear/fusion) or by space-based collection and beaming. This illustrates that accelerating E4 is fundamentally a power-density problem, not only a chemistry problem.

\subsubsection{Economics: minimal energy-cost floor and capex scaling}

Although detailed economics are beyond the scope of this paper, the estimates above imply hard \emph{floors}.
The reversible oxygenation energy corresponds to
\begin{equation}
E_{\min} \approx 3.3\times 10^{18}\ \mathrm{kWh},
\end{equation}
so even at an optimistic electricity price $c_e$ one has an energy-only floor
\begin{equation}
C_{E,\min} \gtrsim (3.3\times 10^{18}\,\mathrm{kWh})\,c_e
\approx 1.7\times 10^{17}\ \mathrm{\$}\,
\left(\frac{c_e}{0.05~\mathrm{\$/kWh}}\right),
\end{equation}
exclusive of capital expenditures, maintenance, and the additional energy required for non-reversible steps (compression, gas separation, transport, and thermal management).

Similarly, if the generation system scales with cost $c_W$ per installed watt, then the capex scaling
for sustaining $\bar P$ is
\begin{equation}
C_{\rm capex} \sim c_W\,\bar P
\approx 3.8\times 10^{14}\ \mathrm{\$}\,
\left(\frac{c_W}{1~\mathrm{\$/W}}\right)
\left(\frac{\bar P}{380~\mathrm{TW}}\right),
\end{equation}
showing that even wildly optimistic $c_W$ values imply civilization-scale investment when $\bar P$
is in the multi-$10^{14}$ W class.

\subsubsection{What industrial activities are actually implied?}

The constraints above map directly onto different industrial primitives depending on the endpoint. For warming-only aerosol pathways, published studies imply source powers and material flows that may lie in the GW-class and \(\sim 10^2\) kg s$^{-1}$ class \cite{Ansari2024Nanoparticles,Richardson2026IRParticles}. By contrast, for century-to-millennial E4 oxygenation and buffer-gas buildout, the relevant primitives escalate to multi-\(10^2\)~TW to PW-class power, very large excavation/beneficiation throughput if feedstock yields are low, and long-duration climate/composition control. The manuscript’s largest throughput and power numbers therefore apply to \emph{E4-scale atmospheric buildout}, not to every warming-only pathway.

\subsection{What would change the feasibility classification?}
\vskip -10pt
The conclusions shift only if one (or more) of the following becomes true:
(i) discovery of orders-of-magnitude larger accessible CO$_2$ and/or fixed-nitrogen inventories;
(ii) a super-greenhouse constituent with strong window absorption, long lifetime, and abundant in-situ feedstock;
(iii) megascale space manufacturing enabling continent-scale mirrors/collectors at low areal mass and manageable station-keeping;
(iv) an industrial ecology capable of sustaining multi-TW power and Gt\,yr$^{-1}$-class throughput for centuries.
Absent such changes, the constraint-based results imply that E1--E2 are near-term plausible, whereas E3--E4 remain planetary-industry, long-horizon projects.

\section{Conclusions}
\label{sec:concl}
\vskip -10pt
Terraforming Mars is constrained by coupled planet-scale budgets in atmospheric inventory, radiative forcing/opacity, industrial throughput and power, and long-term stability. Using transparent order-of-magnitude scalings, we find:

\begin{enumerate}
\item \textit{Pressure targets are exaton-class.} These targets translate directly into atmospheric mass [Eq.~(\ref{eq:mass_pressure})]. E3/E4 endpoints require $10^{17}$--$10^{18}$ kg of gas, far above endogenous CO$_2$ inventories. Hydrostatic balance implies $M_{\rm atm}\simeq 3.89\times 10^{15}$~kg per mbar [Eq.~(\ref{eq:mass_per_mbar})]. Thus, even the Armstrong-limit pressure ($P_s=\SI{6.27}{kPa}$) corresponds to $M_{\rm atm}\approx 2.4\times 10^{17}$~kg (Table~\ref{tab:pressure_endpoints}).

\item \textit{Accessible endogenous CO$_2$ is inventory-limited and cannot deliver melt-class global climates.}
Current evidence supports an \emph{order-of-tens-of-mbar} accessible endogenous CO$_2$ budget rather than a secure multi-bar reservoir (Sec.~\ref{sec:endogenous}). A representative \(\sim\)20 mbar case yields $\lesssim 10$ K warming under present insolation, leaving a large melt-class thermal shortfall to globally stable surface liquid water.

\item \textit{Melt-class $T_s$ requires either $\tau_{\rm IR}\sim 2$--4 at current insolation or extreme direct forcing.} Thermal targets require very large effective forcing.
Accessible CO$_2$ (\(\sim 20\) mbar) yields $\lesssim 10$ K warming, whereas $\sim 60$ K is needed for globally
stable surface liquid water \cite{JakoskyEdwards2018CO2Inventory}.
In a minimal grey-atmosphere approximation, $T_s\sim 250$--273 K at $T_e\approx 210$ K requires $\tau_{\rm IR}\sim 2$--4 (Table~\ref{tab:keyresults}; Eq.~\eqref{eq:tau_required}, Sec.~\ref{sec:greenhouse_closure}). Direct insolation modification requires $\Delta F_{\rm TOA}$ at the
$\mathcal{O}(10^2)\,\mathrm{W\,m^{-2}}$ level for large $T_e$ shifts, implying continent-scale mirrors for $\Delta F_{\rm TOA}\sim \mathcal{O}(10^2)\,\mathrm{W\,m^{-2}}$
(Eq.~\eqref{eq:forcing_Te}; Eq.~\eqref{eq:mirror_area}).

\item \textit{Breathable endpoints are bottlenecked primarily by composition inventories and oxygenation energy.} Earth-like $p_{O_2}=\SI{21}{kPa}$ implies $M_{O_2}\approx 8.2\times 10^{17}$~kg [Eq.~(\ref{eq:MO2})], and modest buffer-gas targets $p_{N_2}=\SI{50}{kPa}$ require comparable or larger masses $M_{N_2}\approx 1.9\times 10^{18}$~kg [Eq.~(\ref{eq:MN2})]. The reversible minimum oxygenation work is $E_{\min}\approx 1.2\times 10^{25}$~J [Eq.~(\ref{eq:Emin})], implying $\bar P_{\min}\sim 0.38$--3.8~PW for $t_{\rm build}=10^3$--$10^2$~yr even before inefficiencies and before filling geochemical sinks (Sec.~\ref{sec:O2_sinks}).
 
\item \textit{Mass-efficient warming levers split into maintenance-dominated and fill-dominated classes, while one-shot forcing trades into extreme structures.}
Aerosol pathways couple to sustained injection set by residence time (Sec.~\ref{sec:nanoparticles}) and are genuinely maintenance-limited. By contrast, global CO$_2$--H$_2$ CIA requires long-term replenishment against escape but is typically fill-dominated during century-to-millennial buildout because a \emph{large, design-dependent global H$_2$ inventory} must be created before the warm state can be held (Sec.~\ref{sec:fill_vs_maint}; Sec.~\ref{sec:H2_escape}). For representative warm-state cases this inventory can correspond to H$_2$ partial pressures ranging from multi-mbar to multi-kPa, depending on the chosen total pressure $P_{\rm tot}$ and mixing ratio $f_{H_2}$. Reflector-based TOA forcing scales to continent-class areas for global $\Delta F_{\rm TOA}$; one-shot mirror pathways are therefore structure- and operations-intensive, while breathable global endpoints still require a civilization-scale industrial base operating for centuries to millennia (Sec.~\ref{sec:industrial_scaling}).

\end{enumerate}

Taken together, these scalings imply a sharp feasibility split. \emph{Regional} habitability gains (E1--E2) are the only broad class of Mars-terraforming pathway that we currently regard as technically plausible. They include local aerogel-style paraterraforming, enclosed agriculture, and potentially larger covered-area or worldhouse-type concepts. Their central advantage is that they scale with covered area and local power rather than with planet-wide atmospheric inventory, and they allow biology to be used where it is strongest: inside controlled environments as an oxygen-recycling, food-production, biomaterials, and bioprocessing subsystem.

By contrast, a \emph{global open-atmosphere} transformation to E3--E4 is not currently a plausible engineering program. With currently inferred accessible volatiles, no surveyed mechanism simultaneously closes the inventory, power, climate-control, sink, and operations/replacement constraints. An Earth-like Mars becomes credible only under explicit conditions: (i) discovery or delivery of exaton-class volatile inventories (especially an N-bearing buffer gas), (ii) sustained \(\mathcal{O}(10^2)\)~TW to PW-class power and \(\mathcal{O}(10^3)\)~Gt\,yr\(^{-1}\) materials handling for centuries to millennia, and (iii) long-duration climate \emph{control authority} (monitoring + actuation) together with sink/retention management and replacement of degraded climate-control assets.

The principal contribution of this paper is a reusable architecture-level framework for Mars terraforming based on endpoint-normalized lower bounds, dimensionless feasibility numbers $(\Pi_M,\Pi_F,\Pi_{\dot M},\Pi_P,\Pi_S)$, and a fill-versus-maintenance crossover criterion $\Lambda_{\rm maint}=t_{\rm build}/\tau_{\rm loss}$. This formulation makes proposed pathways directly comparable, exposes whether a concept is inventory-, radiatively-, power-, throughput-, or retention-limited, and cleanly separates true maintenance-limited approaches (e.g., short-lived aerosols) from pathways whose dominant burden is the initial global fill (e.g., large H$_2$ inventories required to establish CO$_2$--H$_2$ warm states on century-to-millennial schedules). In that sense, the paper does not merely review proposed mechanisms; it supplies a common feasibility language and a minimal control-theoretic structure within which future higher-fidelity 3-D climate, photochemical, and industrial-systems studies can be embedded.

Finally, nothing in the present analysis rules out longer-timescale biologically mediated oxygenation. Rather, the manuscript isolates the lower bounds and bottlenecks for engineering-dominated pathways. A full treatment of biological bootstrapping would require additional constraints on productivity, nutrient cycling, burial efficiency, ecological stability, and sink competition \cite{McKay1982Terraforming,Graham2004Biological,DeBenedictis2025Case}.

Even under optimistic assumptions, the reversible oxygenation work alone is $\sim 3\times 10^{18}$~kWh, which sets an irreducible energy floor before compression, separations, distribution, capital deployment, and long-term maintenance are included. The most technically defensible roadmap is therefore staged: pursue E1--E2 deployments now while using them to demonstrate the industrial primitives---high-capacity power, bulk excavation and processing, global logistics, and closed-loop monitoring/actuation---that are prerequisites for any credible E3--E4 terraforming attempt.


\section*{Acknowledgments}
\vskip -10pt
The work described here was carried out at the Jet Propulsion Laboratory, California Institute of Technology, Pasadena, California, under a contract with the National Aeronautics and Space Administration. 
\vskip -12pt


\begin{thebibliography}{58}%
\makeatletter
\providecommand \@ifxundefined [1]{%
 \@ifx{#1\undefined}
}%
\providecommand \@ifnum [1]{%
 \ifnum #1\expandafter \@firstoftwo
 \else \expandafter \@secondoftwo
 \fi
}%
\providecommand \@ifx [1]{%
 \ifx #1\expandafter \@firstoftwo
 \else \expandafter \@secondoftwo
 \fi
}%
\providecommand \natexlab [1]{#1}%
\providecommand \enquote  [1]{``#1''}%
\providecommand \bibnamefont  [1]{#1}%
\providecommand \bibfnamefont [1]{#1}%
\providecommand \citenamefont [1]{#1}%
\providecommand \href@noop [0]{\@secondoftwo}%
\providecommand \href [0]{\begingroup \@sanitize@url \@href}%
\providecommand \@href[1]{\@@startlink{#1}\@@href}%
\providecommand \@@href[1]{\endgroup#1\@@endlink}%
\providecommand \@sanitize@url [0]{\catcode `\\12\catcode `\$12\catcode
  `\&12\catcode `\#12\catcode `\^12\catcode `\_12\catcode `\%12\relax}%
\providecommand \@@startlink[1]{}%
\providecommand \@@endlink[0]{}%
\providecommand \url  [0]{\begingroup\@sanitize@url \@url }%
\providecommand \@url [1]{\endgroup\@href {#1}{\urlprefix }}%
\providecommand \urlprefix  [0]{URL }%
\providecommand \Eprint [0]{\href }%
\providecommand \doibase [0]{https://doi.org/}%
\providecommand \selectlanguage [0]{\@gobble}%
\providecommand \bibinfo  [0]{\@secondoftwo}%
\providecommand \bibfield  [0]{\@secondoftwo}%
\providecommand \translation [1]{[#1]}%
\providecommand \BibitemOpen [0]{}%
\providecommand \bibitemStop [0]{}%
\providecommand \bibitemNoStop [0]{.\EOS\space}%
\providecommand \EOS [0]{\spacefactor3000\relax}%
\providecommand \BibitemShut  [1]{\csname bibitem#1\endcsname}%
\let\auto@bib@innerbib\@empty
\bibitem [{\citenamefont {McKay}(1982)}]{McKay1982Terraforming}%
  \BibitemOpen
  \bibfield  {author} {\bibinfo {author} {\bibfnamefont {C.~P.}\ \bibnamefont
  {McKay}},\ }\bibfield  {title} {\bibinfo {title} {{Terraforming Mars}},\
  }\href@noop {} {\bibfield  {journal} {\bibinfo  {journal} {JBIS}\ }\textbf
  {\bibinfo {volume} {35}},\ \bibinfo {pages} {427} (\bibinfo {year}
  {1982})}\BibitemShut {NoStop}%
\bibitem [{\citenamefont {McKay}\ \emph {et~al.}(1991)\citenamefont {McKay},
  \citenamefont {Toon},\ and\ \citenamefont
  {Kasting}}]{McKay1991MakingMarsHabitable}%
  \BibitemOpen
  \bibfield  {author} {\bibinfo {author} {\bibfnamefont {C.~P.}\ \bibnamefont
  {McKay}}, \bibinfo {author} {\bibfnamefont {O.~B.}\ \bibnamefont {Toon}},\
  and\ \bibinfo {author} {\bibfnamefont {J.~F.}\ \bibnamefont {Kasting}},\
  }\bibfield  {title} {\bibinfo {title} {{Making {Mars} habitable}},\ }\href
  {https://doi.org/10.1038/352489a0} {\bibfield  {journal} {\bibinfo  {journal}
  {Nature}\ }\textbf {\bibinfo {volume} {352}},\ \bibinfo {pages} {489}
  (\bibinfo {year} {1991})}\BibitemShut {NoStop}%
\bibitem [{\citenamefont {Zubrin}\ and\ \citenamefont
  {McKay}(1993)}]{ZubrinMcKay1993Terraforming}%
  \BibitemOpen
  \bibfield  {author} {\bibinfo {author} {\bibfnamefont {R.}~\bibnamefont
  {Zubrin}}\ and\ \bibinfo {author} {\bibfnamefont {C.~P.}\ \bibnamefont
  {McKay}},\ }\bibfield  {title} {\bibinfo {title} {{Technological Requirements
  for Terraforming {Mars}}},\ }in\ \href {https://doi.org/10.2514/6.1993-2005}
  {\emph {\bibinfo {booktitle} {29th Joint Propulsion Conference and
  Exhibit}}}\ (\bibinfo {year} {1993})\ \bibinfo {note} {{AIAA Paper
  93-2005}}\BibitemShut {NoStop}%
\bibitem [{\citenamefont {Fogg}(1995)}]{Fogg1995Terraforming}%
  \BibitemOpen
  \bibfield  {author} {\bibinfo {author} {\bibfnamefont {M.~J.}\ \bibnamefont
  {Fogg}},\ }\href@noop {} {\emph {\bibinfo {title} {{Terraforming: Engineering
  Planetary Environments}}}}\ (\bibinfo  {publisher} {{SAE International}},\
  \bibinfo {address} {Warrendale, PA},\ \bibinfo {year} {1995})\BibitemShut
  {NoStop}%
\bibitem [{\citenamefont {McKay}\ and\ \citenamefont
  {Marinova}(2001)}]{McKayMarinova2001Habitable}%
  \BibitemOpen
  \bibfield  {author} {\bibinfo {author} {\bibfnamefont {C.~P.}\ \bibnamefont
  {McKay}}\ and\ \bibinfo {author} {\bibfnamefont {M.~M.}\ \bibnamefont
  {Marinova}},\ }\bibfield  {title} {\bibinfo {title} {{The Physics, Biology,
  and Environmental Ethics of Making Mars Habitable}},\ }\href
  {https://doi.org/10.1089/153110701750137477} {\bibfield  {journal} {\bibinfo
  {journal} {Astrobiology}\ }\textbf {\bibinfo {volume} {1}},\ \bibinfo {pages}
  {89} (\bibinfo {year} {2001})}\BibitemShut {NoStop}%
\bibitem [{\citenamefont {Jakosky}\ and\ \citenamefont
  {Edwards}(2018)}]{JakoskyEdwards2018CO2Inventory}%
  \BibitemOpen
  \bibfield  {author} {\bibinfo {author} {\bibfnamefont {B.~M.}\ \bibnamefont
  {Jakosky}}\ and\ \bibinfo {author} {\bibfnamefont {C.~S.}\ \bibnamefont
  {Edwards}},\ }\bibfield  {title} {\bibinfo {title} {{Inventory of {CO$_2$}
  available for terraforming {Mars}}},\ }\href
  {https://doi.org/10.1038/s41550-018-0529-6} {\bibfield  {journal} {\bibinfo
  {journal} {Nature Astronomy}\ }\textbf {\bibinfo {volume} {2}},\ \bibinfo
  {pages} {634} (\bibinfo {year} {2018})}\BibitemShut {NoStop}%
\bibitem [{\citenamefont {Tutolo}\ \emph {et~al.}(2025)\citenamefont {Tutolo},
  \citenamefont {Hausrath}, \citenamefont {Kite}, \citenamefont {Rampe},
  \citenamefont {Bristow}, \citenamefont {Downs} \emph
  {et~al.}}]{Tutolo2025Carbonates}%
  \BibitemOpen
  \bibfield  {author} {\bibinfo {author} {\bibfnamefont {B.~M.}\ \bibnamefont
  {Tutolo}}, \bibinfo {author} {\bibfnamefont {E.~M.}\ \bibnamefont
  {Hausrath}}, \bibinfo {author} {\bibfnamefont {E.~S.}\ \bibnamefont {Kite}},
  \bibinfo {author} {\bibfnamefont {E.~B.}\ \bibnamefont {Rampe}}, \bibinfo
  {author} {\bibfnamefont {T.~F.}\ \bibnamefont {Bristow}}, \bibinfo {author}
  {\bibfnamefont {R.~T.}\ \bibnamefont {Downs}}, \emph {et~al.},\ }\bibfield
  {title} {\bibinfo {title} {{Carbonates Identified by the Curiosity Rover
  Indicate a Carbon Cycle Operated on Ancient Mars}},\ }\href
  {https://doi.org/10.1126/science.ado9966} {\bibfield  {journal} {\bibinfo
  {journal} {Science}\ }\textbf {\bibinfo {volume} {388}},\ \bibinfo {pages}
  {292} (\bibinfo {year} {2025})}\BibitemShut {NoStop}%
\bibitem [{\citenamefont {Buhler}\ \emph {et~al.}(2020)\citenamefont {Buhler},
  \citenamefont {Ingersoll}, \citenamefont {Ehlmann}, \citenamefont {Hayne},\
  and\ \citenamefont {Piqueux}}]{Buhler2020Coevolution}%
  \BibitemOpen
  \bibfield  {author} {\bibinfo {author} {\bibfnamefont {P.~B.}\ \bibnamefont
  {Buhler}}, \bibinfo {author} {\bibfnamefont {A.~P.}\ \bibnamefont
  {Ingersoll}}, \bibinfo {author} {\bibfnamefont {B.~L.}\ \bibnamefont
  {Ehlmann}}, \bibinfo {author} {\bibfnamefont {P.~O.}\ \bibnamefont {Hayne}},\
  and\ \bibinfo {author} {\bibfnamefont {S.}~\bibnamefont {Piqueux}},\
  }\bibfield  {title} {\bibinfo {title} {{Coevolution of Mars's Atmosphere and
  Massive South Polar CO$_2$ Ice Deposit}},\ }\href
  {https://doi.org/10.1038/s41550-019-0976-8} {\bibfield  {journal} {\bibinfo
  {journal} {Nature Astronomy}\ }\textbf {\bibinfo {volume} {4}},\ \bibinfo
  {pages} {364} (\bibinfo {year} {2020})}\BibitemShut {NoStop}%
\bibitem [{\citenamefont {Buhler}\ and\ \citenamefont
  {Piqueux}(2021)}]{BuhlerPiqueux2021Obliquity}%
  \BibitemOpen
  \bibfield  {author} {\bibinfo {author} {\bibfnamefont {P.~B.}\ \bibnamefont
  {Buhler}}\ and\ \bibinfo {author} {\bibfnamefont {S.}~\bibnamefont
  {Piqueux}},\ }\bibfield  {title} {\bibinfo {title} {{Obliquity-Driven CO$_2$
  Exchange Between Mars' Atmosphere, Regolith, and Polar Cap}},\ }\href
  {https://doi.org/10.1029/2020JE006759} {\bibfield  {journal} {\bibinfo
  {journal} {JGR: Planets}\ }\textbf {\bibinfo {volume} {126}},\ \bibinfo
  {pages} {e2020JE006759} (\bibinfo {year} {2021})}\BibitemShut {NoStop}%
\bibitem [{\citenamefont {Broquet}\ \emph {et~al.}(2021)\citenamefont
  {Broquet}, \citenamefont {Wieczorek},\ and\ \citenamefont
  {Fa}}]{Broquet2021SouthPolarCap}%
  \BibitemOpen
  \bibfield  {author} {\bibinfo {author} {\bibfnamefont {A.}~\bibnamefont
  {Broquet}}, \bibinfo {author} {\bibfnamefont {M.~A.}\ \bibnamefont
  {Wieczorek}},\ and\ \bibinfo {author} {\bibfnamefont {W.}~\bibnamefont
  {Fa}},\ }\bibfield  {title} {\bibinfo {title} {{The Composition of the South
  Polar Cap of Mars Derived from Orbital Data}},\ }\href
  {https://doi.org/10.1029/2020JE006730} {\bibfield  {journal} {\bibinfo
  {journal} {JGR: Planets}\ }\textbf {\bibinfo {volume} {126}},\ \bibinfo
  {pages} {e2020JE006730} (\bibinfo {year} {2021})}\BibitemShut {NoStop}%
\bibitem [{\citenamefont {Wordsworth}\ \emph {et~al.}(2019)\citenamefont
  {Wordsworth}, \citenamefont {Kerber},\ and\ \citenamefont
  {Cockell}}]{Wordsworth2019Aerogel}%
  \BibitemOpen
  \bibfield  {author} {\bibinfo {author} {\bibfnamefont {R.}~\bibnamefont
  {Wordsworth}}, \bibinfo {author} {\bibfnamefont {L.}~\bibnamefont {Kerber}},\
  and\ \bibinfo {author} {\bibfnamefont {C.~S.}\ \bibnamefont {Cockell}},\
  }\bibfield  {title} {\bibinfo {title} {Enabling {Martian} habitability with
  silica aerogel via the solid-state greenhouse effect},\ }\href
  {https://doi.org/10.1038/s41550-019-0813-0} {\bibfield  {journal} {\bibinfo
  {journal} {Nature Astronomy}\ }\textbf {\bibinfo {volume} {3}},\ \bibinfo
  {pages} {898} (\bibinfo {year} {2019})}\BibitemShut {NoStop}%
\bibitem [{\citenamefont {Wordsworth}\ and\ \citenamefont
  {Cockell}(2024)}]{WordsworthCockell2024LivingHabitats}%
  \BibitemOpen
  \bibfield  {author} {\bibinfo {author} {\bibfnamefont {R.}~\bibnamefont
  {Wordsworth}}\ and\ \bibinfo {author} {\bibfnamefont {C.}~\bibnamefont
  {Cockell}},\ }\bibfield  {title} {\bibinfo {title} {{Self-Sustaining Living
  Habitats in Extraterrestrial Environments}},\ }\href
  {https://doi.org/10.1089/ast.2024.0080} {\bibfield  {journal} {\bibinfo
  {journal} {Astrobiology}\ }\textbf {\bibinfo {volume} {24}},\ \bibinfo
  {pages} {1187} (\bibinfo {year} {2024})}\BibitemShut {NoStop}%
\bibitem [{\citenamefont {Wordsworth}\ \emph {et~al.}(2025)\citenamefont
  {Wordsworth}, \citenamefont {Quayum}, \citenamefont {Kocharian},
  \citenamefont {Pearson}, \citenamefont {Portillo}, \citenamefont {Yang},
  \citenamefont {Cockell}, \citenamefont {Nangle},\ and\ \citenamefont
  {Church}}]{Wordsworth2025Biomaterials}%
  \BibitemOpen
  \bibfield  {author} {\bibinfo {author} {\bibfnamefont {R.}~\bibnamefont
  {Wordsworth}}, \bibinfo {author} {\bibfnamefont {R.}~\bibnamefont {Quayum}},
  \bibinfo {author} {\bibfnamefont {E.}~\bibnamefont {Kocharian}}, \bibinfo
  {author} {\bibfnamefont {A.}~\bibnamefont {Pearson}}, \bibinfo {author}
  {\bibfnamefont {X.}~\bibnamefont {Portillo}}, \bibinfo {author}
  {\bibfnamefont {M.}~\bibnamefont {Yang}}, \bibinfo {author} {\bibfnamefont
  {C.~S.}\ \bibnamefont {Cockell}}, \bibinfo {author} {\bibfnamefont
  {S.}~\bibnamefont {Nangle}},\ and\ \bibinfo {author} {\bibfnamefont
  {G.}~\bibnamefont {Church}},\ }\bibfield  {title} {\bibinfo {title}
  {{Biomaterials for Organically Generated Habitats Beyond Earth}},\ }\href
  {https://doi.org/10.1126/sciadv.adp4985} {\bibfield  {journal} {\bibinfo
  {journal} {Science Advances}\ }\textbf {\bibinfo {volume} {11}},\ \bibinfo
  {pages} {eadp4985} (\bibinfo {year} {2025})}\BibitemShut {NoStop}%
\bibitem [{\citenamefont {Ansari}\ \emph {et~al.}(2024)\citenamefont {Ansari}
  \emph {et~al.}}]{Ansari2024Nanoparticles}%
  \BibitemOpen
  \bibfield  {author} {\bibinfo {author} {\bibfnamefont {S.}~\bibnamefont
  {Ansari}} \emph {et~al.},\ }\bibfield  {title} {\bibinfo {title} {Feasibility
  of keeping {Mars} warm with nanoparticles},\ }\href
  {https://doi.org/10.1126/sciadv.adn4650} {\bibfield  {journal} {\bibinfo
  {journal} {Science Advances}\ }\textbf {\bibinfo {volume} {10}},\ \bibinfo
  {pages} {eadn4650} (\bibinfo {year} {2024})}\BibitemShut {NoStop}%
\bibitem [{\citenamefont {{Richardson}}\ \emph {et~al.}(2026)\citenamefont
  {{Richardson}}, \citenamefont {{Ansari}}, \citenamefont {{Fan}},
  \citenamefont {{Ramirez}}, \citenamefont {{Mohseni}}, \citenamefont
  {{Mischna}}, \citenamefont {{Hecht}}, \citenamefont {{Steele}}, \citenamefont
  {{Sharipov}},\ and\ \citenamefont {{Kite}}}]{Richardson2026IRParticles}%
  \BibitemOpen
  \bibfield  {author} {\bibinfo {author} {\bibfnamefont {M.~I.}\ \bibnamefont
  {{Richardson}}}, \bibinfo {author} {\bibfnamefont {S.}~\bibnamefont
  {{Ansari}}}, \bibinfo {author} {\bibfnamefont {B.}~\bibnamefont {{Fan}}},
  \bibinfo {author} {\bibfnamefont {R.}~\bibnamefont {{Ramirez}}}, \bibinfo
  {author} {\bibfnamefont {H.}~\bibnamefont {{Mohseni}}}, \bibinfo {author}
  {\bibfnamefont {M.~A.}\ \bibnamefont {{Mischna}}}, \bibinfo {author}
  {\bibfnamefont {M.~H.}\ \bibnamefont {{Hecht}}}, \bibinfo {author}
  {\bibfnamefont {L.~J.}\ \bibnamefont {{Steele}}}, \bibinfo {author}
  {\bibfnamefont {F.}~\bibnamefont {{Sharipov}}},\ and\ \bibinfo {author}
  {\bibfnamefont {E.~S.}\ \bibnamefont {{Kite}}},\ }\bibfield  {title}
  {\bibinfo {title} {{{Atmospheric Dynamics of IR-Active Particles Released
  From Mars' Surface}}},\ }\href {https://doi.org/10.1029/2025GL121051}
  {\bibfield  {journal} {\bibinfo  {journal} {GRL}\ }\textbf {\bibinfo {volume}
  {53}},\ \bibinfo {eid} {e2025GL121051} (\bibinfo {year} {2026})}\BibitemShut
  {NoStop}%
\bibitem [{\citenamefont {DeBenedictis}\ \emph {et~al.}(2025)\citenamefont
  {DeBenedictis}, \citenamefont {Kite}, \citenamefont {Wordsworth},
  \citenamefont {Lanza}, \citenamefont {Cockell}, \citenamefont {Silver},
  \citenamefont {Ramirez}, \citenamefont {Cumbers}, \citenamefont {Mohseni},
  \citenamefont {Mason}, \citenamefont {Fischer},\ and\ \citenamefont
  {McKay}}]{DeBenedictis2025Case}%
  \BibitemOpen
  \bibfield  {author} {\bibinfo {author} {\bibfnamefont {E.~A.}\ \bibnamefont
  {DeBenedictis}}, \bibinfo {author} {\bibfnamefont {E.~S.}\ \bibnamefont
  {Kite}}, \bibinfo {author} {\bibfnamefont {R.~D.}\ \bibnamefont
  {Wordsworth}}, \bibinfo {author} {\bibfnamefont {N.~L.}\ \bibnamefont
  {Lanza}}, \bibinfo {author} {\bibfnamefont {C.~S.}\ \bibnamefont {Cockell}},
  \bibinfo {author} {\bibfnamefont {P.~A.}\ \bibnamefont {Silver}}, \bibinfo
  {author} {\bibfnamefont {R.~M.}\ \bibnamefont {Ramirez}}, \bibinfo {author}
  {\bibfnamefont {J.}~\bibnamefont {Cumbers}}, \bibinfo {author} {\bibfnamefont
  {H.}~\bibnamefont {Mohseni}}, \bibinfo {author} {\bibfnamefont {C.~E.}\
  \bibnamefont {Mason}}, \bibinfo {author} {\bibfnamefont {W.~W.}\ \bibnamefont
  {Fischer}},\ and\ \bibinfo {author} {\bibfnamefont {C.~P.}\ \bibnamefont
  {McKay}},\ }\bibfield  {title} {\bibinfo {title} {{The Case for Mars
  Terraforming Research}},\ }\href {https://doi.org/10.1038/s41550-025-02548-0}
  {\bibfield  {journal} {\bibinfo  {journal} {Nature Astronomy}\ }\textbf
  {\bibinfo {volume} {9}},\ \bibinfo {pages} {634} (\bibinfo {year}
  {2025})}\BibitemShut {NoStop}%
\bibitem [{\citenamefont {{Kite}}\ \emph {et~al.}(2026)\citenamefont {{Kite}},
  \citenamefont {{Essunfeld}}, \citenamefont {{Hecht}}, \citenamefont
  {{Mischna}}, \citenamefont {{Wordsworth}}, \citenamefont {{Mohseni}},
  \citenamefont {{Boies}}, \citenamefont {{Averesch}}, \citenamefont
  {{Ansari}}, \citenamefont {{Richardson}}, \citenamefont {{DeBenedictis}},
  \citenamefont {{Stork}}, \citenamefont {{Bamba}}, \citenamefont {{Handmer}},
  \citenamefont {{Jourdain}}, \citenamefont {{Ramirez}}, \citenamefont
  {{Mason}}, \citenamefont {{Kling}}, \citenamefont {{Braude}}, \citenamefont
  {{Dumitrescu}}, \citenamefont {{Worden}}, \citenamefont {{Cumbers}},
  \citenamefont {{Lanza}}, \citenamefont {{Quayum}},\ and\ \citenamefont
  {{Cockell}}}]{Kite2026Roadmap}%
  \BibitemOpen
  \bibfield  {author} {\bibinfo {author} {\bibfnamefont {E.~S.}\ \bibnamefont
  {{Kite}}}, \bibinfo {author} {\bibfnamefont {A.}~\bibnamefont {{Essunfeld}}},
  \bibinfo {author} {\bibfnamefont {M.~H.}\ \bibnamefont {{Hecht}}}, \bibinfo
  {author} {\bibfnamefont {M.~A.}\ \bibnamefont {{Mischna}}}, \bibinfo {author}
  {\bibfnamefont {R.}~\bibnamefont {{Wordsworth}}}, \bibinfo {author}
  {\bibfnamefont {H.}~\bibnamefont {{Mohseni}}}, \bibinfo {author}
  {\bibfnamefont {A.}~\bibnamefont {{Boies}}}, \bibinfo {author} {\bibfnamefont
  {N.}~\bibnamefont {{Averesch}}}, \bibinfo {author} {\bibfnamefont
  {S.}~\bibnamefont {{Ansari}}}, \bibinfo {author} {\bibfnamefont {M.~I.}\
  \bibnamefont {{Richardson}}}, \bibinfo {author} {\bibfnamefont {E.~A.}\
  \bibnamefont {{DeBenedictis}}}, \bibinfo {author} {\bibfnamefont
  {D.}~\bibnamefont {{Stork}}}, \bibinfo {author} {\bibfnamefont {A.~L.}\
  \bibnamefont {{Bamba}}}, \bibinfo {author} {\bibfnamefont {C.~J.}\
  \bibnamefont {{Handmer}}}, \bibinfo {author} {\bibfnamefont {C.}~\bibnamefont
  {{Jourdain}}}, \bibinfo {author} {\bibfnamefont {R.}~\bibnamefont
  {{Ramirez}}}, \bibinfo {author} {\bibfnamefont {C.~E.}\ \bibnamefont
  {{Mason}}}, \bibinfo {author} {\bibfnamefont {A.}~\bibnamefont {{Kling}}},
  \bibinfo {author} {\bibfnamefont {A.~S.}\ \bibnamefont {{Braude}}}, \bibinfo
  {author} {\bibfnamefont {A.}~\bibnamefont {{Dumitrescu}}}, \bibinfo {author}
  {\bibfnamefont {S.~P.}\ \bibnamefont {{Worden}}}, \bibinfo {author}
  {\bibfnamefont {J.}~\bibnamefont {{Cumbers}}}, \bibinfo {author}
  {\bibfnamefont {N.}~\bibnamefont {{Lanza}}}, \bibinfo {author} {\bibfnamefont
  {R.}~\bibnamefont {{Quayum}}},\ and\ \bibinfo {author} {\bibfnamefont
  {C.~S.}\ \bibnamefont {{Cockell}}},\ }\href
  {https://doi.org/10.48550/arXiv.2604.02242} {\bibinfo {title} {{{A research
  roadmap for assessing the feasibility of warming Mars}}}} (\bibinfo {year}
  {2026})\BibitemShut {NoStop}%
\bibitem [{\citenamefont {Graham}(2004)}]{Graham2004Biological}%
  \BibitemOpen
  \bibfield  {author} {\bibinfo {author} {\bibfnamefont {J.~M.}\ \bibnamefont
  {Graham}},\ }\bibfield  {title} {\bibinfo {title} {{The Biological
  Terraforming of Mars: Planetary Ecosynthesis as Ecological Succession on a
  Global Scale}},\ }\href {https://doi.org/10.1089/153110704323175133}
  {\bibfield  {journal} {\bibinfo  {journal} {Astrobiology}\ }\textbf {\bibinfo
  {volume} {4}},\ \bibinfo {pages} {168} (\bibinfo {year} {2004})}\BibitemShut
  {NoStop}%
\bibitem [{\citenamefont {Hecht}(2002)}]{Hecht2002Metastability}%
  \BibitemOpen
  \bibfield  {author} {\bibinfo {author} {\bibfnamefont {M.~H.}\ \bibnamefont
  {Hecht}},\ }\bibfield  {title} {\bibinfo {title} {{Metastability of Liquid
  Water on Mars}},\ }\href {https://doi.org/10.1006/icar.2001.6794} {\bibfield
  {journal} {\bibinfo  {journal} {Icarus}\ }\textbf {\bibinfo {volume} {156}},\
  \bibinfo {pages} {373} (\bibinfo {year} {2002})}\BibitemShut {NoStop}%
\bibitem [{\citenamefont {Schorghofer}(2020)}]{Schorghofer2020Crocus}%
  \BibitemOpen
  \bibfield  {author} {\bibinfo {author} {\bibfnamefont {N.}~\bibnamefont
  {Schorghofer}},\ }\bibfield  {title} {\bibinfo {title} {{Mars: Quantitative
  Evaluation of Crocus Melting behind Boulders}},\ }\href
  {https://doi.org/10.3847/1538-4357/ab612f} {\bibfield  {journal} {\bibinfo
  {journal} {ApJ}\ }\textbf {\bibinfo {volume} {890}},\ \bibinfo {pages} {49}
  (\bibinfo {year} {2020})}\BibitemShut {NoStop}%
\bibitem [{\citenamefont {Lange}\ and\ \citenamefont
  {Forget}(2026)}]{LangeForget2026Gullies}%
  \BibitemOpen
  \bibfield  {author} {\bibinfo {author} {\bibfnamefont {L.}~\bibnamefont
  {Lange}}\ and\ \bibinfo {author} {\bibfnamefont {F.}~\bibnamefont {Forget}},\
  }\bibfield  {title} {\bibinfo {title} {{On the Possibility of Melting Water
  Ice During the Recent Past of Mars: Implications for the Formation of
  Gullies}},\ }\href {https://doi.org/10.1029/2025JE009512} {\bibfield
  {journal} {\bibinfo  {journal} {JRG: Planets}\ }\textbf {\bibinfo {volume}
  {131}},\ \bibinfo {pages} {e2025JE009512} (\bibinfo {year}
  {2026})}\BibitemShut {NoStop}%
\bibitem [{\citenamefont {{International Association for the Properties of
  Water and Steam (IAPWS)}}(2011)}]{IAPWS2011}%
  \BibitemOpen
  \bibfield  {author} {\bibinfo {author} {\bibnamefont {{International
  Association for the Properties of Water and Steam (IAPWS)}}},\ }\href
  {https://www.iapws.org/relguide/MeltSub2011.pdf} {\bibinfo {title} {{Revised
  Release on the Pressure along the Melting and Sublimation Curves of Ordinary
  Water Substance}}} (\bibinfo {year} {2011}),\ \bibinfo {note} {iAPWS Release
  (R14-08(2011))}\BibitemShut {NoStop}%
\bibitem [{\citenamefont {{NASA Office of the Chief Health and Medical
  Officer}}(2023)}]{NASAHabitableAtmosphere2023}%
  \BibitemOpen
  \bibfield  {author} {\bibinfo {author} {\bibnamefont {{NASA Office of the
  Chief Health and Medical Officer}}},\ }\href
  {https://www.nasa.gov/wp-content/uploads/2023/12/ochmo-tb-003-habitable-atmosphere.pdf}
  {\bibinfo {title} {{Habitable Atmosphere (OCHMO-TB-003 Rev A)}}},\ \bibinfo
  {howpublished} {{NASA-STD-3001 Technical Brief}} (\bibinfo {year}
  {2023})\BibitemShut {NoStop}%
\bibitem [{\citenamefont {Alexander}(2001)}]{Alexander2001MarsEnv}%
  \BibitemOpen
  \bibfield  {author} {\bibinfo {author} {\bibfnamefont {M.}~\bibnamefont
  {Alexander}},\ }\href {https://ntrs.nasa.gov/citations/20010046858} {\emph
  {\bibinfo {title} {{Mars Transportation Environment Definition Document}}}},\
  \bibinfo {type} {Tech. Rep.}\ \bibinfo {number} {{NASA/TM-2001-210935}}\
  (\bibinfo  {institution} {NASA Marshall Space Flight Center},\ \bibinfo
  {year} {2001})\ \bibinfo {note} {{NTRS Document ID: 20010046858}}\BibitemShut
  {NoStop}%
\bibitem [{\citenamefont {Dartnell}\ \emph {et~al.}(2007)\citenamefont
  {Dartnell}, \citenamefont {Desorgher}, \citenamefont {Ward},\ and\
  \citenamefont {Coates}}]{Dartnell2007Radiation}%
  \BibitemOpen
  \bibfield  {author} {\bibinfo {author} {\bibfnamefont {L.~R.}\ \bibnamefont
  {Dartnell}}, \bibinfo {author} {\bibfnamefont {L.}~\bibnamefont {Desorgher}},
  \bibinfo {author} {\bibfnamefont {J.~M.}\ \bibnamefont {Ward}},\ and\
  \bibinfo {author} {\bibfnamefont {A.~J.}\ \bibnamefont {Coates}},\ }\bibfield
   {title} {\bibinfo {title} {{Modelling the Surface and Subsurface Martian
  Radiation Environment: Implications for Astrobiology}},\ }\href
  {https://doi.org/10.1029/2006GL027494} {\bibfield  {journal} {\bibinfo
  {journal} {GRL}\ }\textbf {\bibinfo {volume} {34}},\ \bibinfo {pages}
  {L02207} (\bibinfo {year} {2007})}\BibitemShut {NoStop}%
\bibitem [{\citenamefont {Moores}\ and\ \citenamefont
  {Schuerger}(2008)}]{Moores2007UVShielding}%
  \BibitemOpen
  \bibfield  {author} {\bibinfo {author} {\bibfnamefont {J.~E.}\ \bibnamefont
  {Moores}}\ and\ \bibinfo {author} {\bibfnamefont {A.~C.}\ \bibnamefont
  {Schuerger}},\ }\bibfield  {title} {\bibinfo {title} {{The Shielding Effect
  of Small-Scale Martian Surface Geometry on Ultraviolet Flux}},\ }\href
  {https://doi.org/10.1016/j.icarus.2007.11.008} {\bibfield  {journal}
  {\bibinfo  {journal} {Icarus}\ }\textbf {\bibinfo {volume} {195}},\ \bibinfo
  {pages} {131} (\bibinfo {year} {2008})}\BibitemShut {NoStop}%
\bibitem [{\citenamefont {Stern}\ \emph {et~al.}(2015)\citenamefont {Stern}
  \emph {et~al.}}]{Stern2015Nitrates}%
  \BibitemOpen
  \bibfield  {author} {\bibinfo {author} {\bibfnamefont {J.~C.}\ \bibnamefont
  {Stern}} \emph {et~al.},\ }\bibfield  {title} {\bibinfo {title} {{Evidence
  for indigenous nitrogen in sedimentary and aeolian deposits from the
  Curiosity rover investigations at Gale crater, {Mars}}},\ }\href
  {https://doi.org/10.1073/pnas.1420932112} {\bibfield  {journal} {\bibinfo
  {journal} {Proc. of the National Academy of Sciences of the United States of
  America}\ }\textbf {\bibinfo {volume} {112}},\ \bibinfo {pages} {4245}
  (\bibinfo {year} {2015})}\BibitemShut {NoStop}%
\bibitem [{\citenamefont {{National Institute of Standards and Technology
  (NIST)}}(2026)}]{NISTCO2WebBook}%
  \BibitemOpen
  \bibfield  {author} {\bibinfo {author} {\bibnamefont {{National Institute of
  Standards and Technology (NIST)}}},\ }\href
  {https://webbook.nist.gov/cgi/cbook.cgi?ID=C124389&Units=SI} {\bibinfo
  {title} {{NIST Chemistry WebBook, SRD 69: Carbon dioxide (thermophysical
  properties)}}},\ \bibinfo {howpublished} {Online database} (\bibinfo {year}
  {2026})\BibitemShut {NoStop}%
\bibitem [{\citenamefont {Gerstell}\ \emph {et~al.}(2001)\citenamefont
  {Gerstell}, \citenamefont {Francisco}, \citenamefont {Yung}, \citenamefont
  {Boxe},\ and\ \citenamefont {Aaltonee}}]{Gerstell2001SuperGHG}%
  \BibitemOpen
  \bibfield  {author} {\bibinfo {author} {\bibfnamefont {M.~F.}\ \bibnamefont
  {Gerstell}}, \bibinfo {author} {\bibfnamefont {J.~S.}\ \bibnamefont
  {Francisco}}, \bibinfo {author} {\bibfnamefont {Y.~L.}\ \bibnamefont {Yung}},
  \bibinfo {author} {\bibfnamefont {C.}~\bibnamefont {Boxe}},\ and\ \bibinfo
  {author} {\bibfnamefont {E.~T.}\ \bibnamefont {Aaltonee}},\ }\bibfield
  {title} {\bibinfo {title} {{Keeping {Mars} warm with new super greenhouse
  gases}},\ }\href {https://doi.org/10.1073/pnas.051511598} {\bibfield
  {journal} {\bibinfo  {journal} {Proc. of the National Academy of Sciences of
  the United States of America}\ }\textbf {\bibinfo {volume} {98}},\ \bibinfo
  {pages} {2154} (\bibinfo {year} {2001})}\BibitemShut {NoStop}%
\bibitem [{\citenamefont {Marinova}\ \emph {et~al.}(2005)\citenamefont
  {Marinova}, \citenamefont {McKay},\ and\ \citenamefont
  {Hashimoto}}]{Marinova2005PFCMars}%
  \BibitemOpen
  \bibfield  {author} {\bibinfo {author} {\bibfnamefont {M.~M.}\ \bibnamefont
  {Marinova}}, \bibinfo {author} {\bibfnamefont {C.~P.}\ \bibnamefont
  {McKay}},\ and\ \bibinfo {author} {\bibfnamefont {H.}~\bibnamefont
  {Hashimoto}},\ }\bibfield  {title} {\bibinfo {title} {{Radiative-convective
  model of warming {Mars} with artificial greenhouse gases}},\ }\href
  {https://doi.org/10.1029/2004JE002306} {\bibfield  {journal} {\bibinfo
  {journal} {JGR: Planets}\ }\textbf {\bibinfo {volume} {110}},\ \bibinfo
  {pages} {E03002} (\bibinfo {year} {2005})}\BibitemShut {NoStop}%
\bibitem [{\citenamefont {Ramirez}\ \emph {et~al.}(2014)\citenamefont
  {Ramirez}, \citenamefont {Kopparapu}, \citenamefont {Zugger}, \citenamefont
  {Robinson}, \citenamefont {Freedman},\ and\ \citenamefont
  {Kasting}}]{Ramirez2014WarmEarlyMars}%
  \BibitemOpen
  \bibfield  {author} {\bibinfo {author} {\bibfnamefont {R.~M.}\ \bibnamefont
  {Ramirez}}, \bibinfo {author} {\bibfnamefont {R.}~\bibnamefont {Kopparapu}},
  \bibinfo {author} {\bibfnamefont {M.~E.}\ \bibnamefont {Zugger}}, \bibinfo
  {author} {\bibfnamefont {T.~D.}\ \bibnamefont {Robinson}}, \bibinfo {author}
  {\bibfnamefont {R.}~\bibnamefont {Freedman}},\ and\ \bibinfo {author}
  {\bibfnamefont {J.~F.}\ \bibnamefont {Kasting}},\ }\bibfield  {title}
  {\bibinfo {title} {Warming early {Mars} with {CO$_2$} and {H$_2$}},\ }\href
  {https://doi.org/10.1038/ngeo2000} {\bibfield  {journal} {\bibinfo  {journal}
  {Nature Geoscience}\ }\textbf {\bibinfo {volume} {7}},\ \bibinfo {pages} {59}
  (\bibinfo {year} {2014})}\BibitemShut {NoStop}%
\bibitem [{\citenamefont {Turbet}\ \emph {et~al.}(2019)\citenamefont {Turbet},
  \citenamefont {Tran}, \citenamefont {Pirali}, \citenamefont {Forget},
  \citenamefont {Boulet},\ and\ \citenamefont {Hartmann}}]{Turbet2019FarIR}%
  \BibitemOpen
  \bibfield  {author} {\bibinfo {author} {\bibfnamefont {M.}~\bibnamefont
  {Turbet}}, \bibinfo {author} {\bibfnamefont {H.}~\bibnamefont {Tran}},
  \bibinfo {author} {\bibfnamefont {O.}~\bibnamefont {Pirali}}, \bibinfo
  {author} {\bibfnamefont {F.}~\bibnamefont {Forget}}, \bibinfo {author}
  {\bibfnamefont {C.}~\bibnamefont {Boulet}},\ and\ \bibinfo {author}
  {\bibfnamefont {J.-M.}\ \bibnamefont {Hartmann}},\ }\bibfield  {title}
  {\bibinfo {title} {{Far Infrared Measurements of Absorptions by CH$_4$ +
  CO$_2$ and H$_2$ + CO$_2$ Mixtures and Implications for Greenhouse Warming on
  Early Mars}},\ }\href {https://doi.org/10.1016/j.icarus.2018.11.021}
  {\bibfield  {journal} {\bibinfo  {journal} {Icarus}\ }\textbf {\bibinfo
  {volume} {321}},\ \bibinfo {pages} {189} (\bibinfo {year}
  {2019})}\BibitemShut {NoStop}%
\bibitem [{\citenamefont {Godin}\ \emph {et~al.}(2020)\citenamefont {Godin},
  \citenamefont {Ramirez}, \citenamefont {Campbell}, \citenamefont {Wizenberg},
  \citenamefont {Nguyen}, \citenamefont {Strong},\ and\ \citenamefont
  {Moores}}]{Godin2020CIA}%
  \BibitemOpen
  \bibfield  {author} {\bibinfo {author} {\bibfnamefont {P.~J.}\ \bibnamefont
  {Godin}}, \bibinfo {author} {\bibfnamefont {R.~M.}\ \bibnamefont {Ramirez}},
  \bibinfo {author} {\bibfnamefont {C.~L.}\ \bibnamefont {Campbell}}, \bibinfo
  {author} {\bibfnamefont {T.}~\bibnamefont {Wizenberg}}, \bibinfo {author}
  {\bibfnamefont {T.~G.}\ \bibnamefont {Nguyen}}, \bibinfo {author}
  {\bibfnamefont {K.}~\bibnamefont {Strong}},\ and\ \bibinfo {author}
  {\bibfnamefont {J.~E.}\ \bibnamefont {Moores}},\ }\bibfield  {title}
  {\bibinfo {title} {{Collision-Induced Absorption of CH$_4$--CO$_2$ and
  H$_2$--CO$_2$ Complexes and Their Effect on the Ancient Martian
  Atmosphere}},\ }\href {https://doi.org/10.1029/2019JE006357} {\bibfield
  {journal} {\bibinfo  {journal} {JGR: Planets}\ }\textbf {\bibinfo {volume}
  {125}},\ \bibinfo {pages} {e2019JE006357} (\bibinfo {year}
  {2020})}\BibitemShut {NoStop}%
\bibitem [{\citenamefont {McInnes}(2010)}]{McInnes2010MarsReflectors}%
  \BibitemOpen
  \bibfield  {author} {\bibinfo {author} {\bibfnamefont {C.~R.}\ \bibnamefont
  {McInnes}},\ }\bibfield  {title} {\bibinfo {title} {{Mars Climate Engineering
  using Orbiting Solar Reflectors}},\ }in\ \href
  {https://doi.org/10.1007/978-3-642-03629-3_25} {\emph {\bibinfo {booktitle}
  {{Mars: Prospective Energy and Material Resources}}}}\ (\bibinfo  {publisher}
  {Springer},\ \bibinfo {address} {Berlin, Heidelberg},\ \bibinfo {year}
  {2010})\ pp.\ \bibinfo {pages} {645--659}\BibitemShut {NoStop}%
\bibitem [{\citenamefont {Handmer}(2024)}]{Handmer2024MarsSails}%
  \BibitemOpen
  \bibfield  {author} {\bibinfo {author} {\bibfnamefont {C.~J.}\ \bibnamefont
  {Handmer}},\ }\href@noop {} {\bibinfo {title} {{Mars climate engineering with
  mass-produced solar sails}}},\ \bibinfo {howpublished} {Presented at the 10th
  International Conference on Mars} (\bibinfo {year} {2024}),\ \bibinfo {note}
  {conference presentation; replace with the exact proceedings record used by
  the author, if available}\BibitemShut {NoStop}%
\bibitem [{\citenamefont {Turyshev}\ \emph {et~al.}(2023)\citenamefont
  {Turyshev}, \citenamefont {Garber}, \citenamefont {Friedman}, \citenamefont
  {Hein}, \citenamefont {Barnes}, \citenamefont {Batygin}, \citenamefont
  {Brown}, \citenamefont {Cronin}, \citenamefont {Davoyan}, \citenamefont
  {Dubill}, \citenamefont {Eubanks}, \citenamefont {Gibson}, \citenamefont
  {Hassler}, \citenamefont {Izenberg}, \citenamefont {Kervella}, \citenamefont
  {Mauskopf}, \citenamefont {Murphy}, \citenamefont {Nutter}, \citenamefont
  {Porco}, \citenamefont {Riccobono}, \citenamefont {Schalkwyk}, \citenamefont
  {Stevenson}, \citenamefont {Sykes}, \citenamefont {Sultana}, \citenamefont
  {Toth}, \citenamefont {Velli},\ and\ \citenamefont
  {Worden}}]{Turyshev2023Smallsats}%
  \BibitemOpen
  \bibfield  {author} {\bibinfo {author} {\bibfnamefont {S.~G.}\ \bibnamefont
  {Turyshev}}, \bibinfo {author} {\bibfnamefont {D.}~\bibnamefont {Garber}},
  \bibinfo {author} {\bibfnamefont {L.~D.}\ \bibnamefont {Friedman}}, \bibinfo
  {author} {\bibfnamefont {A.~M.}\ \bibnamefont {Hein}}, \bibinfo {author}
  {\bibfnamefont {N.}~\bibnamefont {Barnes}}, \bibinfo {author} {\bibfnamefont
  {K.}~\bibnamefont {Batygin}}, \bibinfo {author} {\bibfnamefont {M.~E.}\
  \bibnamefont {Brown}}, \bibinfo {author} {\bibfnamefont {L.}~\bibnamefont
  {Cronin}}, \bibinfo {author} {\bibfnamefont {A.}~\bibnamefont {Davoyan}},
  \bibinfo {author} {\bibfnamefont {A.}~\bibnamefont {Dubill}}, \bibinfo
  {author} {\bibfnamefont {T.~M.}\ \bibnamefont {Eubanks}}, \bibinfo {author}
  {\bibfnamefont {S.}~\bibnamefont {Gibson}}, \bibinfo {author} {\bibfnamefont
  {D.~M.}\ \bibnamefont {Hassler}}, \bibinfo {author} {\bibfnamefont {N.~R.}\
  \bibnamefont {Izenberg}}, \bibinfo {author} {\bibfnamefont {P.}~\bibnamefont
  {Kervella}}, \bibinfo {author} {\bibfnamefont {P.~D.}\ \bibnamefont
  {Mauskopf}}, \bibinfo {author} {\bibfnamefont {N.}~\bibnamefont {Murphy}},
  \bibinfo {author} {\bibfnamefont {A.}~\bibnamefont {Nutter}}, \bibinfo
  {author} {\bibfnamefont {C.}~\bibnamefont {Porco}}, \bibinfo {author}
  {\bibfnamefont {D.}~\bibnamefont {Riccobono}}, \bibinfo {author}
  {\bibfnamefont {J.}~\bibnamefont {Schalkwyk}}, \bibinfo {author}
  {\bibfnamefont {K.~B.}\ \bibnamefont {Stevenson}}, \bibinfo {author}
  {\bibfnamefont {M.~V.}\ \bibnamefont {Sykes}}, \bibinfo {author}
  {\bibfnamefont {M.}~\bibnamefont {Sultana}}, \bibinfo {author} {\bibfnamefont
  {V.~T.}\ \bibnamefont {Toth}}, \bibinfo {author} {\bibfnamefont
  {M.}~\bibnamefont {Velli}},\ and\ \bibinfo {author} {\bibfnamefont {S.~P.}\
  \bibnamefont {Worden}},\ }\bibfield  {title} {\bibinfo {title} {{Science
  opportunities with solar sailing smallsats}},\ }\href
  {https://doi.org/10.1016/j.pss.2023.105744} {\bibfield  {journal} {\bibinfo
  {journal} {Planetary and Space Science}\ }\textbf {\bibinfo {volume} {235}},\
  \bibinfo {pages} {105744} (\bibinfo {year} {2023})}\BibitemShut {NoStop}%
\bibitem [{\citenamefont {Turyshev}\ \emph {et~al.}(2026)\citenamefont
  {Turyshev} \emph {et~al.}}]{Turyshev2026SGLTrades}%
  \BibitemOpen
  \bibfield  {author} {\bibinfo {author} {\bibfnamefont {S.~G.}\ \bibnamefont
  {Turyshev}} \emph {et~al.},\ }\bibfield  {title} {\bibinfo {title}
  {{Propulsion Trades for a 2035--2040 Solar Gravitational Lens Mission}},\
  }\href@noop {} {\bibfield  {journal} {\bibinfo  {journal} {arXiv e-prints}\ }
  (\bibinfo {year} {2026})},\ \Eprint {https://arxiv.org/abs/2602.04198}
  {arXiv:2602.04198 [astro-ph.IM]} \BibitemShut {NoStop}%
\bibitem [{\citenamefont {Wilkie}\ \emph {et~al.}(2021)\citenamefont {Wilkie}
  \emph {et~al.}}]{Wilkie2021ACS3}%
  \BibitemOpen
  \bibfield  {author} {\bibinfo {author} {\bibfnamefont {W.~K.}\ \bibnamefont
  {Wilkie}} \emph {et~al.},\ }\bibfield  {title} {\bibinfo {title} {{The {NASA}
  Advanced Composite Solar Sail System ({ACS3}) Project Overview}},\ }in\
  \href@noop {} {\emph {\bibinfo {booktitle} {35th Annual Small Satellite
  Conference}}}\ (\bibinfo {year} {2021})\ \bibinfo {note} {nASA SmallSat
  Conference paper; total spacecraft mass $\sim$16 kg and sail area $\sim$80
  m$^2$}\BibitemShut {NoStop}%
\bibitem [{\citenamefont {Wilkie}\ \emph {et~al.}(2023)\citenamefont {Wilkie}
  \emph {et~al.}}]{Wilkie2023ACS3Update}%
  \BibitemOpen
  \bibfield  {author} {\bibinfo {author} {\bibfnamefont {K.}~\bibnamefont
  {Wilkie}} \emph {et~al.},\ }\bibfield  {title} {\bibinfo {title} {{Advanced
  Composite Solar Sail System ({ACS3}) Mission Update}},\ }in\ \href@noop {}
  {\emph {\bibinfo {booktitle} {International Symposium on Solar Sailing}}}\
  (\bibinfo {year} {2023})\ \bibinfo {note} {effective characteristic
  acceleration at 1 AU $a_c \simeq 0.045$ mm s$^{-2}$}\BibitemShut {NoStop}%
\bibitem [{\citenamefont {Johnson}\ \emph {et~al.}(2025)\citenamefont {Johnson}
  \emph {et~al.}}]{Johnson2025SailRoadmap}%
  \BibitemOpen
  \bibfield  {author} {\bibinfo {author} {\bibfnamefont {L.}~\bibnamefont
  {Johnson}} \emph {et~al.},\ }\href@noop {} {\emph {\bibinfo {title} {{Solar
  Photon Sails Roadmap}}}},\ \bibinfo {type} {Tech. Rep.}\ (\bibinfo
  {institution} {NASA},\ \bibinfo {year} {2025})\ \bibinfo {note} {roadmap
  targets include deploy-and-control areas $>2.5\times10^3$ m$^2$ (Generation
  2) and $>10^4$ m$^2$ (Generation 3a)}\BibitemShut {NoStop}%
\bibitem [{\citenamefont {Kang}\ \emph {et~al.}(2020)\citenamefont {Kang} \emph
  {et~al.}}]{Kang2020SolarSailDurability}%
  \BibitemOpen
  \bibfield  {author} {\bibinfo {author} {\bibfnamefont {J.~H.}\ \bibnamefont
  {Kang}} \emph {et~al.},\ }\bibfield  {title} {\bibinfo {title} {{Durability
  Characterization of Mechanical Interfaces in Solar Sail Membrane
  Structures}},\ }in\ \href@noop {} {\emph {\bibinfo {booktitle} {AIAA SciTech
  Forum}}}\ (\bibinfo {year} {2020})\ \bibinfo {note} {nASA materials study
  emphasizing membrane/interface durability, radiation-driven adhesive
  degradation, and reflectivity loss}\BibitemShut {NoStop}%
\bibitem [{\citenamefont {Chase}(1998)}]{Chase1998JANAF}%
  \BibitemOpen
  \bibfield  {author} {\bibinfo {author} {\bibfnamefont {J.}~\bibnamefont
  {Chase}, \bibfnamefont {Malcolm~W.}},\ }\href@noop {} {\emph {\bibinfo
  {title} {{NIST-JANAF Thermochemical Tables}}}},\ \bibinfo {edition} {4th}\
  ed.\ (\bibinfo  {publisher} {{American Chemical Society and American
  Institute of Physics for the National Institute of Standards and
  Technology}},\ \bibinfo {address} {Washington, DC},\ \bibinfo {year} {1998})\
  \bibinfo {note} {journal of Physical and Chemical Reference Data, Monograph
  No. 9}\BibitemShut {NoStop}%
\bibitem [{\citenamefont {Field}\ \emph {et~al.}(1998)\citenamefont {Field},
  \citenamefont {Behrenfeld}, \citenamefont {Randerson},\ and\ \citenamefont
  {Falkowski}}]{Field1998BiosphereNPP}%
  \BibitemOpen
  \bibfield  {author} {\bibinfo {author} {\bibfnamefont {C.~B.}\ \bibnamefont
  {Field}}, \bibinfo {author} {\bibfnamefont {M.~J.}\ \bibnamefont
  {Behrenfeld}}, \bibinfo {author} {\bibfnamefont {J.~T.}\ \bibnamefont
  {Randerson}},\ and\ \bibinfo {author} {\bibfnamefont {P.}~\bibnamefont
  {Falkowski}},\ }\bibfield  {title} {\bibinfo {title} {{Primary production of
  the biosphere: Integrating terrestrial and oceanic components}},\ }\href
  {https://doi.org/10.1126/science.281.5374.237} {\bibfield  {journal}
  {\bibinfo  {journal} {Science}\ }\textbf {\bibinfo {volume} {281}},\ \bibinfo
  {pages} {237} (\bibinfo {year} {1998})}\BibitemShut {NoStop}%
\bibitem [{\citenamefont {Hoffman}\ \emph {et~al.}(2022)\citenamefont {Hoffman}
  \emph {et~al.}}]{Hoffman2022MOXIE}%
  \BibitemOpen
  \bibfield  {author} {\bibinfo {author} {\bibfnamefont {J.~A.}\ \bibnamefont
  {Hoffman}} \emph {et~al.},\ }\bibfield  {title} {\bibinfo {title} {{Mars
  Oxygen {ISRU} Experiment ({MOXIE})---Preparing for human {Mars}
  exploration}},\ }\href {https://doi.org/10.1126/sciadv.abp8636} {\bibfield
  {journal} {\bibinfo  {journal} {Science Advances}\ }\textbf {\bibinfo
  {volume} {8}},\ \bibinfo {pages} {eabp8636} (\bibinfo {year} {2022})},\
  \bibinfo {note} {open-access via PubMed Central: PMC9432831}\BibitemShut
  {NoStop}%
\bibitem [{\citenamefont {{NASA}}(2023)}]{NASAMOXIE2023}%
  \BibitemOpen
  \bibfield  {author} {\bibinfo {author} {\bibnamefont {{NASA}}},\ }\href
  {https://www.nasa.gov/missions/mars-2020-perseverance/perseverance-rover/nasas-oxygen-generating-experiment-moxie-completes-mars-mission/}
  {\bibinfo {title} {{NASA’s Oxygen-Generating Experiment {MOXIE} Completes
  Mars Mission}}} (\bibinfo {year} {2023})\BibitemShut {NoStop}%
\bibitem [{\citenamefont {Taylor}(2013)}]{Taylor2013BulkMars}%
  \BibitemOpen
  \bibfield  {author} {\bibinfo {author} {\bibfnamefont {G.~J.}\ \bibnamefont
  {Taylor}},\ }\bibfield  {title} {\bibinfo {title} {{The bulk composition of
  {Mars}}},\ }\href {https://doi.org/10.1016/j.chemer.2013.09.006} {\bibfield
  {journal} {\bibinfo  {journal} {Chemie der Erde -- Geochemistry}\ }\textbf
  {\bibinfo {volume} {73}},\ \bibinfo {pages} {401} (\bibinfo {year}
  {2013})}\BibitemShut {NoStop}%
\bibitem [{\citenamefont {McSween}\ \emph {et~al.}(2009)\citenamefont
  {McSween}, \citenamefont {Taylor},\ and\ \citenamefont
  {Wyatt}}]{McSween2009Crust}%
  \BibitemOpen
  \bibfield  {author} {\bibinfo {author} {\bibfnamefont {J.}~\bibnamefont
  {McSween}, \bibfnamefont {Harry~Y.}}, \bibinfo {author} {\bibfnamefont
  {G.~J.}\ \bibnamefont {Taylor}},\ and\ \bibinfo {author} {\bibfnamefont
  {M.~B.}\ \bibnamefont {Wyatt}},\ }\bibfield  {title} {\bibinfo {title}
  {Elemental composition of the {Martian} crust},\ }\href
  {https://doi.org/10.1126/science.1165871} {\bibfield  {journal} {\bibinfo
  {journal} {Science}\ }\textbf {\bibinfo {volume} {324}},\ \bibinfo {pages}
  {736} (\bibinfo {year} {2009})}\BibitemShut {NoStop}%
\bibitem [{\citenamefont {Jakosky}\ \emph {et~al.}(2018)\citenamefont {Jakosky}
  \emph {et~al.}}]{Jakosky2018AtmosLoss}%
  \BibitemOpen
  \bibfield  {author} {\bibinfo {author} {\bibfnamefont {B.~M.}\ \bibnamefont
  {Jakosky}} \emph {et~al.},\ }\bibfield  {title} {\bibinfo {title} {{Loss of
  the {Martian} atmosphere to space: Present-day loss rates determined from
  {MAVEN} observations and integrated loss through time}},\ }\href
  {https://doi.org/10.1016/j.icarus.2018.05.030} {\bibfield  {journal}
  {\bibinfo  {journal} {Icarus}\ }\textbf {\bibinfo {volume} {315}},\ \bibinfo
  {pages} {146} (\bibinfo {year} {2018})}\BibitemShut {NoStop}%
\bibitem [{\citenamefont {Green}\ \emph {et~al.}(2017)\citenamefont {Green},
  \citenamefont {Hollingsworth}, \citenamefont {Brain}, \citenamefont
  {Airapetian}, \citenamefont {Glocer}, \citenamefont {Pulkkinen},
  \citenamefont {Dong},\ and\ \citenamefont
  {Bamford}}]{Green2017FutureMarsEnv}%
  \BibitemOpen
  \bibfield  {author} {\bibinfo {author} {\bibfnamefont {J.~L.}\ \bibnamefont
  {Green}}, \bibinfo {author} {\bibfnamefont {J.~L.}\ \bibnamefont
  {Hollingsworth}}, \bibinfo {author} {\bibfnamefont {D.}~\bibnamefont
  {Brain}}, \bibinfo {author} {\bibfnamefont {V.}~\bibnamefont {Airapetian}},
  \bibinfo {author} {\bibfnamefont {A.}~\bibnamefont {Glocer}}, \bibinfo
  {author} {\bibfnamefont {A.}~\bibnamefont {Pulkkinen}}, \bibinfo {author}
  {\bibfnamefont {C.}~\bibnamefont {Dong}},\ and\ \bibinfo {author}
  {\bibfnamefont {R.}~\bibnamefont {Bamford}},\ }\bibfield  {title} {\bibinfo
  {title} {{A Future {Mars} Environment for Science and Exploration}},\ }in\
  \href {https://www.hou.usra.edu/meetings/V2050/pdf/8250.pdf} {\emph {\bibinfo
  {booktitle} {Planetary Science Vision 2050 Workshop}}}\ (\bibinfo {year}
  {2017})\ \bibinfo {note} {lPI Contribution No. 1989, abstract
  \#8250}\BibitemShut {NoStop}%
\bibitem [{\citenamefont {Hunten}(1973)}]{Hunten1973}%
  \BibitemOpen
  \bibfield  {author} {\bibinfo {author} {\bibfnamefont {D.~M.}\ \bibnamefont
  {Hunten}},\ }\bibfield  {title} {\bibinfo {title} {{The Escape of Light Gases
  from Planetary Atmospheres}},\ }\href
  {https://doi.org/10.1175/1520-0469(1973)030<1481:TEOLGF>2.0.CO;2} {\bibfield
  {journal} {\bibinfo  {journal} {J. Atmospheric Sciences}\ }\textbf {\bibinfo
  {volume} {30}},\ \bibinfo {pages} {1481} (\bibinfo {year}
  {1973})}\BibitemShut {NoStop}%
\bibitem [{\citenamefont {Catling}\ and\ \citenamefont
  {Kasting}(2017)}]{CatlingKasting2017}%
  \BibitemOpen
  \bibfield  {author} {\bibinfo {author} {\bibfnamefont {D.~C.}\ \bibnamefont
  {Catling}}\ and\ \bibinfo {author} {\bibfnamefont {J.~F.}\ \bibnamefont
  {Kasting}},\ }\href {https://doi.org/10.1017/9781139020558} {\emph {\bibinfo
  {title} {{Atmospheric Evolution on Inhabited and Lifeless Worlds}}}}\
  (\bibinfo  {publisher} {Cambridge University Press},\ \bibinfo {year}
  {2017})\ \bibinfo {note} {eISBN: 9781139020558}\BibitemShut {NoStop}%
\bibitem [{\citenamefont {Forget}\ \emph {et~al.}(2013)\citenamefont {Forget},
  \citenamefont {Wordsworth}, \citenamefont {Millour}, \citenamefont
  {Madeleine}, \citenamefont {Kerber}, \citenamefont {Leconte}, \citenamefont
  {Marcq},\ and\ \citenamefont {Haberle}}]{Forget2013EarlyMarsCO2}%
  \BibitemOpen
  \bibfield  {author} {\bibinfo {author} {\bibfnamefont {F.}~\bibnamefont
  {Forget}}, \bibinfo {author} {\bibfnamefont {R.}~\bibnamefont {Wordsworth}},
  \bibinfo {author} {\bibfnamefont {E.}~\bibnamefont {Millour}}, \bibinfo
  {author} {\bibfnamefont {J.-B.}\ \bibnamefont {Madeleine}}, \bibinfo {author}
  {\bibfnamefont {L.}~\bibnamefont {Kerber}}, \bibinfo {author} {\bibfnamefont
  {J.}~\bibnamefont {Leconte}}, \bibinfo {author} {\bibfnamefont
  {E.}~\bibnamefont {Marcq}},\ and\ \bibinfo {author} {\bibfnamefont {R.~M.}\
  \bibnamefont {Haberle}},\ }\bibfield  {title} {\bibinfo {title} {{3D
  modelling of the early martian climate under a denser {CO$_2$} atmosphere:
  Temperatures and {CO$_2$} ice clouds}},\ }\href
  {https://doi.org/10.1016/j.icarus.2012.10.019} {\bibfield  {journal}
  {\bibinfo  {journal} {Icarus}\ }\textbf {\bibinfo {volume} {222}},\ \bibinfo
  {pages} {81} (\bibinfo {year} {2013})}\BibitemShut {NoStop}%
\bibitem [{\citenamefont {Taylor}(2001)}]{Taylor2001Worldhouse}%
  \BibitemOpen
  \bibfield  {author} {\bibinfo {author} {\bibfnamefont {R.~L.~S.}\
  \bibnamefont {Taylor}},\ }\bibfield  {title} {\bibinfo {title} {{The Mars
  Atmosphere Problem: Paraterraforming --- The Worldhouse Solution}},\
  }\href@noop {} {\bibfield  {journal} {\bibinfo  {journal} {JBIS}\ }\textbf
  {\bibinfo {volume} {54}},\ \bibinfo {pages} {236} (\bibinfo {year}
  {2001})}\BibitemShut {NoStop}%
\bibitem [{\citenamefont {Rzymski}\ \emph {et~al.}(2024)\citenamefont
  {Rzymski}, \citenamefont {Losiak}, \citenamefont {Heinz}, \citenamefont
  {Szukalska}, \citenamefont {Florek}, \citenamefont {Poniedzia{\l}ek},
  \citenamefont {Kaczmarek},\ and\ \citenamefont
  {Schulze-Makuch}}]{Rzymski2024Perchlorates}%
  \BibitemOpen
  \bibfield  {author} {\bibinfo {author} {\bibfnamefont {P.}~\bibnamefont
  {Rzymski}}, \bibinfo {author} {\bibfnamefont {A.}~\bibnamefont {Losiak}},
  \bibinfo {author} {\bibfnamefont {J.}~\bibnamefont {Heinz}}, \bibinfo
  {author} {\bibfnamefont {M.}~\bibnamefont {Szukalska}}, \bibinfo {author}
  {\bibfnamefont {E.}~\bibnamefont {Florek}}, \bibinfo {author} {\bibfnamefont
  {B.}~\bibnamefont {Poniedzia{\l}ek}}, \bibinfo {author} {\bibfnamefont
  {{\L}.}~\bibnamefont {Kaczmarek}},\ and\ \bibinfo {author} {\bibfnamefont
  {D.}~\bibnamefont {Schulze-Makuch}},\ }\bibfield  {title} {\bibinfo {title}
  {{Perchlorates on Mars: Occurrence and implications for putative life on the
  Red Planet}},\ }\href {https://doi.org/10.1016/j.icarus.2024.116246}
  {\bibfield  {journal} {\bibinfo  {journal} {Icarus}\ }\textbf {\bibinfo
  {volume} {421}},\ \bibinfo {pages} {116246} (\bibinfo {year}
  {2024})}\BibitemShut {NoStop}%
\bibitem [{\citenamefont {Wadsworth}\ and\ \citenamefont
  {Cockell}(2017)}]{Wadsworth2017Perchlorates}%
  \BibitemOpen
  \bibfield  {author} {\bibinfo {author} {\bibfnamefont {J.}~\bibnamefont
  {Wadsworth}}\ and\ \bibinfo {author} {\bibfnamefont {C.~S.}\ \bibnamefont
  {Cockell}},\ }\bibfield  {title} {\bibinfo {title} {{Perchlorates on Mars
  enhance the bacteriocidal effects of {UV} light}},\ }\href
  {https://doi.org/10.1038/s41598-017-04910-3} {\bibfield  {journal} {\bibinfo
  {journal} {Scientific Reports}\ }\textbf {\bibinfo {volume} {7}},\ \bibinfo
  {pages} {4662} (\bibinfo {year} {2017})}\BibitemShut {NoStop}%
\bibitem [{\citenamefont {{United Nations Environment Programme}}\ and\
  \citenamefont {{International Resource Panel}}(2024)}]{UNEP_GRO2024}%
  \BibitemOpen
  \bibfield  {author} {\bibinfo {author} {\bibnamefont {{United Nations
  Environment Programme}}}\ and\ \bibinfo {author} {\bibnamefont
  {{International Resource Panel}}},\ }\href
  {https://www.unep.org/resources/Global-Resource-Outlook-2024} {\emph
  {\bibinfo {title} {Global Resources Outlook 2024}}},\ \bibinfo {type} {Tech.
  Rep.}\ (\bibinfo  {institution} {UNEP / International Resource Panel},\
  \bibinfo {address} {Nairobi, Kenya},\ \bibinfo {year} {2024})\BibitemShut
  {NoStop}%
\bibitem [{\citenamefont {{World Steel
  Association}}(2024)}]{worldsteel_figures_2024}%
  \BibitemOpen
  \bibfield  {author} {\bibinfo {author} {\bibnamefont {{World Steel
  Association}}},\ }\href
  {https://worldsteel.org/wp-content/uploads/World-Steel-in-Figures-2024.pdf}
  {\emph {\bibinfo {title} {{World Steel in Figures 2024}}}},\ \bibinfo {type}
  {Tech. Rep.}\ (\bibinfo  {institution} {World Steel Association},\ \bibinfo
  {address} {Brussels, Belgium},\ \bibinfo {year} {2024})\BibitemShut {NoStop}%
\bibitem [{\citenamefont {{Energy Institute}}(2024)}]{EI_StatReview_2024}%
  \BibitemOpen
  \bibfield  {author} {\bibinfo {author} {\bibnamefont {{Energy Institute}}},\
  }\href
  {https://assets.kpmg.com/content/dam/kpmg/pe/Campa%C3%B1as-2025/enero/Statistical-Review-of-World-Energy.pdf}
  {\emph {\bibinfo {title} {{Statistical Review of World Energy 2024}}}},\
  \bibinfo {type} {Tech. Rep.}\ (\bibinfo  {institution} {Energy Institute},\
  \bibinfo {year} {2024})\BibitemShut {NoStop}%
\end{thebibliography}
%

\end{document}